\documentclass[linenumbers, twocolumn]{aastex631}

\shorttitle{molecular clouds in G60 region }
\shortauthors{Chunxue Li et al.}
\graphicspath{{./}{figures/}}

\begin{document}

\title{Molecular Clouds in the Galactic Plane from $l$ = [59.75$^\circ$, 74.75$^\circ$] and $b$ = [$-$5.25$^\circ$, +5.25$^\circ$]}

\author[0000-0002-8409-7908]{Chunxue Li}
\affiliation{Purple Mountain Observatory and Key Laboratory of Radio Astronomy, Chinese Academy of Sciences, 10 Yuanhua Road, Qixia District, Nanjing 210033, China}
\affiliation{School of Astronomy and Space Science, University of Science and Technology of China, 96 Jinzhai Road, Hefei 230026, China}

\author[0000-0003-0746-7968]{Hongchi Wang}
\affiliation{Purple Mountain Observatory and Key Laboratory of Radio Astronomy, Chinese Academy of Sciences, 10 Yuanhua Road, Qixia District, Nanjing 210033, China}
\affiliation{School of Astronomy and Space Science, University of Science and Technology of China, 96 Jinzhai Road, Hefei 230026,  China}

\correspondingauthor{Hongchi Wang}
\email{hcwang@pmo.ac.cn}

\author[0000-0002-8051-5228]{Yuehui Ma}
\affiliation{Purple Mountain Observatory and Key Laboratory of Radio Astronomy, Chinese Academy of Sciences, 10 Yuanhua Road, Qixia District, Nanjing 210033, China}

\author[0000-0001-8926-2215]{Lianghao Lin}
\affiliation{Max-Planck-Institut f\"ur Radioastronomie, Auf dem H\"ugel 69, 53121 Bonn, Germany\\}

\author[0000-0002-0197-470X]{Yang Su}
\affiliation{Purple Mountain Observatory and Key Laboratory of Radio Astronomy, Chinese Academy of Sciences, 10 Yuanhua Road, Qixia District, Nanjing 210033, China}

\author[0000-0003-2218-3437]{Chong Li}
\affiliation{School of Astronomy and Space Science, Nanjing University, 163 Xianlin Avenue, Nanjing 210023, China \\}
\affiliation{Key Laboratory of Modern Astronomy and Astrophysics (Nanjing University), Ministry of Education, Nanjing 210023, China\\}

\author[0000-0002-3904-1622]{Yan Sun}
\affiliation{Purple Mountain Observatory and Key Laboratory of Radio Astronomy, Chinese Academy of Sciences, 10 Yuanhua Road, Qixia District, Nanjing 210033, China}

\author[0000-0003-2418-3350]{Xin Zhou}
\affiliation{Purple Mountain Observatory and Key Laboratory of Radio Astronomy, Chinese Academy of Sciences, 10 Yuanhua Road, Qixia District, Nanjing 210033, China}

\author[0000-0001-7768-7320]{Ji Yang}
\affiliation{Purple Mountain Observatory and Key Laboratory of Radio Astronomy, Chinese Academy of Sciences, 10 Yuanhua Road, Qixia District, Nanjing 210033, China}






\begin{abstract}

In this paper we present the distribution of molecular gas in the  Milky Way Galactic plane from $l$ = [59.75, 74.75]$^{\circ}$ and $b$ = [${-}$5.25, +5.25]$^{\circ}$, using the MWISP $^{12}$CO/$^{13}$CO/$\rm {C}^{18}{O}$ emission line data. The molecular gas in  this region can be mainly attributed to the Local spur, Local arm, Perseus arm, and Outer arm. Statistics of the physical properties of the molecular gas in each arm, such as excitation temperature, optical depth, and column density, are presented. Using the DBSCAN algorithm, we identified 15 extremely distant molecular clouds with kinematic distances of 14.72$-$17.77 kpc and masses of 363$-$520 M$_{\odot}$, which we find could be part of the Outer  Scutum-Centaurus (OSC) arm  identified by \cite{2011ApJ...734L..24D} and \cite{2015ApJ...798L..27S}. It is also possible that, 12 of these 15 extremely distant molecular clouds constitute an independent structure between the Outer and the OSC arms or a spur. There exist two Gaussian components in the vertical distribution of the molecular gas in the Perseus spiral arm. These two Gaussian components  correspond to two giant filaments parallel to the Galactic plane. We find an upward warping of the molecular gas in the Outer spiral arm with a displacement of around 270 pc with respect to the Galactic mid-plane.

\end{abstract}



\section{Introduction} \label{sec:intro}
Star formation occurs in molecular clouds, the coldest and densest regions of the interstellar medium (ISM). It is challenging to observe H$_2$ molecules directly via the rotational transition lines  because the H$_2$ molecules do not have permanent dipole moments. Therefore, CO, the second most abundant molecule next to H$_2$ in the ISM, becomes the most common tracer for molecular clouds \citep{1970ApJ...161L..43W}. Among the three common isotopologues, $^{12}$CO can effectively probe the diffuse envelopes of molecular clouds of densities of $\sim10^2$ cm$^{-3}$, while  $^{13}$CO  can trace regions with intermediate densities of a few 10$^2$ to 10$^3$ cm$^{-3}$. In contrast, C$^{18}$O is a good tracer of dense molecular cores with densities as high as 10$^4$ cm$^{-3}$ (e.g., \citealt{2017PASJ...69...78U}). So far, our understanding of the physical properties and spatial distribution of molecular clouds in the Milky Way mainly comes from large-scale CO surveys, such as the early survey by \cite{1987ApJ...319..730S}, the CfA-Chile survey \citep{2001ApJ...547..792D,2016ApJ...822...52R}, the FCRAO Galactic Ring Survey and Outer Galaxy Survey \citep{2001ApJ...551..852H, 2010ApJ...723..492R}, and the SEDIGISM (Structure, Excitation, and Dynamics of the Inner Galactic Interstellar Medium) survey \citep{2021MNRAS.500.3064S}. These surveys helped to build a panoramic view about the Galactic molecular clouds. For example, it is found that the mass of molecular clouds ranges from  $\sim$10$^{1}$ to 10$^{6}$ M$_{\odot}$, while their sizes lie in the range from 10 to 150 pc, and their number densities are about 50$-$10$^{4}$ cm$^{-3}$. Most of the molecular gas is concentrated in the so-called giant molecular clouds (GMCs) which have mass greater than 10$^{4}$ M$_{\odot}$. The velocity dispersions of clouds are about 1 km s$^{-1}$ \citep{2010ApJ...723..492R}. The mass function of molecular cloud is found to follow a power law, with exponents  within $\sim $ $-$1.4 to $\sim $ $-$2.2 \citep{1993prpl.conf..125B,1998A&A...329..249K,2016ApJ...822...52R}. There also exists a  power law scaling relation between the size of molecular clouds and the line width, which is thought to originate from the turbulent motions of the molecular gas \citep{1981MNRAS.194..809L}.

The ongoing Milky Way Imaging Scroll Painting (MWISP) project provides us  an  opportunity to get a much deeper understanding of  molecular gas distribution in the Milky Way, benefiting from its wide sky coverage (l = [$12\arcdeg, 230\arcdeg$]), multiple spectral lines ($^{12}$CO, $^{13}$CO, and C$^{18}$O), and high-sensitivity (0.5 K per channel of 0.17 km $^{-1}$ for $^{12}$CO) \citep{2019ApJS..240....9S}. The sky area of Galactic longitude from 59.75$^{\circ}$ to 74.75$^{\circ}$ and Galactic latitude from $-$5.25$^{\circ}$ to 5.25$^{\circ}$ (hereafter the ``G60 region''), part of the MWISP survey, was fully observed lately. This region has not been thoroughly studied before. The CfA 1.2 m telescope survey \citep{2001ApJ...547..792D} has observed this region in the $^{12}$CO molecular line, but its spatial resolution is quite low. The Exeter-FCRAO (EXFC) survey  \citep{2016ApJ...818..144R} observed two regions, l = [55, 100]$^\circ$, b = [$-$1.4, 1.9]$^\circ$, and l = [135, 195]$^\circ$, b = [$-$3.6, 3.6]$^\circ$. The distribution in the face-on view of the Milky Way of molecular gas in the first region was presented in Figure 9 of \cite{2016ApJ...818..144R}. This region contains active past and ongoing star formation, as shown by the presence of supernova remnants (SNRs) \citep{2019JApA...40...36G} and \ion{H}{2} regions/candidates \citep{2014ApJS..212....1A}. We focus in this work on the molecular gas distribution and physical properties using the $^{12}$CO, $^{13}$CO, and the C$^{18}$O J = 1$-$0 line data from the MWISP project. The high-sensitivity of the MWISP data also provides an opportunity to identify extremely distant molecular clouds (e.g., \citealt{2015ApJ...798L..27S}).  \cite{2011ApJ...734L..24D} found a segment of the Outer Scutum-Centaurus arm  at a Galactocentric distance of about 15 kpc using the data from the CfA 1.2 m telescope survey.  We will search for distant molecular clouds corresponding to the OSC arm in the part of the sky investigated in this work.

\begin{figure*}[h]
\plotone{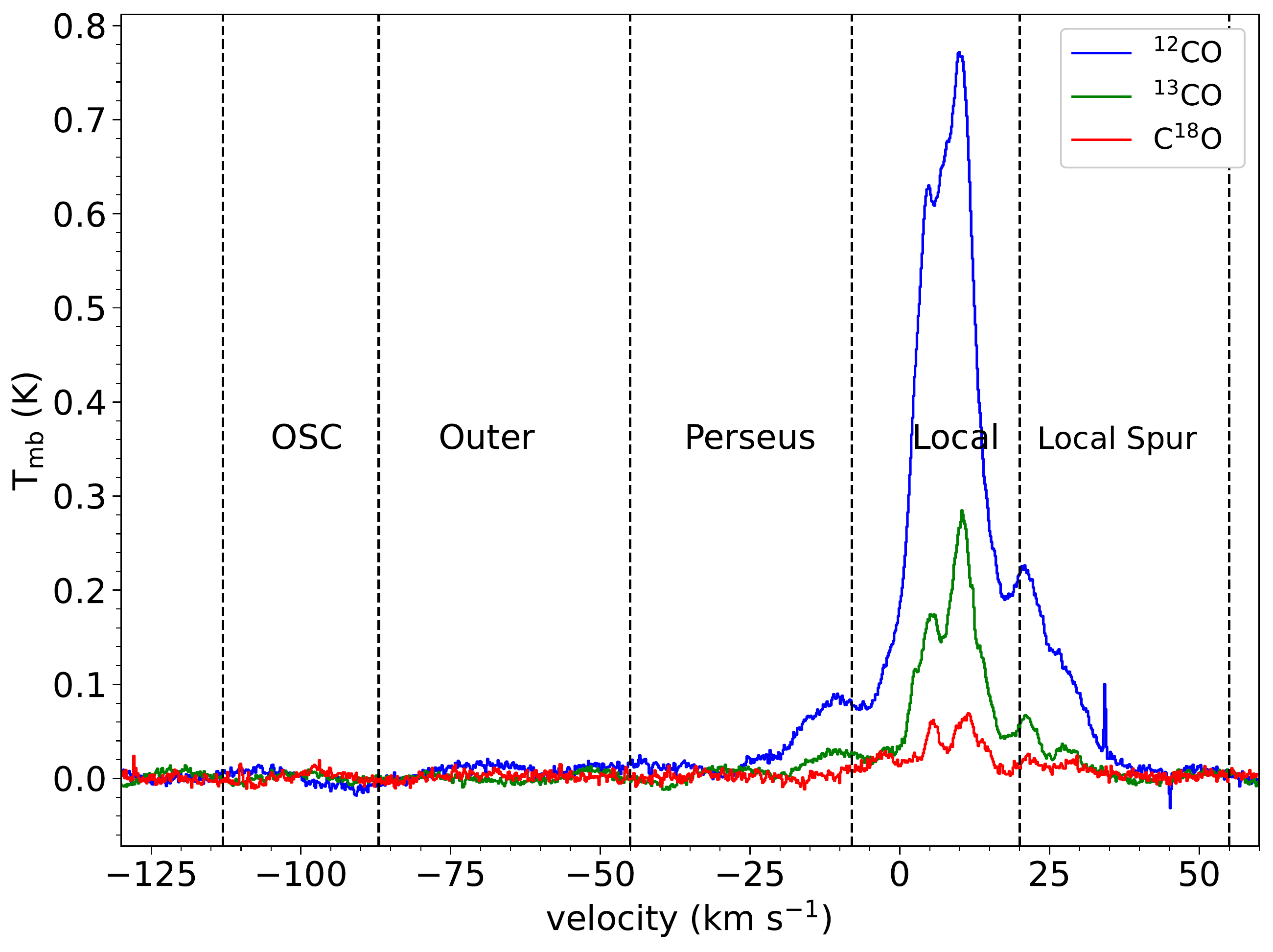}
\caption{Average spectra of the G60 region, with blue, green, and red showing the $^{12}$CO, $^{13}$CO, and C$^{18}$O J = 1${-}$0 line emission, respectively. The $^{12}$CO spectrum contains spikes at 35 km s$^{-1}$ and 45 km s$^{-1}$, the $^{13}$CO spectrum at ${-}$74 km s$^{-1}$, and the C$^{18}$O spectrum  at ${-}$127 and ${-}$109 km s$^{-1}$. These spikes are caused by bad channels. According to \cite{2019ApJ...885..131R}, spiral arms, from near to far, in the portion of the Galactic plane investigated in this work, are the Local arm, Local Spur, Perseus arm, Outer arm, and OSC arm. The velocity ranges of these arms are marked with vertical dashed lines. The $^{12}$CO, $^{13}$CO, and C$^{18}$O spectrum are averaged on the positions that show $^{12}$CO, $^{13}$CO, and C$^{18}$O emission brightness above two times their RMS noise level in at least five, four, and three contiguous channels, respectively.
\label{fig:fig1}}
\end{figure*}

The paper is organized as follows. The observations are presented in Section \ref{sec:Obs}, and the results are given in Section \ref{sec:results}. The discussion about the vertical distribution of molecular gas in the direction vertical to the Galactic plane is presented in Section \ref{sec:Vertical distribution}, and we give our summary in Section \ref{sec:Summary}.

\section{Observations} \label{sec:Obs}
\subsection{CO data}
The observation of the G60 region is part of the MWISP survey and was taken from 2016 to 2020, using the 13.7 m millimeter-wavelength telescope of the Purple Mountain Observatory (PMO-13.7m) at Delingha, China. The telescope is equipped with the Superconducting Spectroscopic Array Receiver (SSAR), with 3$\times$3 beams. The $^{12}$CO, $^{13}$CO, and C$^{18}$O (J = 1$-$0) line emission were observed simultaneously with the $^{12}$CO (J = 1$-$0) emission being contained in the upper sideband, while the  $^{13}$CO (J = 1$-$0) and C$^{18}$O (J = 1$-$0) emission in the lower sideband. The back end for each sideband is a fast Fourier transform spectrometer (FFTS) which has a total bandwidth of 1 GHz and 16,384 frequency channels, providing a spectral resolution of 61 kHz, corresponding to velocity resolution of 0.158 km s$^{-1}$ at 115 GHz and 0.167 km s$^{-1}$ at 110 GHz, respectively \citep{2012ITTST...2..593S}.

\begin{figure*}[htb!]
\gridline{\fig{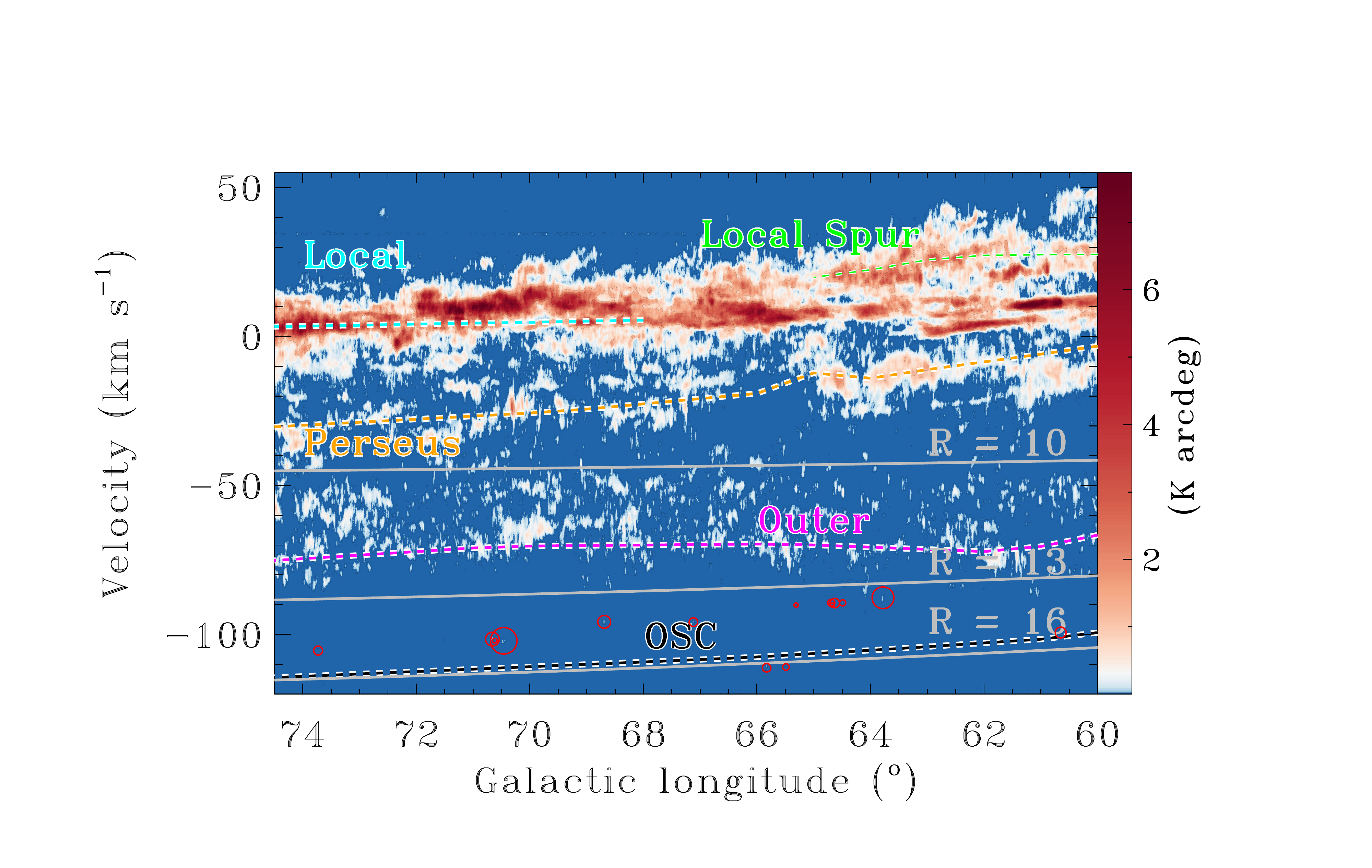}{0.80\textwidth}{(a)}}
\gridline{\fig{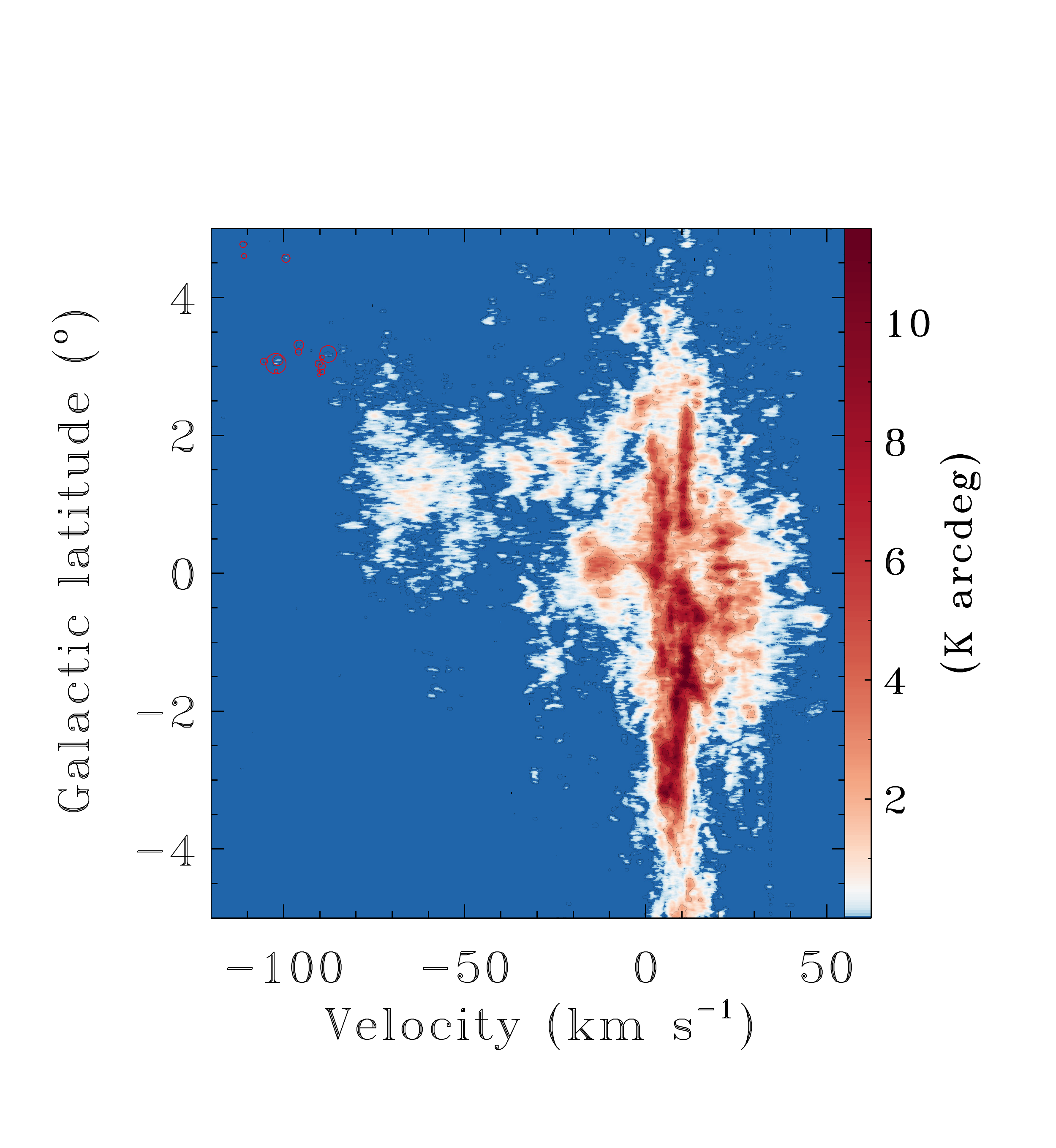}{0.55\textwidth}{(b)}}            
\caption{(a) Longitude-velocity diagram of the $^{12}$CO emission, integrated within b = $-$5$^\circ$ to 5$^\circ$. The colored dashed lines represent the l-v position of the Local Spur, Local, Perseus, Outer, and Outer-Scutum-Centaurus (OSC) arms from \cite{2019ApJ...885..131R}. Constant Galactocentric distances of 10, 13, and 16 kpc are indicated with grey lines. These lines are calculated with the rotation curve of \cite{2019ApJ...885..131R}. (b) Latitude-velocity diagram of the $^{12}$CO emission, integrated within l = 60$^\circ$ to 74.5$^\circ$. The red circles mark molecular clouds located in the OSC arm (for details, see Section \ref{subsec:OSC Cloud}). 
\label{fig:fig2}}
\end{figure*}

\begin{figure*}[ht!]
\plotone{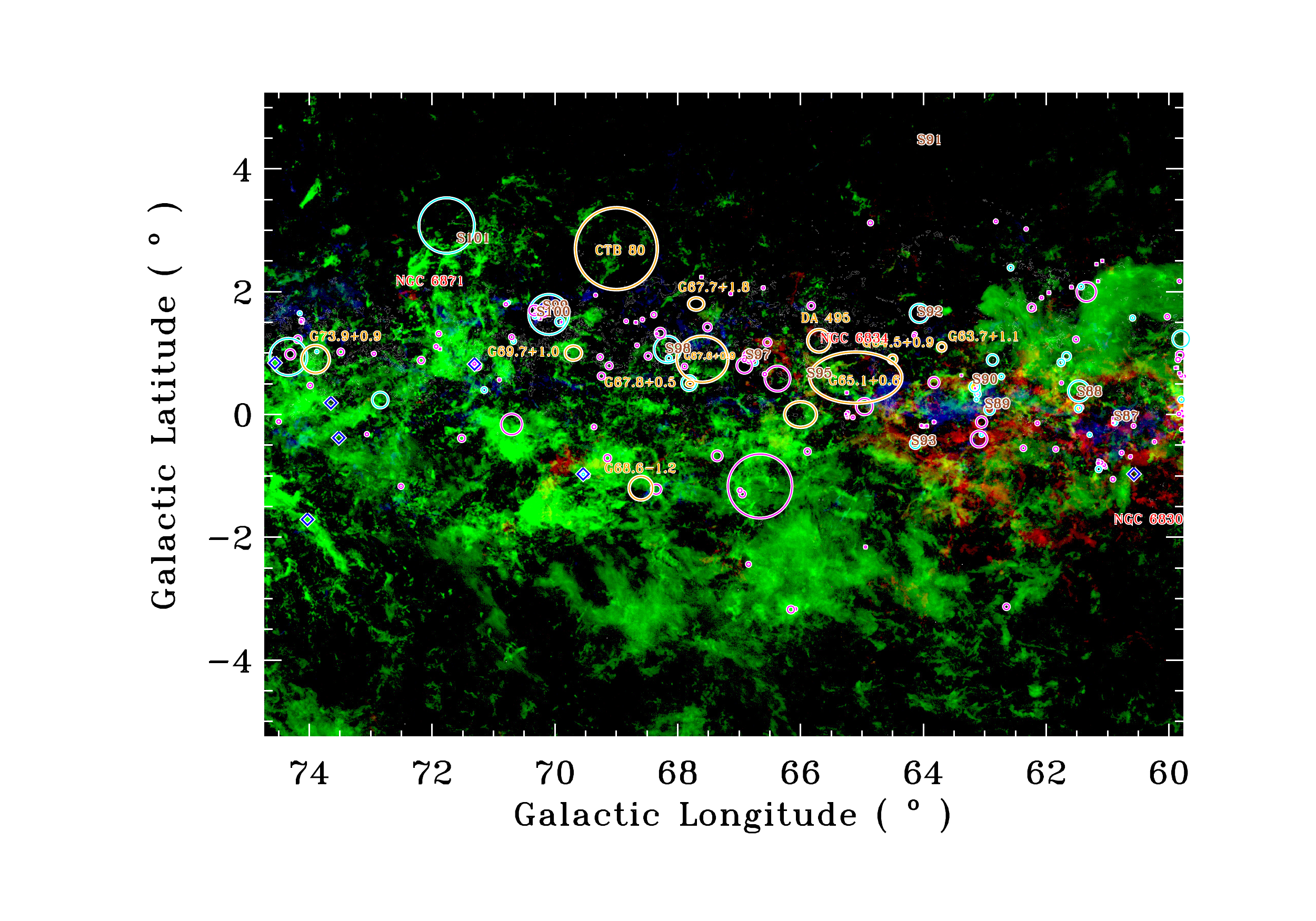}
\caption{Color-coded intensity maps of $^{12}$CO emission in different velocity ranges. The integrated intensity of $^{12}$CO emission in velocity ranges from 20 to 55, $-$8 to 20, and $-$45 to $-$8 km s$^{-1}$ are shown by red, green, and blue colors, respectively. White contours represent intensity integrated in [$-$87,$-$45] km s$^{-1}$. The orange ellipses show the locations of known SNRs \citep{2019JApA...40...36G} in this region. The magenta circles and the cyan circles show the locations of the \ion{H}{2} region and \ion{H}{2} candidates from the WISE catalog \citep{2014ApJS..212....1A}. The names of the \ion{H}{2} regions in the Sharpless catalog \citep{1959ApJS....4..257S} are represented by brown letters, and the names of the SNRs are denoted by orange. The masers \citep{2019ApJ...885..131R} are indicated by blue diamonds. Three active star-forming regions, NGC 6830, NGC 6834, and NGC 6871 are shown in red letters.
 \label{fig:fig3}}
\end{figure*}

The observations cover the sky region of Galactic longitude of l = [59.75, 74.75]$^{\circ}$ and Galactic latitude of b = [$-$5.25, 5.25]$^{\circ}$, and were made in the position-switch on-the-fly (OTF) mode with a typical scan rate and dump time of 50 (or 75) $\arcsec$/s and 0.3 (or 0.2) s, respectively. The surveyed region is divided into individual cells of 30$\arcmin$ $\times$ 30$\arcmin$. Each cell was scanned at least twice, along the directions of Galactic longitude and Galactic latitude, to reduce scanning effects. The pointing accuracy of the telescope is about 5$\arcsec$ in all observational epochs, and the half-power beam widths are about 55$\arcsec$ and 52$\arcsec$ at 110 GHz and 115 GHz, respectively. The antenna temperature ($T\rm _{A}^{\ast}$) is corrected to the main beam brightness temperature (T$\rm _{mb}$) using T$\rm _{mb}$ = $T\rm_{A}^{\ast}$ / $\eta\rm _{mb}$ during the observation, where the value of the main beam efficiency ($\eta\rm _{mb}$) can be obtained from the annual status report\footnote{http://english.dlh.pmo.cas.cn/fs/} of the telescope. All the raw data were reduced using the GILDAS-CLASS\footnote{http://www.iram.fr/IRAMFR/GILDAS/} package. The final data were regridded into 30$\arcsec$ $\times$ 30$\arcsec$ pixels in the directions of Galactic longitude and Galactic latitude and converted to FITS files. The median RMS noise level of the molecular line is  $\sim$0.5 K per channel at the $^{12}$CO wavelength and $\sim$0.3 K per channel at the $^{13}$CO and C$^{18}$O wavelengths. For more details about the MWISP project, please refer to \cite{2019ApJS..240....9S}.

\section{Results} \label{sec:results}

\subsection{Average Spectra}
\label{subsec:average spectra}

Figure \ref{fig:fig1} shows the average spectra of the $^{12}$CO, $^{13}$CO, and C$^{18}$O J = 1${-}$0 line emission in the whole G60 region.  The $^{12}$CO spectrum shows strong emission within the velocity range from $\sim-$30 to 40 km s$^{-1}$, while the $^{13}$CO emission is mainly contained within a narrower velocity range, i.e., [$-$25, 25] km s$^{-1}$. The average spectrum of  C$^{18}$O shows emission between $-$25 and 40 km s$^{-1}$. The spikes at 35 km s$^{-1}$ and 45 km s$^{-1}$ in the $^{12}$CO spectrum, $-$74 km s$^{-1}$ in $^{13}$CO, ${-}$127 and ${-}$109 km s$^{-1}$ in C$^{18}$O are all due to bad channels. Only spectra with peak intensities above 3$\sigma_{RMS}$ and containing adequate number of contiguous channels that have intensities above 2$\sigma_{RMS}$ are considered as the spectra of detection. For $^{12}$CO, $^{13}$CO, and C$^{18}$O, this number of contiguous channels is five, four, and three, respectively.

\begin{figure*}[ht]
\gridline{\fig{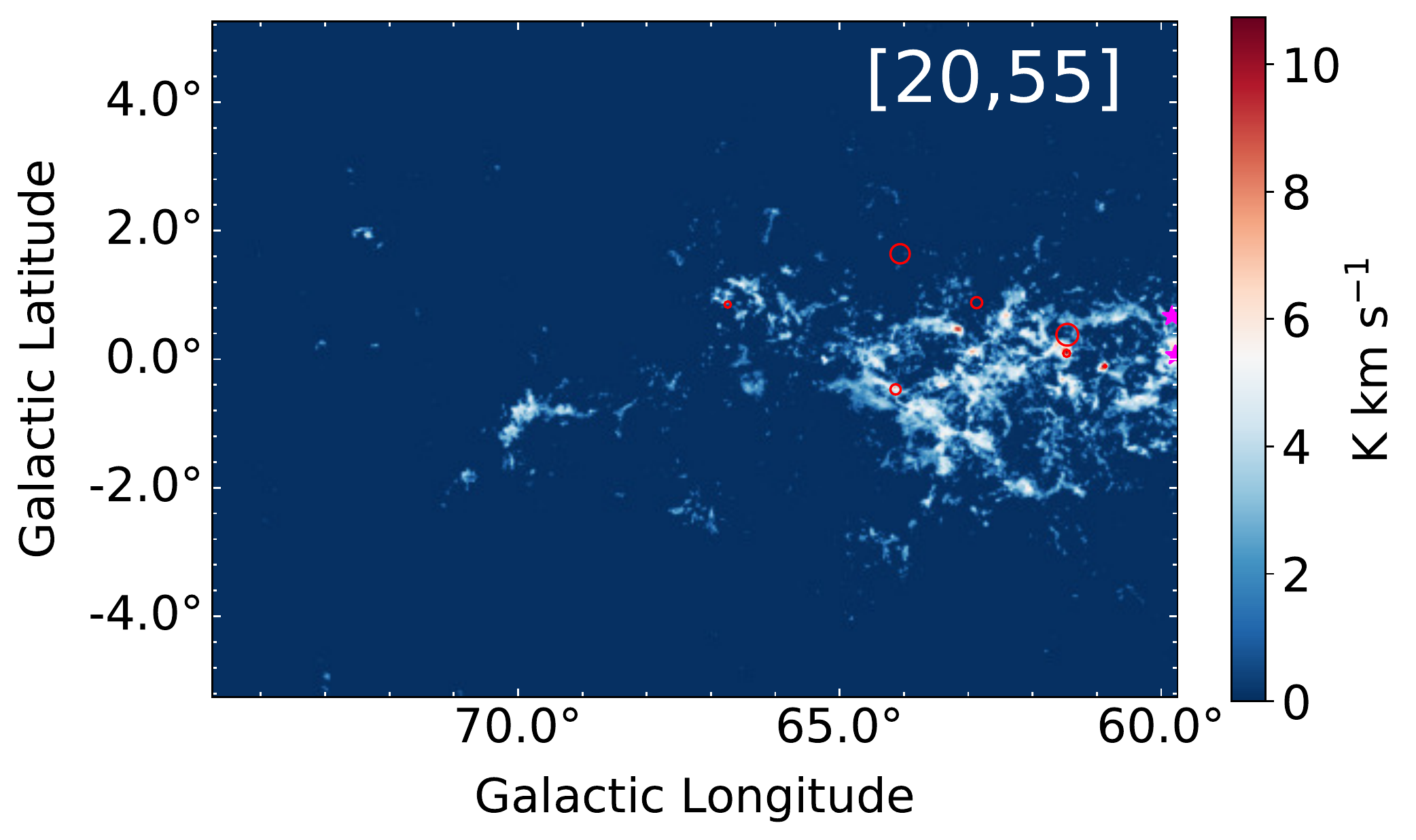}{0.47\textwidth}{(a)}
          \fig{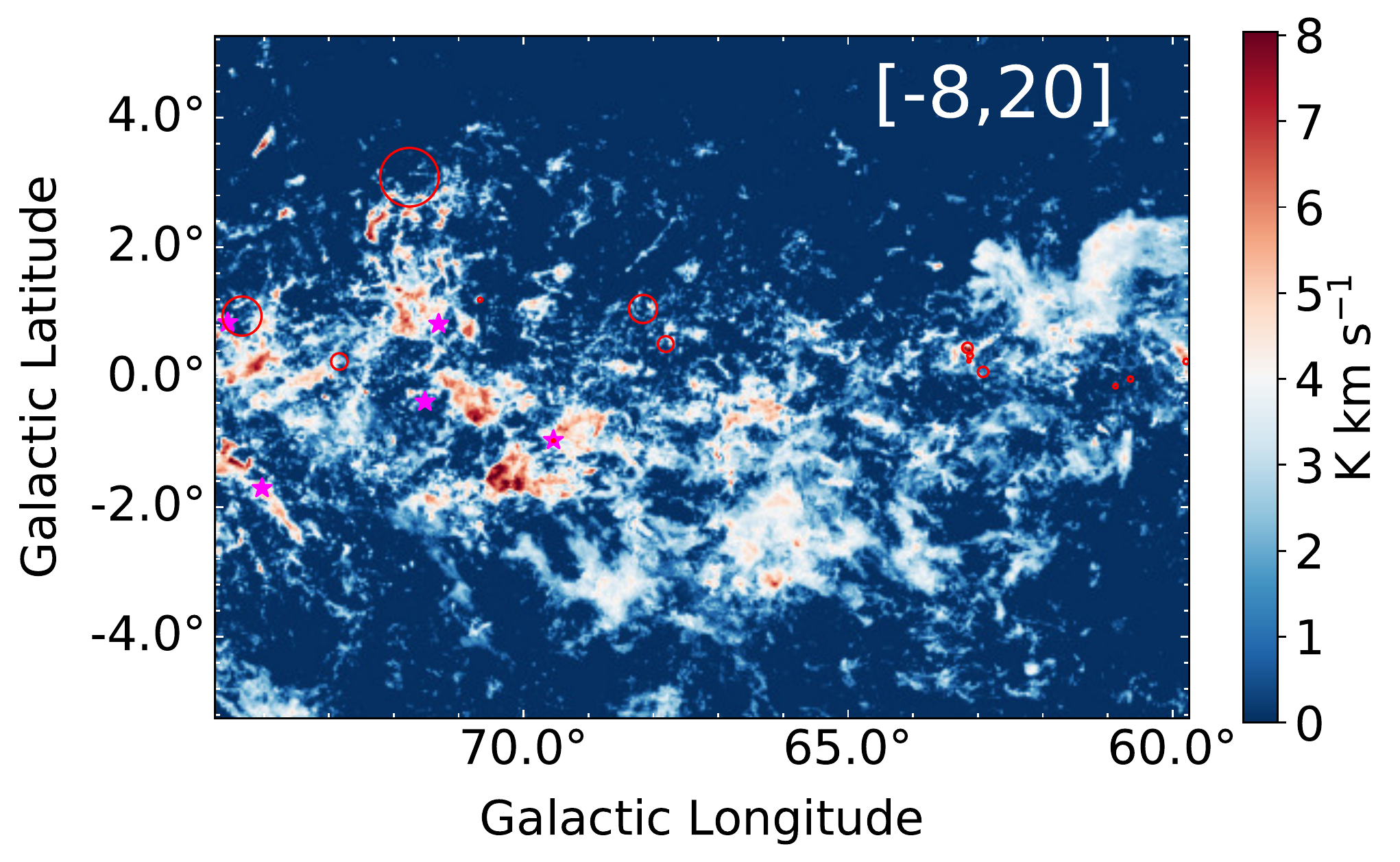}{0.47\textwidth}{(b)}
          }
          
\gridline{\fig{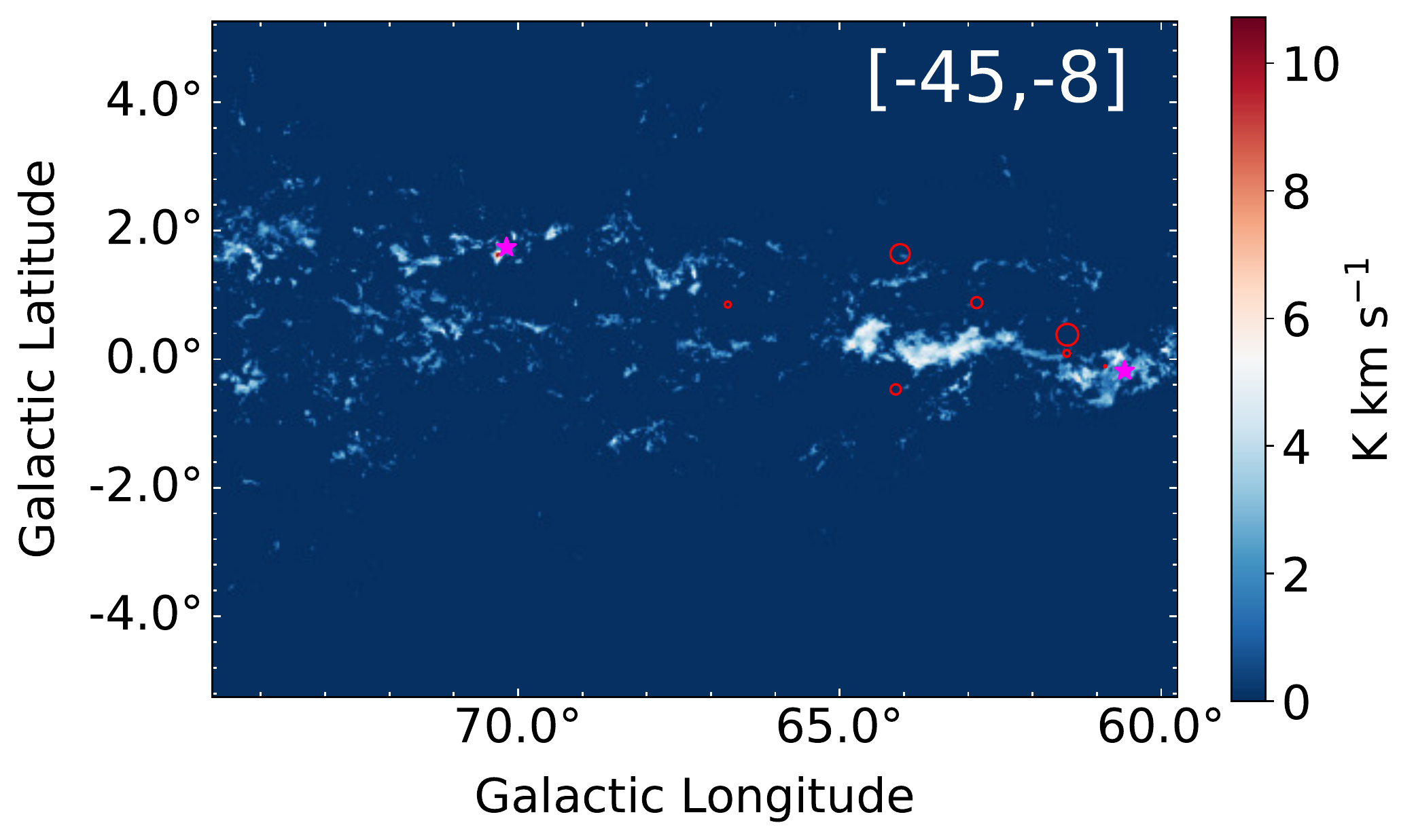}{0.47\textwidth}{(c)}
          \fig{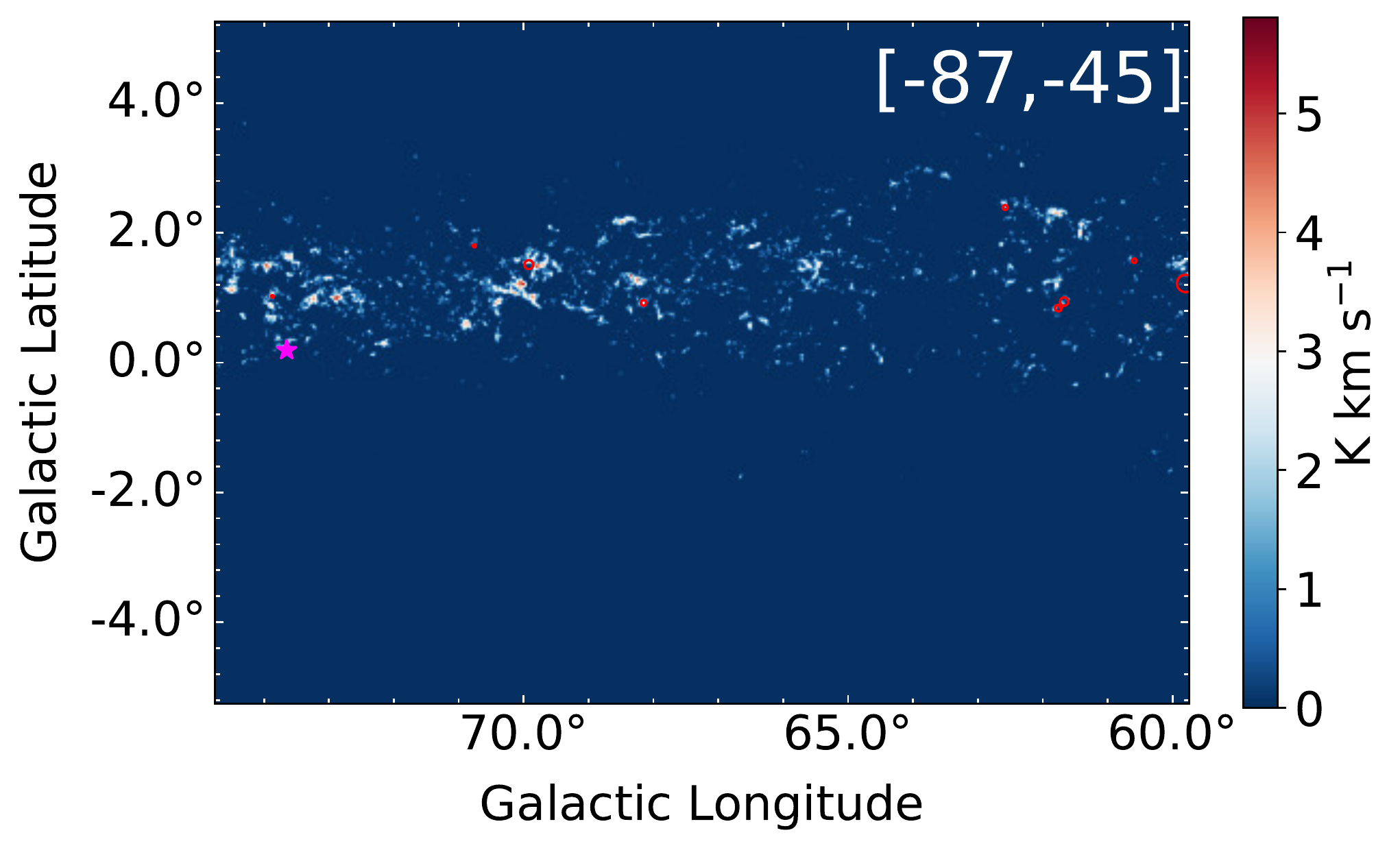}{0.47\textwidth}{(d)}
          }
\caption{Integrated intensity maps of $^{12}$CO emission in the (a) Local Spur, (b) Local arm, (c) Perseus arm, and (d) Outer arm. The magenta pentagrams mark the positions of the associated masers from \cite{2019ApJ...885..131R}. The velocity range of each arm is given in the top right corner of each panel. The red circles indicate the locations of \ion{H}{2} region and candidates in the corresponding velocity range. }
\label{fig:fig4}
\end{figure*}

\clearpage
\subsection{Position-Velocity Distribution and Velocity Ranges of Spiral Arms}
The l-v and b-v diagrams of the $^{12}$CO line emission are shown in Figure \ref{fig:fig2}. The integrated range of the l-v diagram is between b = $-$5$^\circ$ and b = 5$^\circ$, while that for the b-v diagram is between l = 60$^\circ$ and l = 74.5$^\circ$. In the two diagrams, the noise has been clipped using the criteria mentioned in Section \ref{subsec:average spectra}. The 34 km s$^{-1}$ bad channels show as thin horizontal and vertical lines in the l-v and b-v diagrams, respectively. The traces of the Galactic spiral arms, adopted from \cite{2019ApJ...885..131R}, are marked as dashed lines in the l-v map. Based on the rotation curve from \cite{2019ApJ...885..131R}, we calculated Galactocentric distances (R). The lines for R = 10, 13, and 16 kpc are shown in the l-v diagram. The molecular clouds located in the OSC arm are indicated by red circles (for details, see Section \ref{subsec:OSC Cloud}). According to the l-v distribution and also the average spectra in Figure~\ref{fig:fig1}, the molecular emission contained in velocity intervals of [20, 55], [${-}$8, 20], [${-}$45, ${-}$8], [$-$87, $-$45], [${-}$119, ${-}$87] km s$^{-1}$ is attributed to molecular gas located in the Local Spur, Local, Perseus, Outer, and OSC spiral arms, respectively. From Figure \ref{fig:fig2} (a), we can see that the molecular gas in this section of the Galactic disk well follows the spiral arms or the Local spur. Both the molecular gas and the spiral arms run nearly horizontal, i.e., along the direction of Galactic longitude, with the Local spur, the Perseus arm, and the OSC arm being diagonally oriented slightly. Ideally, we should cut the l-v diagram into diagonal strips to attribute the molecular gas to each arm or the Local spur. However, it is hard to exactly define the boundary between the Local spur and the Local arm, and the one between the Local arm and the Perseus arm. For simplicity, we just used the horizontal cuts as given above. From Figure \ref{fig:fig2} (a), we can see that this provides a clean enough separation for most of the molecular gas, although there are some ambiguity between the Local spur and the Local arm, and between the Local arm and the Perseus arm. Significant $^{12}$CO emission is concentrated in the Local Spur, Local arm, and the Perseus arm. Molecular clouds in the Outer arm exhibit fragmented distribution. As shown in the b-v diagram (Figure~\ref{fig:fig2} b), the molecular emission in the Outer spiral arm is mainly distributed above b = 0\arcdeg, which may be the result of warping in the gas disk of our Galaxy \citep{1957BAN....13..201W,1990A&A...230...21W}. The l-v and b-v diagrams of the $^{13}$CO emission in the G60 region are presented in Figure \ref{fig:figA2} in the Appendix.

\begin{figure*}[ht!]
\plotone{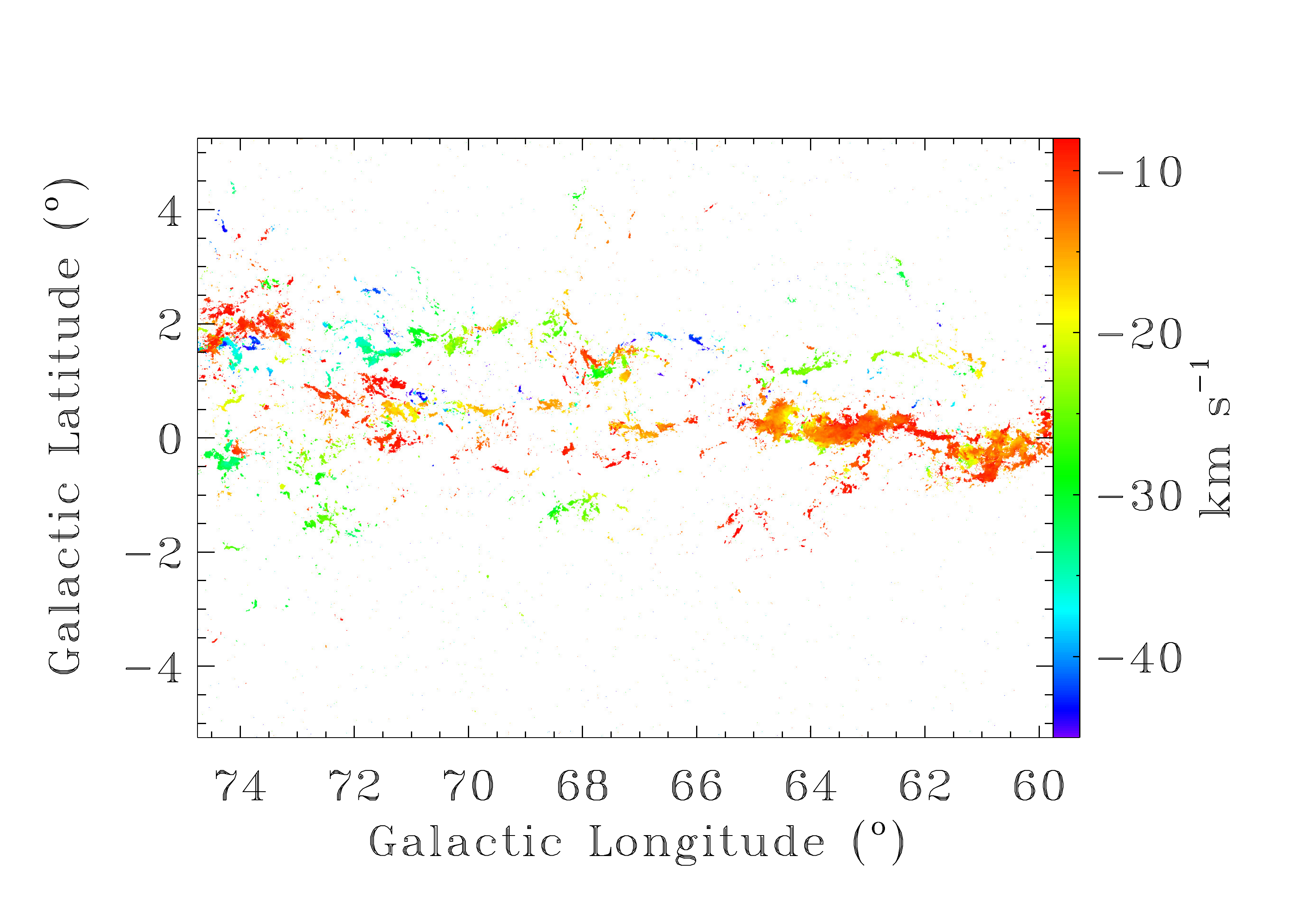}
\caption{Intensity-weighted centroid velocity map of $^{12}$CO emission in the veloctity range between $-$45 and $-$8 km s$^{-1}$.
 \label{fig:U_perseus_m1}}
\end{figure*}

\begin{figure*}[ht!]
\plotone{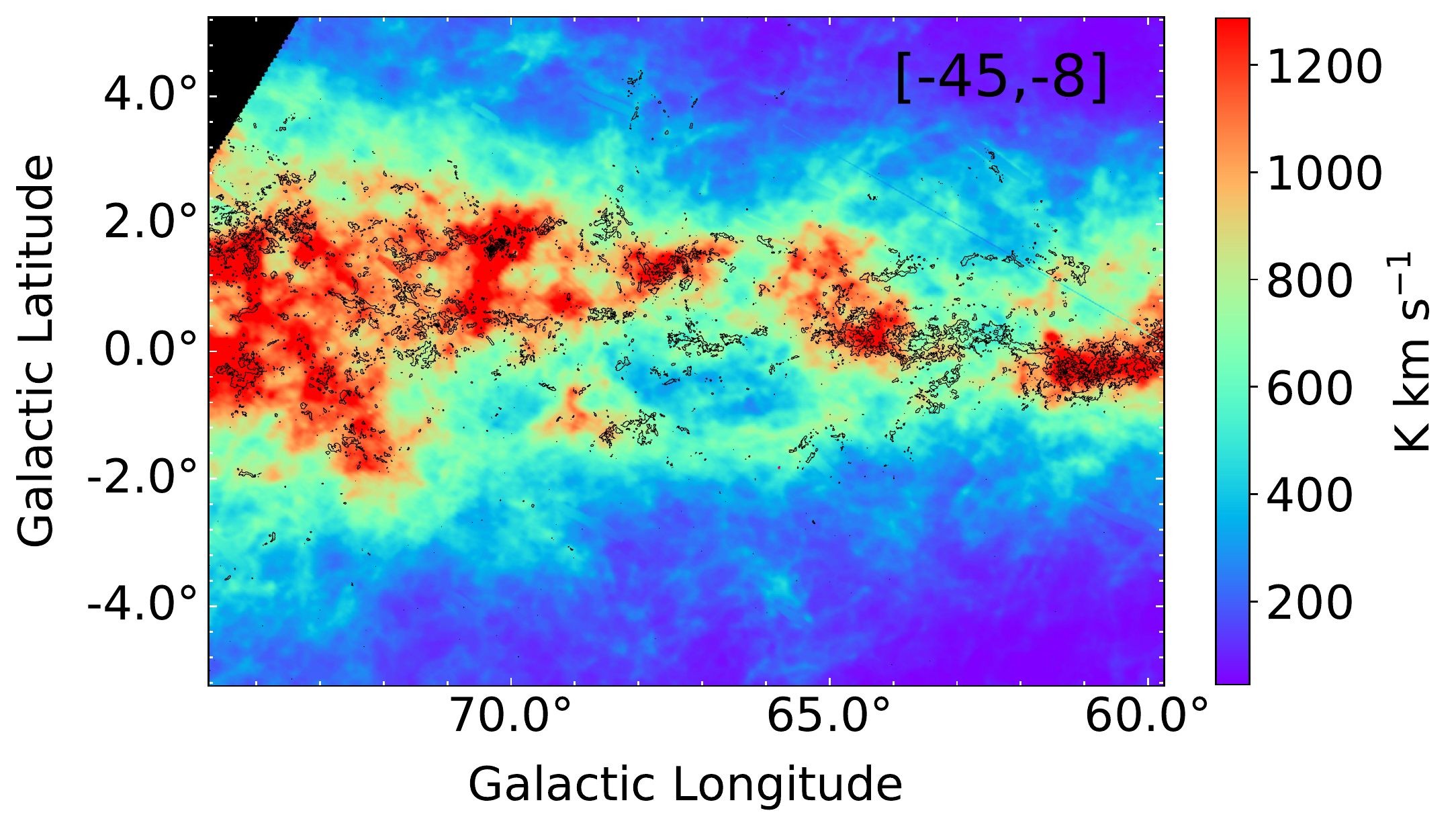}
\caption{\ion{H}{1} integrated intensity map overlaid with the contours of $\rm^{12}$CO emission in the Perseus arm. The integrated velocity range is from ${-}$45 to 8 km s$^{-1}$ for both \ion{H}{1} and $\rm^{12}$CO.
\label{fig:34_HI_CO}}
\end{figure*}

\subsection{Spatial Distribution}
The sky distribution of the molecular gas in the G60 region is presented in Figure \ref{fig:fig3}. Red, green, and blue colors correspond to the $^{12}$CO intensities integrated in the velocity range of the Local Spur, Local arm, and Perseus arm, respectively, and the white contours represent the integrated intensity of molecular gas in the Outer arm. The masers \citep{2019ApJ...885..131R} are marked with blue diamonds. The known SNRs \citep{2019JApA...40...36G} are indicated with orange ellipses. The \ion{H}{2} regions and the \ion{H}{2} region candidates from \cite{2014ApJS..212....1A} are shown with magenta circles and cyan circles, respectively. The three prominent star-forming regions in the G60 region, NGC 6830, NGC 6834, and NGC 6871 are shown with red letters. The names of the SNRs are denoted with orange letters, and the names of \ion{H}{2} regions from the Sharpless catalog \citep{1959ApJS....4..257S} with brown letters. In this figure, the most conspicuous component is the Local arm (green). The molecular gas in the Local Spur is mainly distributed in regions of l${\leq}$ 67$^{\arcdeg}$ and exhibits more clumpiness than gas in the Local arm. The \ion{H}{2} regions and candidates are more spatially coincident with the molecular gas in the Local Spur and the Perseus/Outer arm regions than with the molecular gas in the Local arm. 

For a closer inspection, we individually present the integrated $\rm^{12}$CO emission intensity maps of the Local Spur, Local arm, Perseus arm, and Outer arm in Figure \ref{fig:fig4}. The differences of molecular gas in the distribution and morphology between the Local spur and Local arm are obvious. The most prominent feature of the molecular gas in the Perseus arm is that it can be divided into two nearly parallel layers along the Galactic longitude direction, i.e., a layer at $b$ = 0$^\circ$ and another layer at around $b$ = 2$^\circ$. For the Outer arm, molecular gas is all located above $b$ = 0$^\circ$ and it is concentrated in small compact cores. The masers \citep{2019ApJ...885..131R} associated with the molecular gas in each arm are indicated with magenta pentagrams in Figure \ref{fig:fig4}. The integrated intensity map of the $^{13}$CO emission for each arm is presented in Figure \ref{fig:fig4_13CO} in Appendix. The velocity channel maps of the region for the $\rm^{12}$CO emission are presented in Figures \ref{fig:fig33channel}$-$\ref{fig:fig33channel_3} in the Appendix. As the $\rm^{13}$CO emission is much weaker compared with the $\rm^{12}$CO emission, we do not present the $\rm^{13}$CO velocity channel maps. 

 Figure \ref{fig:U_perseus_m1} presents the intensity-weighted centroid velocity map of the $\rm ^{12}$CO emission from the Perseus arm. As shown in Figure \ref{fig:U_perseus_m1}, the two layers of molecular gas in the Perseus arm located at around b = 0$^\circ$ and b = 2$^\circ$ also exhibit velocity coherence. The layer at b = 2$^\circ$ possesses velocities from $-$30 to $-$20 km s$^{-1}$, while the layer at b = 0$^\circ$ has velocities from $-$20 and $-$10 km s$^{-1}$. These two layers show a systematic velocity difference of about 10 km s$^{-1}$, with the upper layer (b = 2$^\circ$) being blue-shifted with respect to the lower one (b = 0$^\circ$). From the centroid velocity map, we can also see that the lower layer runs up the Galactic plane gradually from the west to the east, and then merges with the upper layer at the position around [74, 2]$^\circ$.
 
 As these two layers possess both good spatial and velocity coherence, we propose that they are two giant molecular filaments. Taking the distance of the Perseus arm to be 7.34 kpc (see Section \ref{subsec:Distance} for the estimation of distance of each arm in this area), the lengths of these two giant molecular filaments are roughly 1.9 kpc, and their separation is roughly 0.12 kpc if we take their average Galactic latitude separation to be 1$^\circ$.
 
 To explore the possible origins of the two parallel filaments, we overlaid in Figure \ref{fig:34_HI_CO} the CO contours to the \ion{H}{1} 21 cm line integrated intensity map over the same velocity range, from ${-}$45 to ${-}$8 km s$^{-1}$. The \ion{H}{1} data used in this work are obtained from the data release 2 of the Galactic Arecibo L-Band Feed Array \ion{H}{1} (GALFA-\ion{H}{1}) survey \citep{2018ApJS..234....2P}. The survey covers the region of 0$\rm^h$ $\leq$ RA $\leq$ 24$\rm^h$ and $-$1$\arcdeg$17$\arcmin$ $\leq$ Dec $\leq$ +37$\arcdeg$57$\arcmin$, i.e., 32\% of the sky. The angular resolution is 4$\arcmin$ and the velocity resolution is 0.184 km s$^{-1}$. The median rms noise is 352 mK per channel. From Figure \ref{fig:34_HI_CO}, we can see that, overall, the molecular gas distribution correlates well with that of the \ion{H}{1} gas. The \ion{H}{1} gas shows a series of cavities between the two parallel giant molecular filaments which have sizes comparable to the separation of the filaments. This fact suggests that the \ion{H}{1} gas cavities may be super bubbles and the two parallel giant molecular filaments may be caused by the compression of the supper bubbles, with the upper filament representing the near-side layer of compressed molecular gas and the lower filament being the far-side one. Surely, more observations of this region, such as radio continuum and OB star clusters, are needed to support this scenario.

\subsection{Detection of Isotopologues in Different Arms}
\label{detec_mask}

\begin{figure*}[htb!]
\gridline{\fig{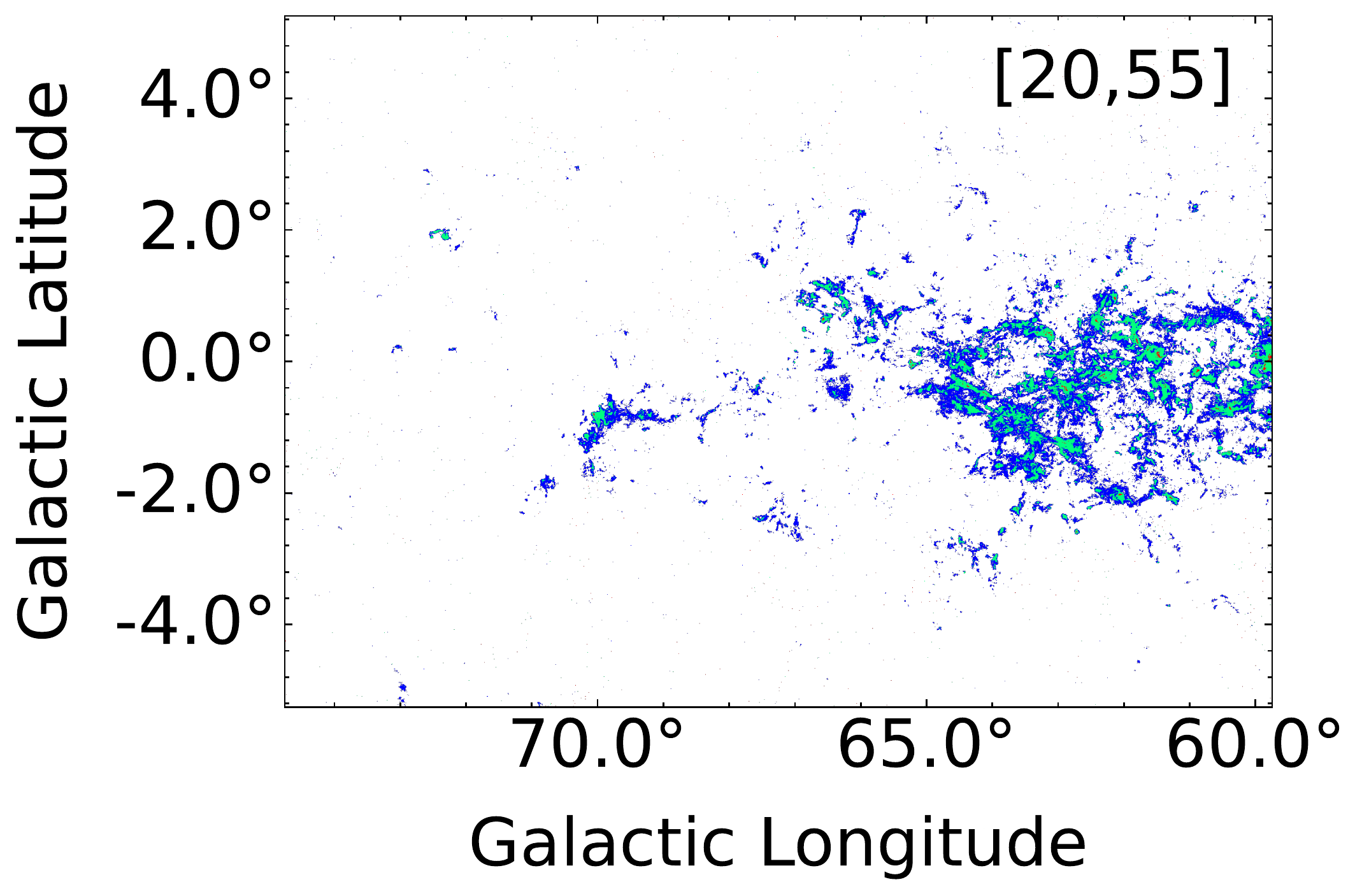}{0.48\textwidth}{(a)}
          \fig{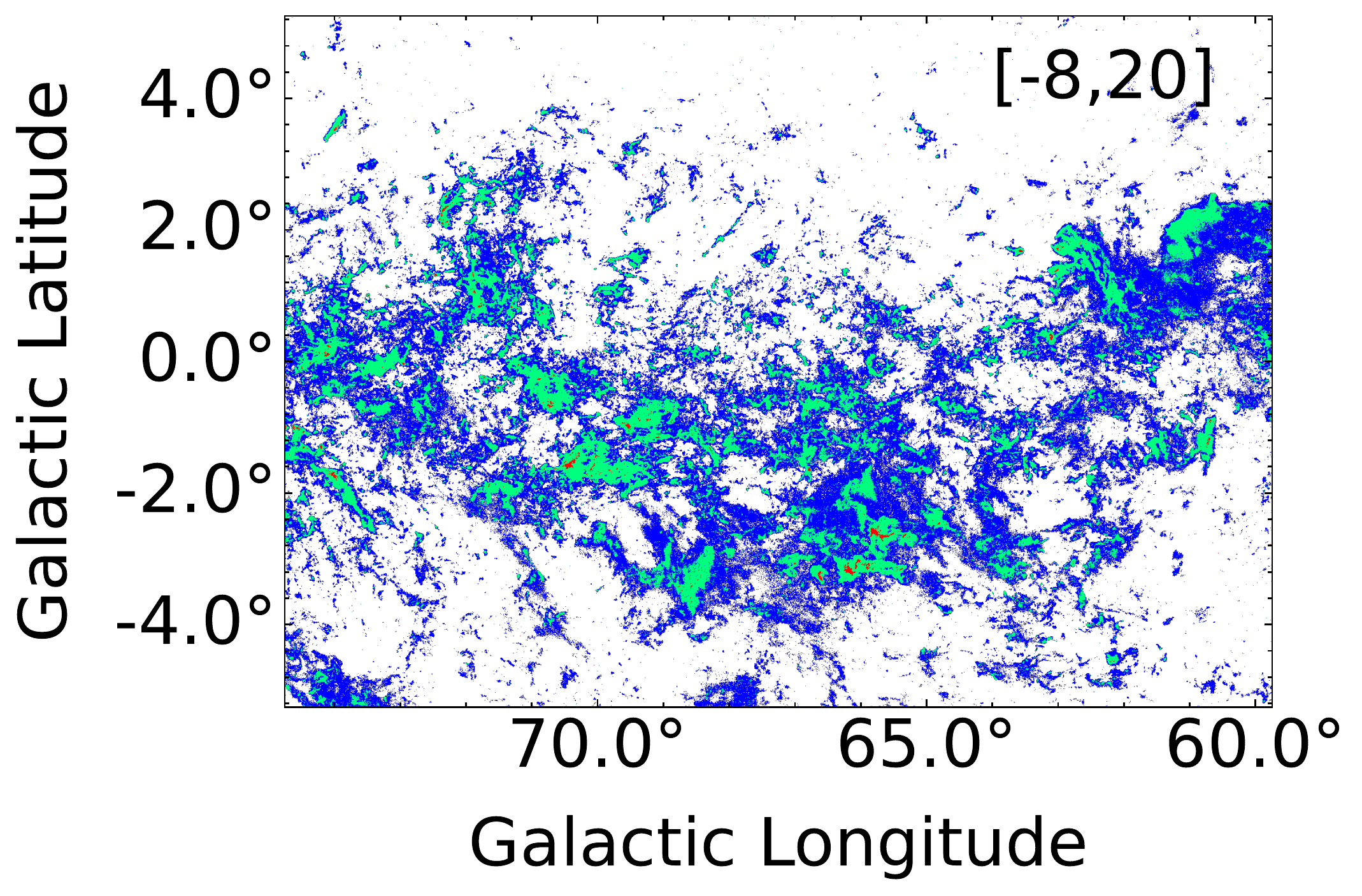}{0.48\textwidth}{(b)}
          }
          
\gridline{\fig{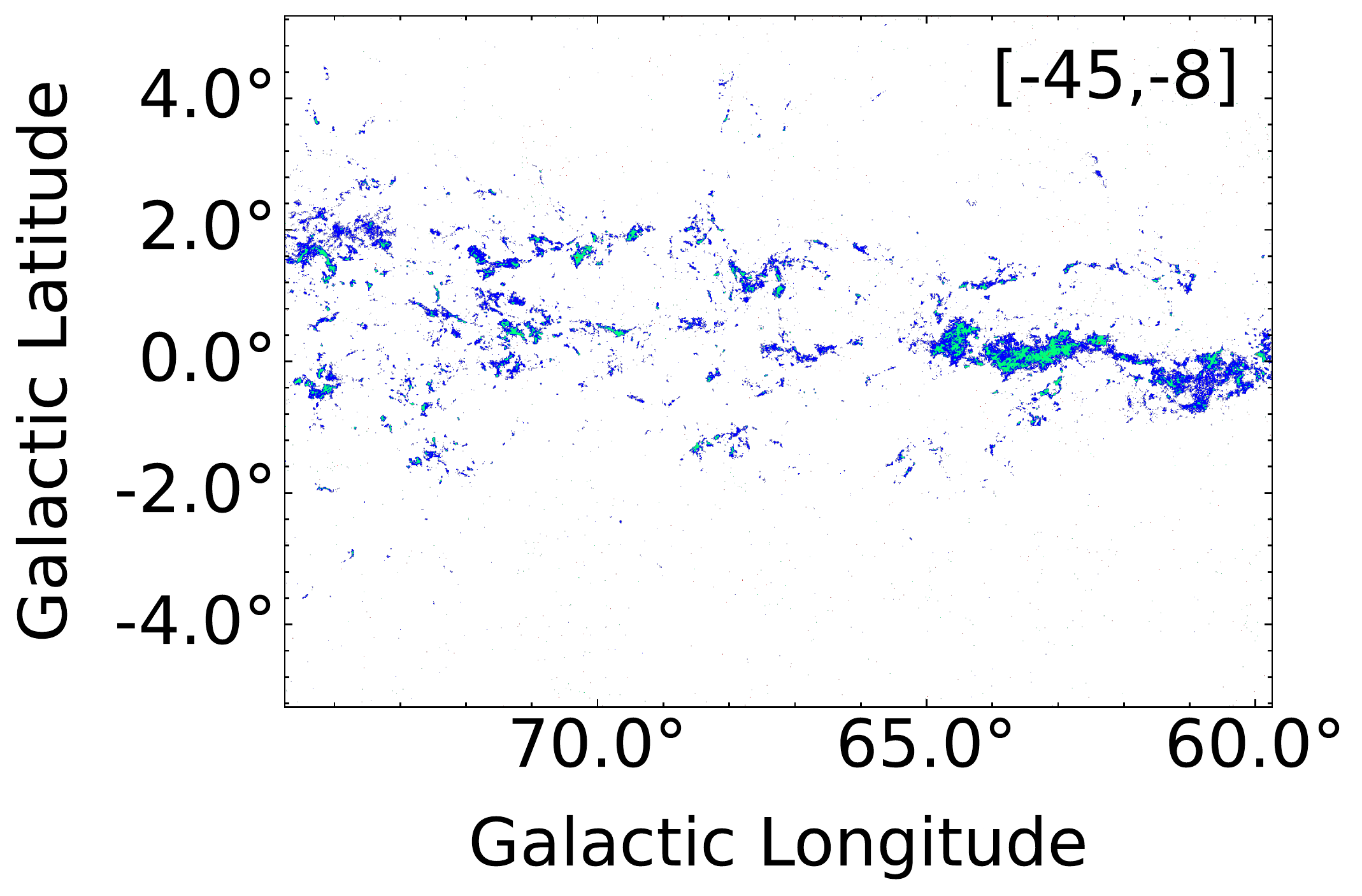}{0.48\textwidth}{(c)}
          \fig{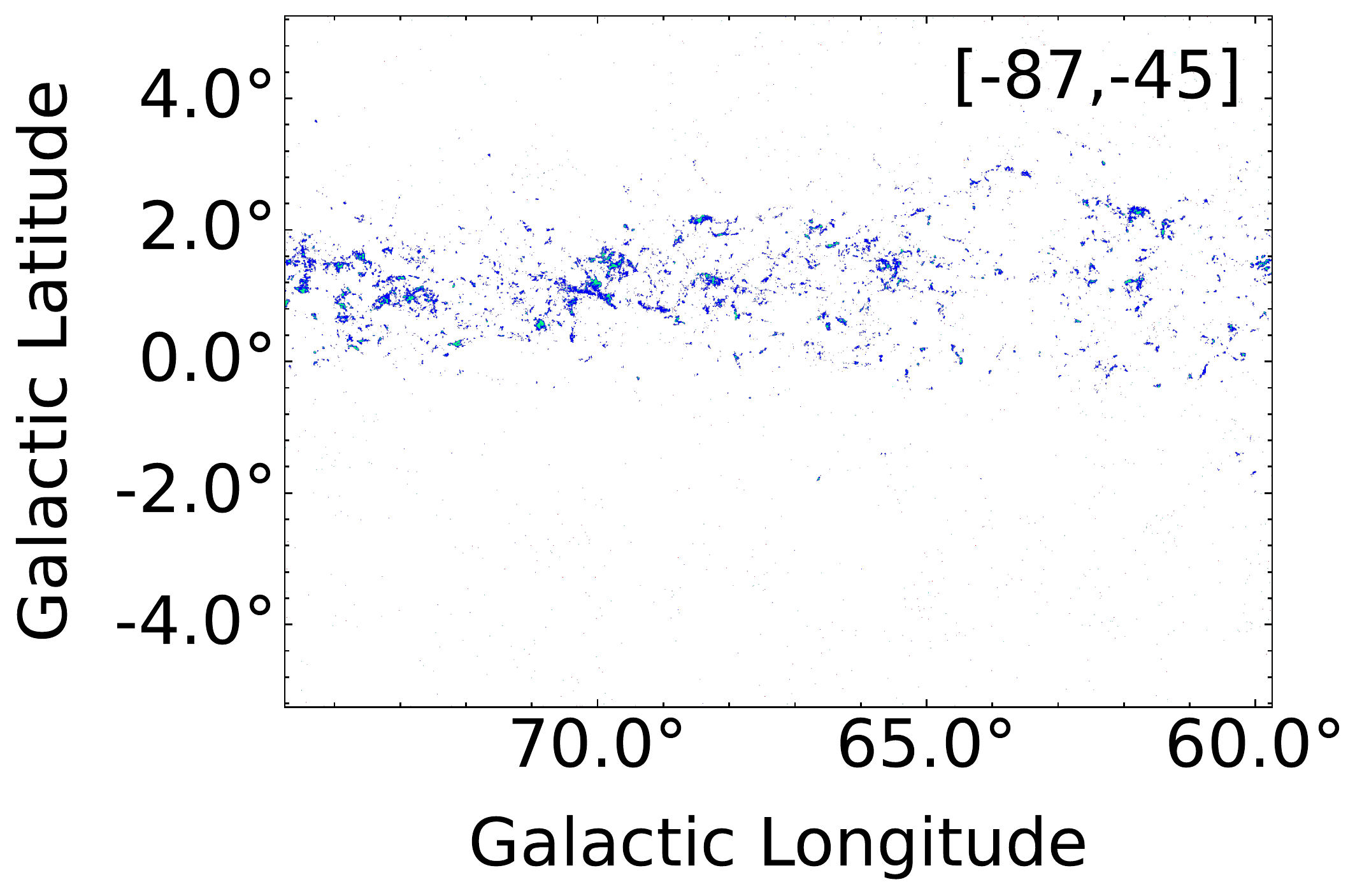}{0.48\textwidth}{(d)}
          }
\caption{Spatial distributions of $^{12}$CO, $^{13}$CO, and C$^{18}$O emission in four velocity ranges. The blue, green,  and red colors indicate the regions where $^{12}$CO, $^{13}$CO, and C$^{18}$O emission is detected, respectively. The (a) to (d) panels correspond to the Local Spur, Local, Perseus, and Outer arms, respectively. The velocity range of each arm is given in the top right corner of each panel.}
\label{fig:fig5}
\end{figure*}

In the G60 region, the molecular gas in different arms
can be further divided into three types of masks, according to the detection of $^{12}$CO, $^{13}$CO, and C$^{18}$O. The regions with $^{12}$CO detection are defined as Mask 1,  and the regions with both $^{12}$CO and $^{13}$CO detection are defined as Mask 2, while the regions with all three isotopologues  being detected are defined as Mask 3. This means that the Mask 1 region contains Masks 2 and 3 regions, and the Mask 2 regions contain Mask 3 regions. The spatial distributions of Masks 1${-}$3 in the four velocity ranges are shown in Figure \ref{fig:fig5}, with Mask 1$-$3 regions in blue, green, and red colors, respectively. $^{12}$CO and $^{13}$CO are detected in all spiral arms, while C$^{18}$O is only detected in a few pixels of the Local Spur and the Local arm. Table \ref{tab1} lists the total number and percentages of the detection of the $^{12}$CO, $^{13}$CO, and C$^{18}$O emission in different arms and different Masks. $^{13}$CO is detected among 26${\%}$ pixels with $^{12}$CO detection in the Local Spur and Local arm together. This percentage is 20${\%}$ for the Perseus arm, and 12${\%}$ for the Outer arm. The C$^{18}$O emission is detected among only about 0.27${\%}$ pixels with $^{12}$CO detection in the Local Spur and Local arm together. 

\begin{longrotatetable}
\begin{deluxetable*}{llcllllccccc}
\tablecaption{Statistical Properties of Molecular Gas in Different Spiral Arms of the G60 Region\label{tab1}}
\tablewidth{1000pt}
\tabletypesize{\tiny}
\tablehead{
\colhead{Arm Name} & \colhead{Velocity Range} &
\colhead{Mask Name} & \colhead{Pixel Numer(per)} &
\colhead{Area} & \colhead{T$\rm _{ex}$} &
\colhead{$\rm \tau_{(^{13}CO)}$} & \colhead{$\rm \tau_ {({C}^{18}{O})}$} &
\colhead{log(N$\rm_{^{12}CO}$)} & \colhead{log(N$\rm _{^{13}CO})$} & \colhead{log(N$\rm_ {C^{18}O}$)} \\
\colhead{} & \colhead{(km s$^{-1}$)} & \colhead{} & \colhead{$\rm \%$} &
\colhead{(deg$^{2}$)} & \colhead{(K)} & \colhead{} &
\colhead{} & \colhead{(cm$^{-2}$)} & \colhead{(cm$^{-2}$)} & \colhead{(cm$^{-2}$)}
}
\startdata
Local Spur & [20,55] & Mask 1 & 684890(30.15)&47.56& 3.7$-$38.8(6.8) &  & \nodata  & 15.46$-$18.32(17.09) &\nodata & \nodata \\
 &  & Mask 2 & 181702 (8.00) & 12.61 & 4.1$-$38.8(8.6) & 0.07$-$8.61(0.31) & \nodata & \nodata & 14.34$-$17.27(15.97) & \nodata
 \\
 &  & Mask 3 & 1868 ( 0.08) & 0.13 & 4.1$-$38.8(12.6)  & \nodata & 0.02$-$1.35(0.14) & \nodata & \nodata  & 14.00$-$15.75(14.71)\\
Local Arm & [$-$8,20]  & Maks 1 & 124950 ( 5.50) & 8.67 & 3.7$-$47.6(6.7) &  \nodata& \nodata & 15.46$-$18.57(17.05) & \nodata  & \nodata\\
 &  & Mask 2 & 33028 (1.45) & 2.29 & 4.1$-$47.6(8.8) & 0.07$-$3.26(0.31) & \nodata & \nodata & 13.97$-$17.76(15.99) & \nodata
 \\
 &  & Mask 3 & 345 (0.015) & 0.02 & 5.2$-$47.6(13.8)  & \nodata & 0.02$-$0.72(0.13) & \nodata & \nodata  & 14.01$-$15.9(14.79)\\
Perseus Arm & [$-$45,$-$8]  & Mask 1 & 73867 (3.25) & 5.13 & 3.8$-$31.9(6.1)  & \nodata & \nodata & 15.46$-$18.69(16.95) & \nodata  & \nodata\\
 &  & Mask 2 & 14548 (0.64) & 1.01 & 4.2$-$31.9(7.5) & 0.10$-$1.8(0.33) & \nodata  &  & 14.06$-$17.7(15.87) & \nodata
 \\
 &  & Mask 3 & 53(0.002) & \nodata  & 5.3$-$29.9(10.1) & \nodata & 0.04$-$0.76(0.19) & \nodata & \nodata & 14.03$-$15.73(14.55) \\
Outer Arm& [$-$87,$-$45] & Mask 1 & 31860(1.4)  & 2.21 & 3.6$-$22.0(5.9)  & \nodata & \nodata & 15.46$-$18.04(16.84)
& \nodata & \nodata\\
 &  & Mask 2 & 3925  (0.17) & 0.27 & 4.1$-$22.0(7.5)  & 0.10$-$1.07(0.30) & \nodata & \nodata& 14.02$-$16.9(15.75) & \nodata
 \\
 & & Mask 3 & 22 (0.001) & \nodata & 5.1$-$7.3(6.6) & \nodata & 0.16$-$0.46(0.29) & \nodata & \nodata  & \nodata\\
\enddata
\tablecomments{Columns 1$-$3: name of the arms, arm velocity range, and mask name (see Section \ref{subsec:average spectra}). Columns 4$-$5: number of pixels with detection and detection percentage. Column 6, range and the median value (in parenthesis) of the
excitation temperature. Columns 7$-$8: ranges and median values (in parenthesis) of the optical depths of $^{13}$CO and C$^{18}$O emission, respectively. Columns 9$-$11: ranges and median values (in parenthesis) of the column
density of $^{12}$CO, $^{13}$CO, and C$^{18}$O, respectively. }
\end{deluxetable*}
\end{longrotatetable}

\subsection{Statistical Properties of Molecular Gas in Different Spiral Arms}\label{subsec:Distance}
The distance of a molecular cloud is essential to estimate the physical parameters of the cloud. For molecular clouds located at relatively large distances (d $>$ 3 kpc), the distances of the individual clouds can be appropriately determined using the Galactic rotation curve and the  radial velocities of the individual clouds (e.g. \citealt{2021ApJS..254....3M}). For molecular clouds located at relatively near distances, the above method may introduce large relative errors in the obtained distances. For these clouds, the extinction map can be used to derive reliable distances for the individual clouds (e.g. \citealt{2019ApJ...879..125Z,2021ApJS..254....3M}). In this work, we just attempt to estimate the statistical properties of molecular gas in different spiral arms. For this purpose, we do not try to determine the distances of individual clouds in each spiral arm, but use the measured parallaxes of masers located in each arm from the BeSSel survey as the reference distance to the molecular gas in each spiral arm. The association between the molecular gas and the masers is determined from the position-position-velocity information of the MWISP spectra and that of the masers. We assigned the mean parallax distance of masers in a spiral arm. Finally, the reference distances for the Local Spur, Local, Perseus, and Outer arms in the section within the Galactic longitudes studied in this work are 3.15, 3.01, 7.34, and 13.33 kpc, respectively. 

With the reference distances, we can estimate the total mass of molecular hydrogen in each spiral arm. The total mass can be estimated through the summation of H$_2$ column densities according to Eq. \ref{eq1}, 
\begin{equation}
\label{eq1}
M = \mu m_{H} D^{2} \int N_{H_{2}} d\Omega ,
\end{equation} 
where $N_{H_2}$ is the column density of molecular hydrogen, D is the distance, $m\rm _H$ is the mass of the hydrogen atom, $\Omega$ is the total angular area the emission occupies, and $\mu=2.8$ is the mean molecular weight. The pixel-by-pixel $N_{H_2}$ for each spiral arm can be calculated from the integrated $^{12}$CO intensity using the ``X$\rm _{CO}$'' method or from the integrated $^{13}$CO intensity using the ``LTE'' method, as introduced in \cite{2021ApJS..254....3M}. We use X$\rm _{CO}$ = 2.0$\times$10$^{20}$ cm$^{-2}$ (K km s$^{-1}$)$^{-1}$ \citep{2013ARA&A..51..207B} to convert the integrated intensity of $^{12}$CO emission into the H$_{2}$ column density. Finally, the mass derived for the Local Spur, Local arm, Perseus arm, and Outer arms is 1.02$\times$10$^{6}$, 5.29$\times$10$^{6}$, 2.63$\times$10$^{6}$, and 2.75$\times$10$^{6}$ M$_{\odot }$, respectively, using the ``X$\rm _{CO}$'' method, while the mass is  2.28$\times$10$^{5}$, 9.40$\times$10$^{5}$, 3.21$\times$10$^{5}$, and 2.24$\times$10$^{5}$ M$_{\odot }$, respectively, using the LTE method. 

\begin{figure*}[htb!]
\gridline{\fig{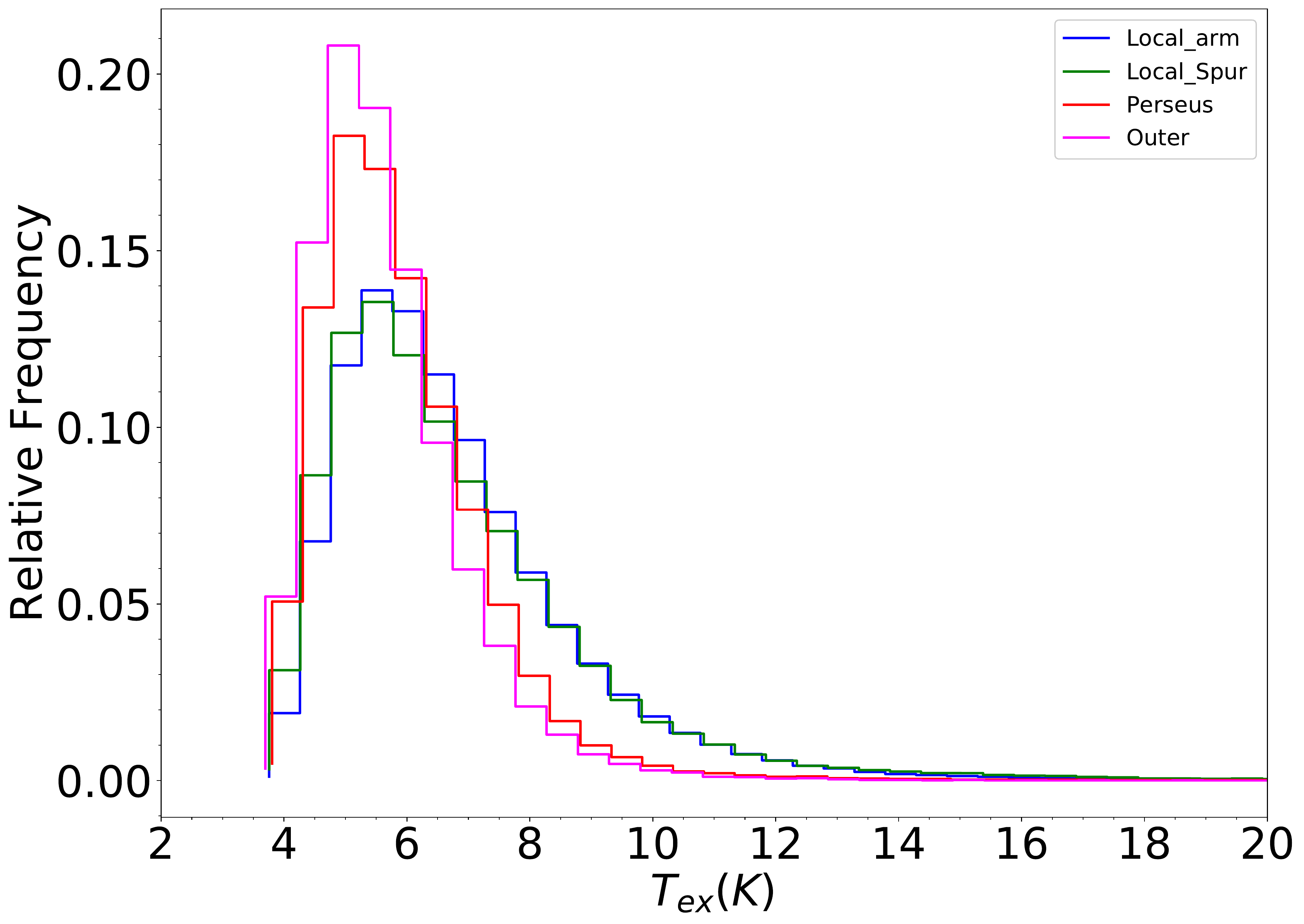}{0.33\textwidth}{(a)}
          \fig{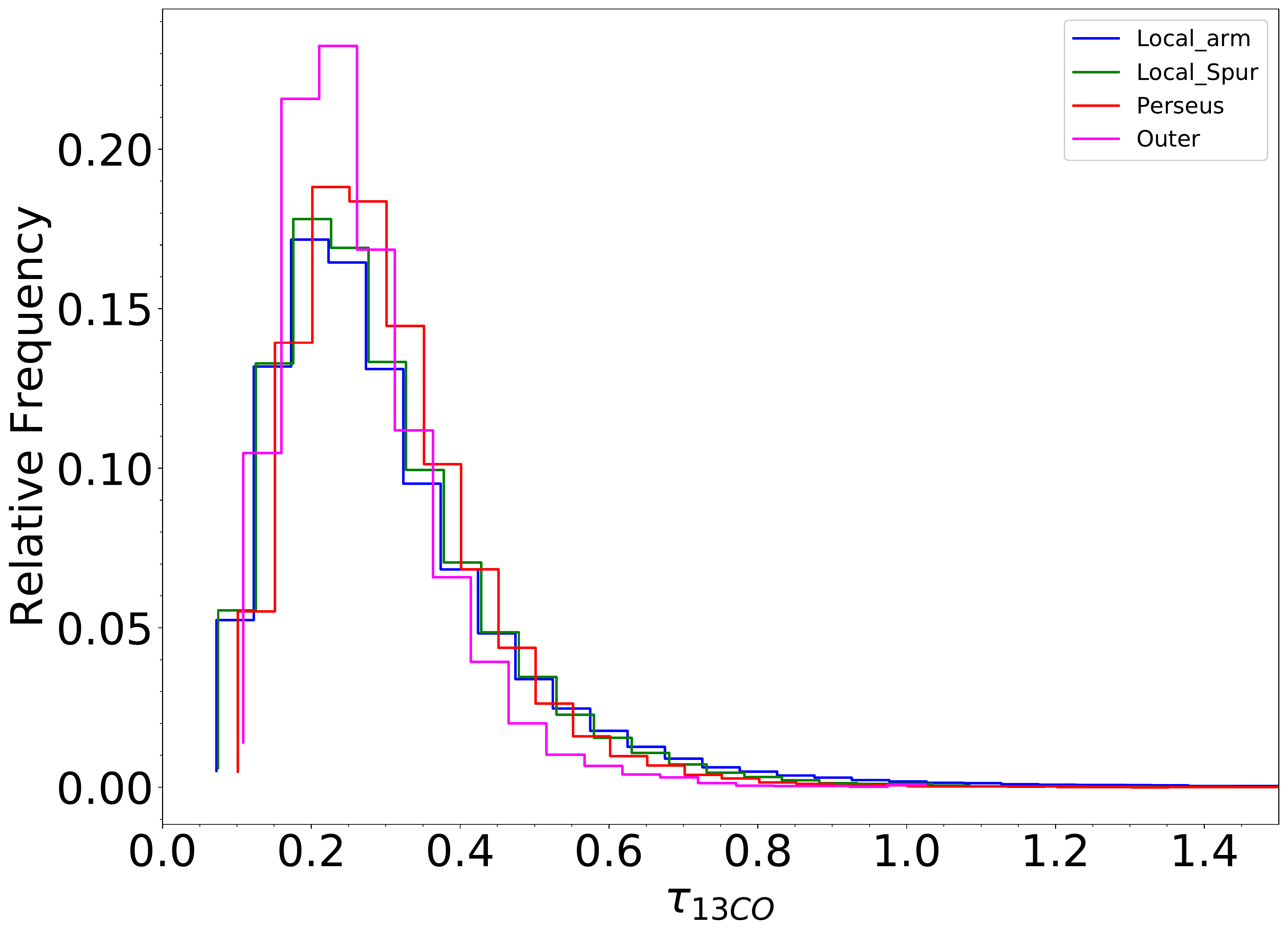}{0.32\textwidth}{(b)}
          \fig{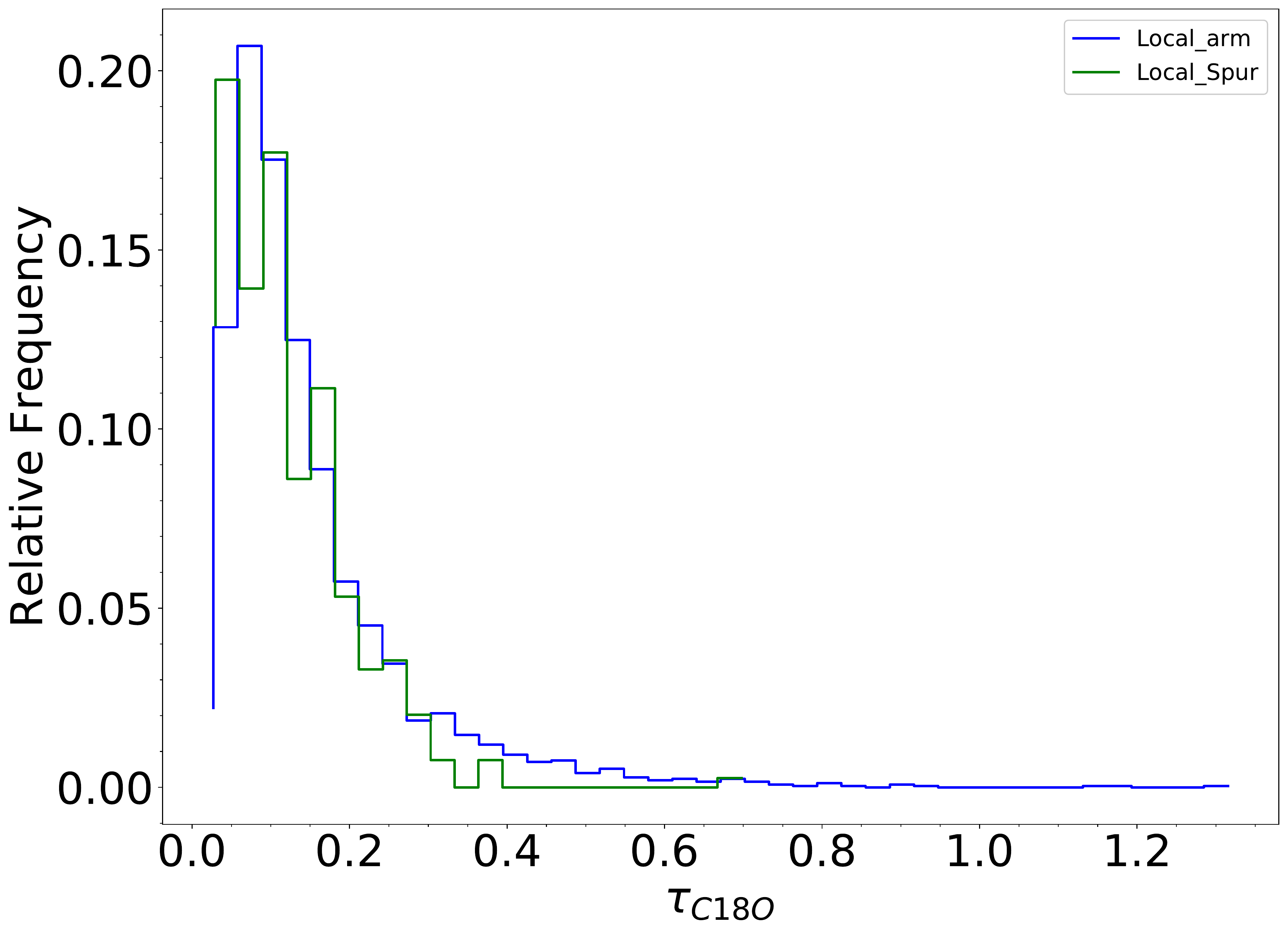}{0.32\textwidth}{(c)}
          }
          
\gridline{\fig{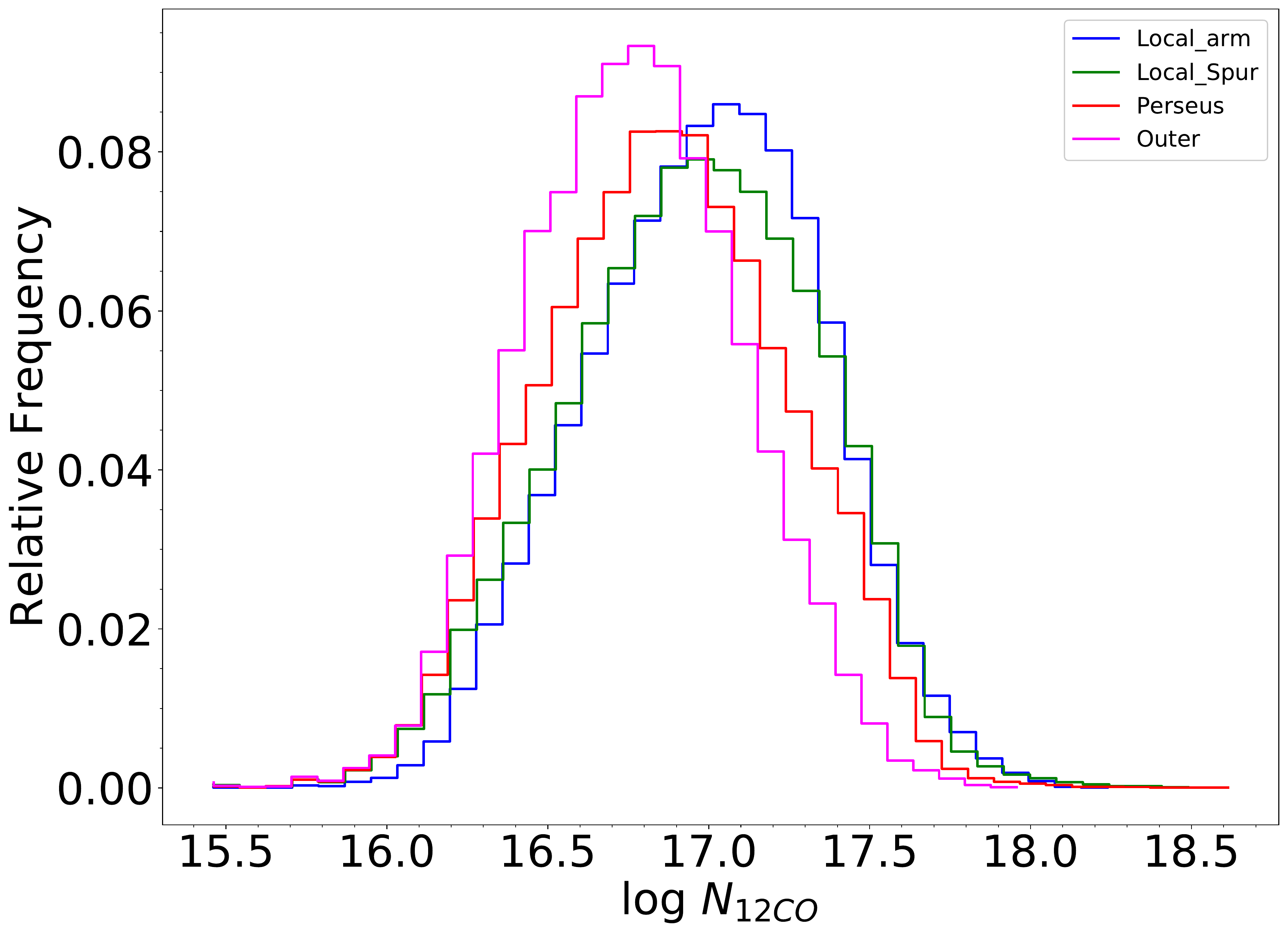}{0.33\textwidth}{(d)}
          \fig{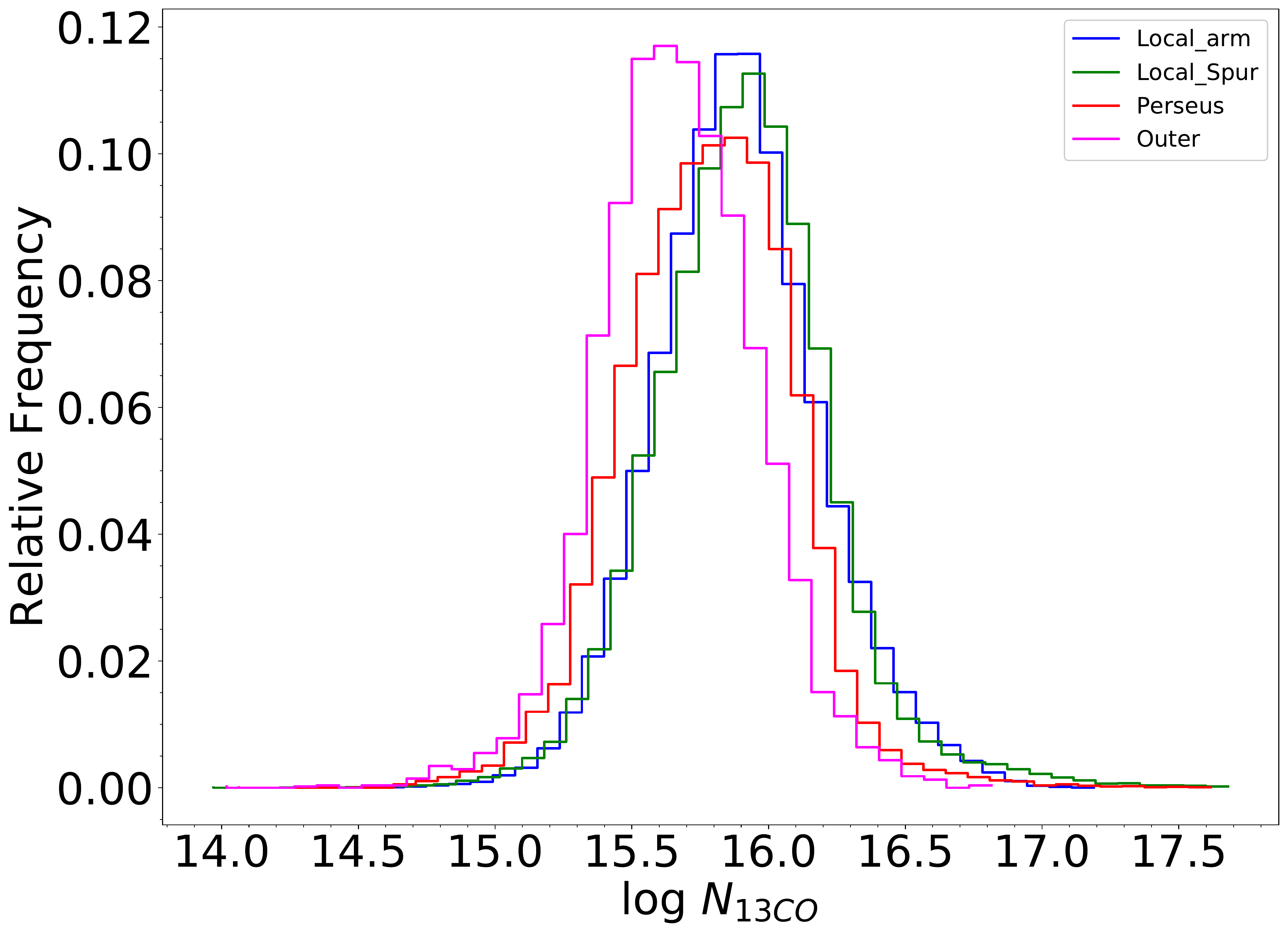}{0.32\textwidth}{(e)}
          \fig{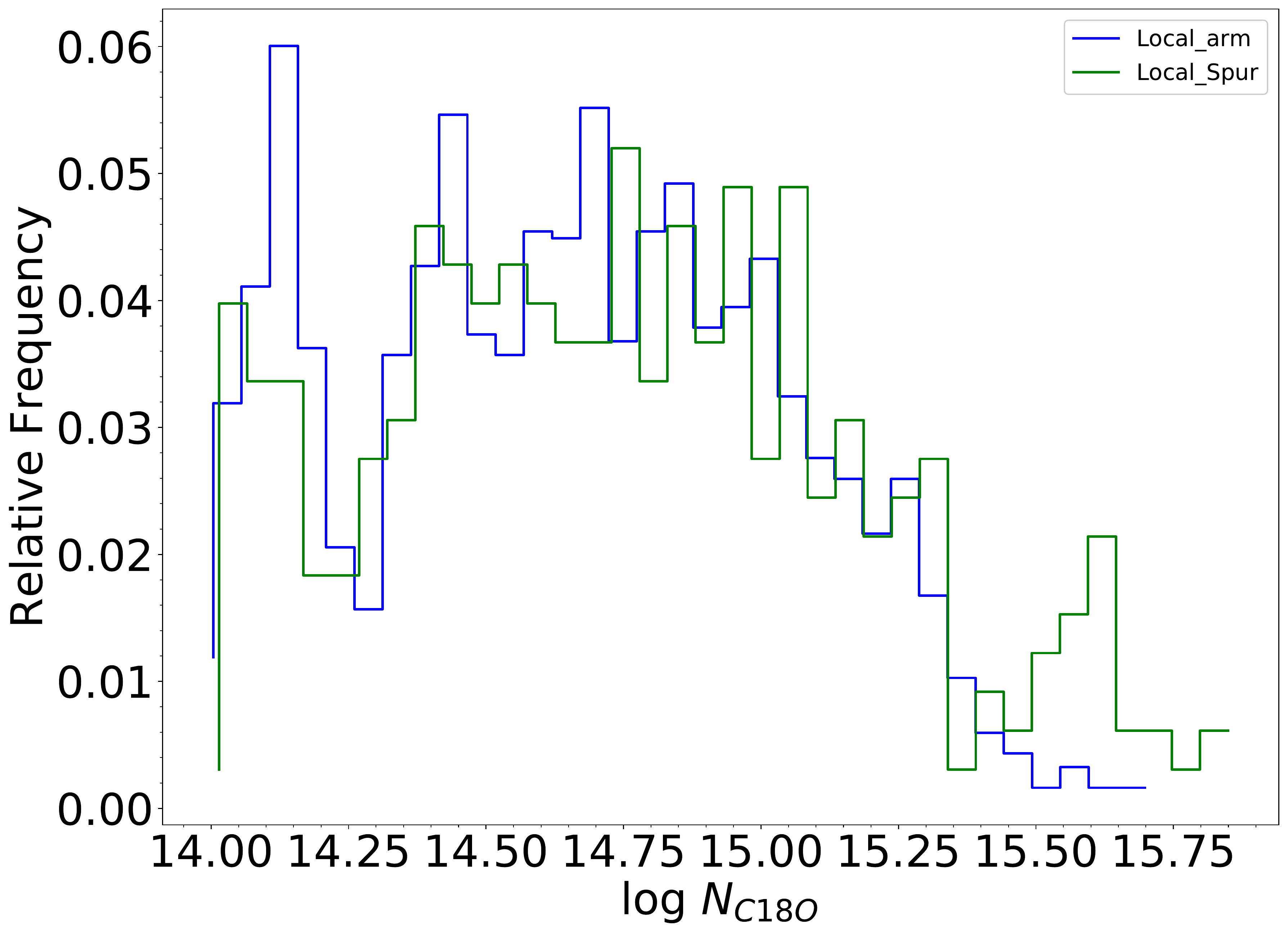}{0.32\textwidth}{(f)}
          }
\caption{Histograms of excitation temperature in Mask 1 regions (a), optical depth of $^{13}$CO in Mask 2 regions (b), optical depth of $\rm {C}^{18}{O}$ in Mask 3 regions (c), column density of $^{12}$CO in Mask 1 regions (d), column density of $^{13}$CO in Mask 2 regions (e), column density of $\rm {C}^{18}{O}$ in Mask 3 regions (f) in different spiral arms. Different colors represent different spiral arms. }
\label{fig:fig6}
\end{figure*}

Calculation of the excitation temperature, optical depth of the $^{13}$CO and C$^{18}$O emission, column density of the $^{12}$CO, $^{13}$CO, and C$^{18}$O line emission for each pixel in Masks 1${-}$3 in different spiral arms are the same as in \cite{2021ApJS..254....3M}. The statistics of the above physical quantities are presented in Table \ref{tab1} and Figure \ref{fig:fig6}. For all the spiral arms, the Masks 1$-$2 regions have similar median T$_{ex}$ of $\sim$ 6$-$8 K, while the median T$_{ex}$ is higher in the Mask 3 region, which is around 10$-$14 K. Statistically, the Perseus and Outer arms contain relatively higher fractions of pixels with T$_{ex}$ higher than $\sim$7 K (see Figure~\ref{fig:fig6} (a)). The $^{13}$CO emission is relatively optically thin, with median $\tau_{13CO}$ around 0.3, in all spiral arms, though there are pixels with $^{13}$CO optical depth up to 1.5. The C$^{18}$O emission is optically thin in all spiral arms, although $\tau$$_{C18O}$ can be up to 1.35 in some pixels.

The column densities of $^{12}$CO and $^{13}$CO in different arms have similar distributions, as shown in Figures \ref{fig:fig6} (d) and (e). The median column densities of $^{12}$CO in the Local Spur and Local arm are greater than that found in the Perseus and Outer arm by a factor of ${\sim}$1.5. For C$^{18}$O, its median column density in the Local Spur is almost the same as that in the Local arm. We also present in Figure \ref{fig:figA3} the distributions of $^{12}$CO and $^{13}$CO column densities in the Appendix for each arm individually in the log-log space and the log-normal functions with parameters estimated from the data are overlaid. Log-normal distribution functions of column densitiy (i.e., LN N-PDFs) usually indicate turbulence dominance in molecular gas \citep{Kritsuk_2011,Federrath_2013}, while excesses from LN distributions imply self-gravity dominance, and they typically coincide to star formation activities \citep{2009A&A...508L..35K}. In our results, the N-PDFs of $^{12}$CO emission in the Local Spur, Local arm, and Outer arm are well described by LN functions, while that of the Perseus arm has a slight excess at the high column density end. In contrast, the $^{13}$CO N-PDFs for the Local Spur and the Perseus arm have significant excesses above LN distribution at their high column density ends. The regions in the Local spur and the Perseus arm where the $^{13}$CO column density exceeds the value at which significant excess above LN distribution appears are indicated with orange boxes in Figure \ref{fig:fig4_13CO} (a) and (c). As shown in Figure \ref{fig:fig4_13CO} (a) for Local spur, four regions have high column density, among which three hosts \ion{H}{2} regions (G060.873-00.108, G061.473+00.094) or maser (l = 59.78$^\circ$, b = 0.06$^\circ$). For the Perseus arm in Figure \ref{fig:fig4_13CO} (c), one region possesses high column density and it hosts a maser (l = 70.18$^\circ$, b = 1.74$^\circ$). Therefore, we can see that the majority of the high-column density regions host ongoing star formation.

\subsection{Clouds in the Outer-Scutum-Centaurus Arm} 
\label{subsec:OSC Cloud}

\begin{figure*}
\gridline{\fig{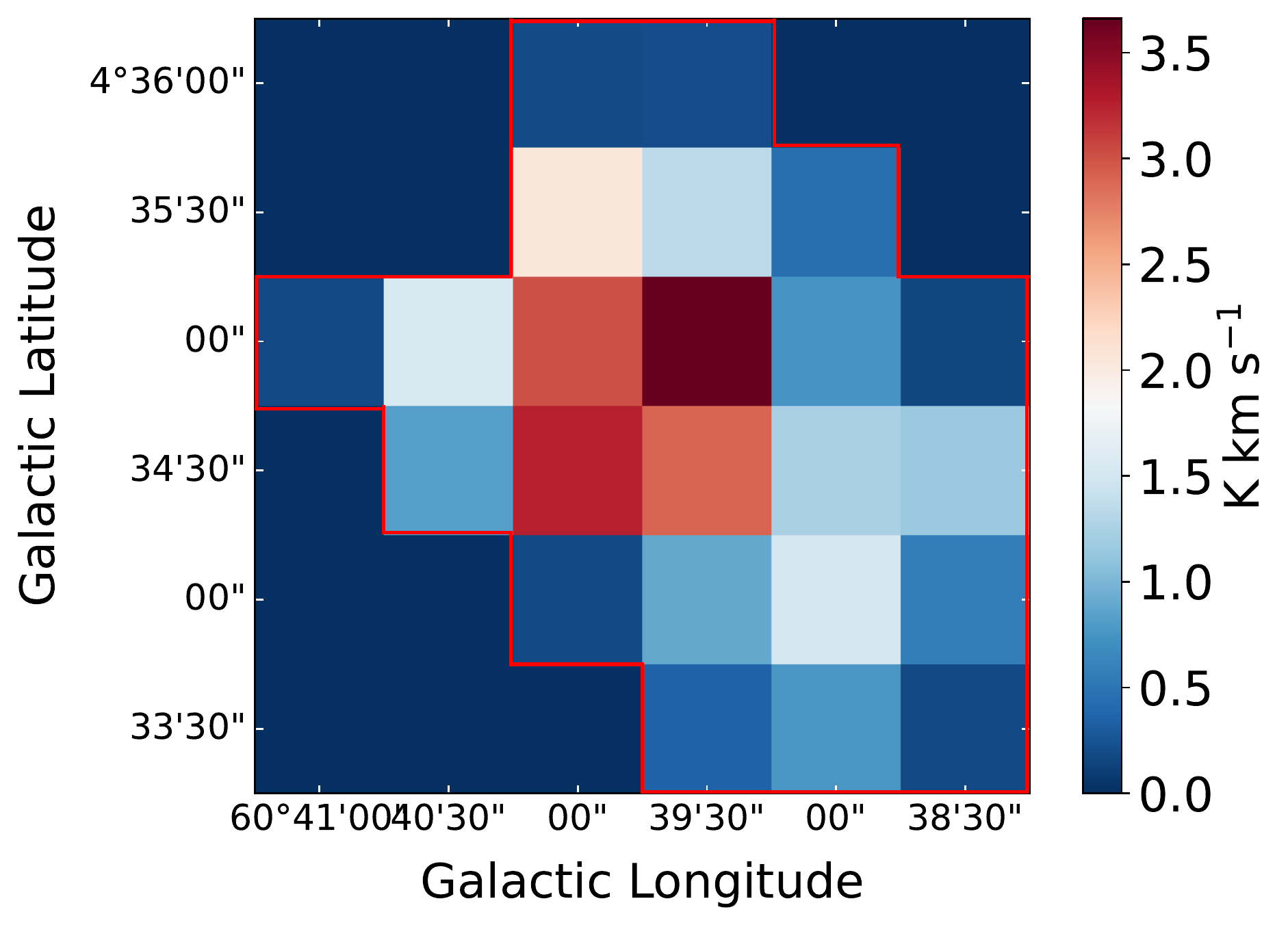}{0.4\textwidth}{(a1)}
          \fig{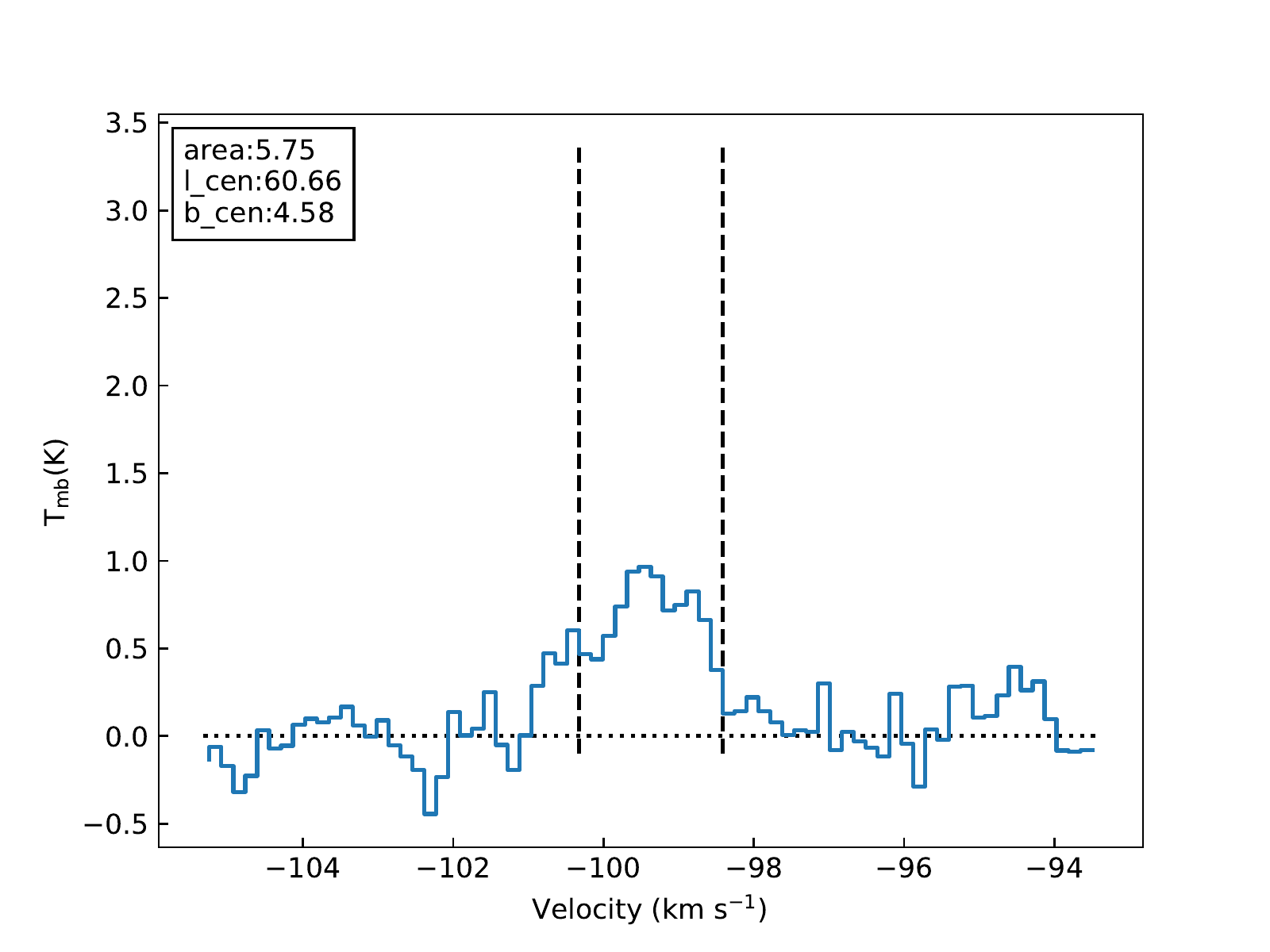}{0.4\textwidth}{(a2)}
          }
          
\gridline{\fig{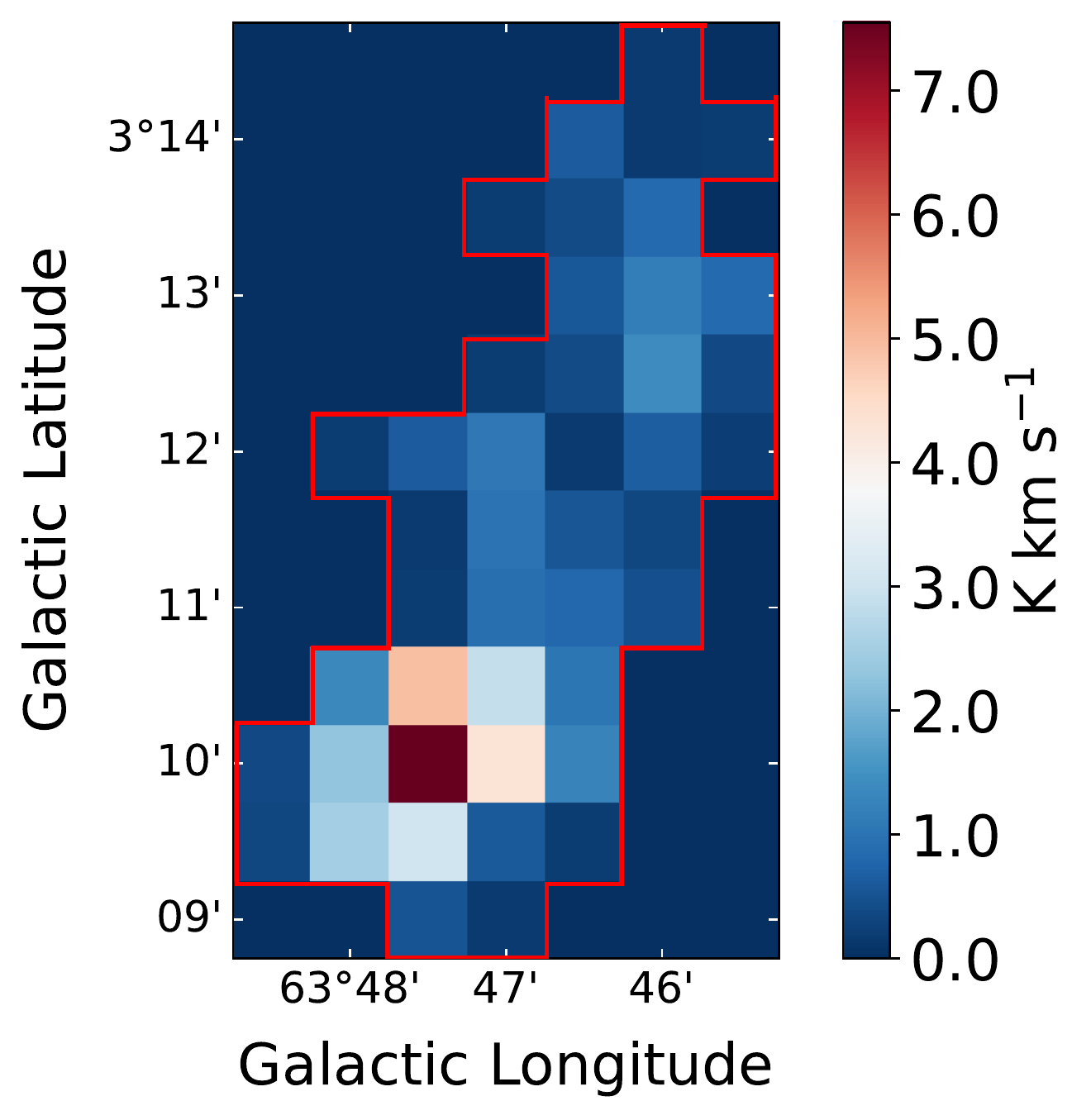}{0.4\textwidth}{(b1)}
          \fig{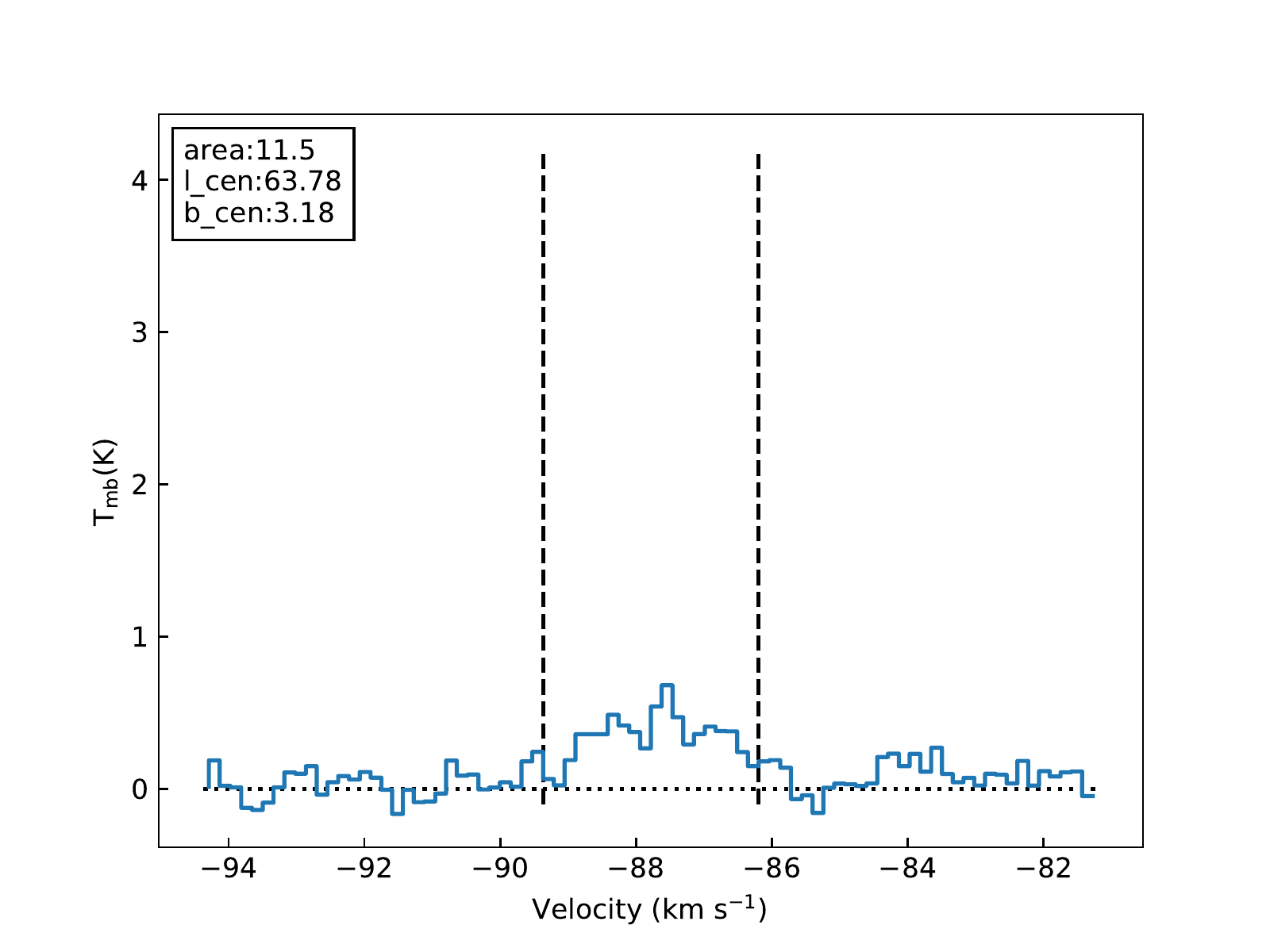}{0.4\textwidth}{(b2)}
          }
          
\gridline{\fig{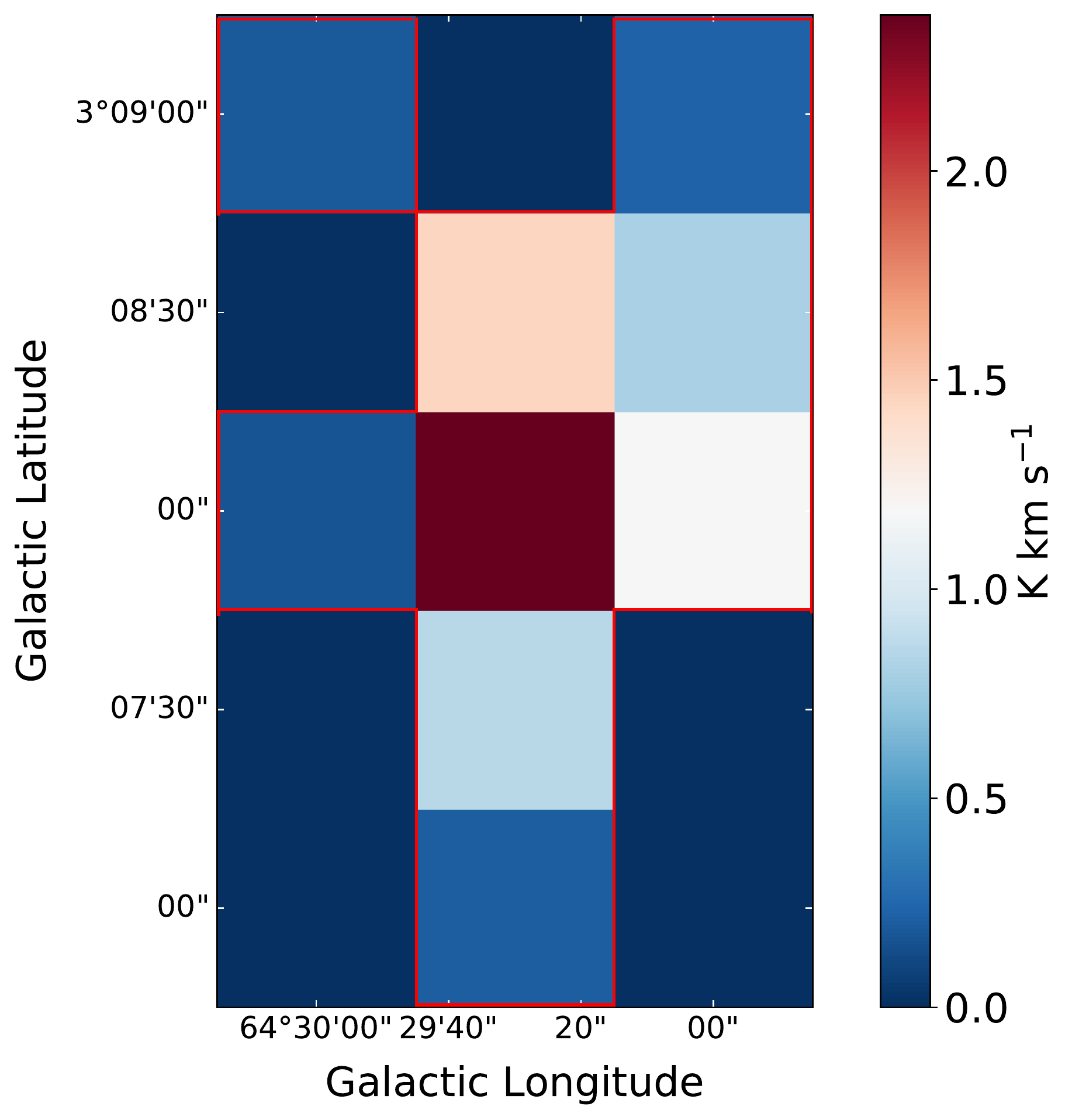}{0.4\textwidth}{(c1)}
          \fig{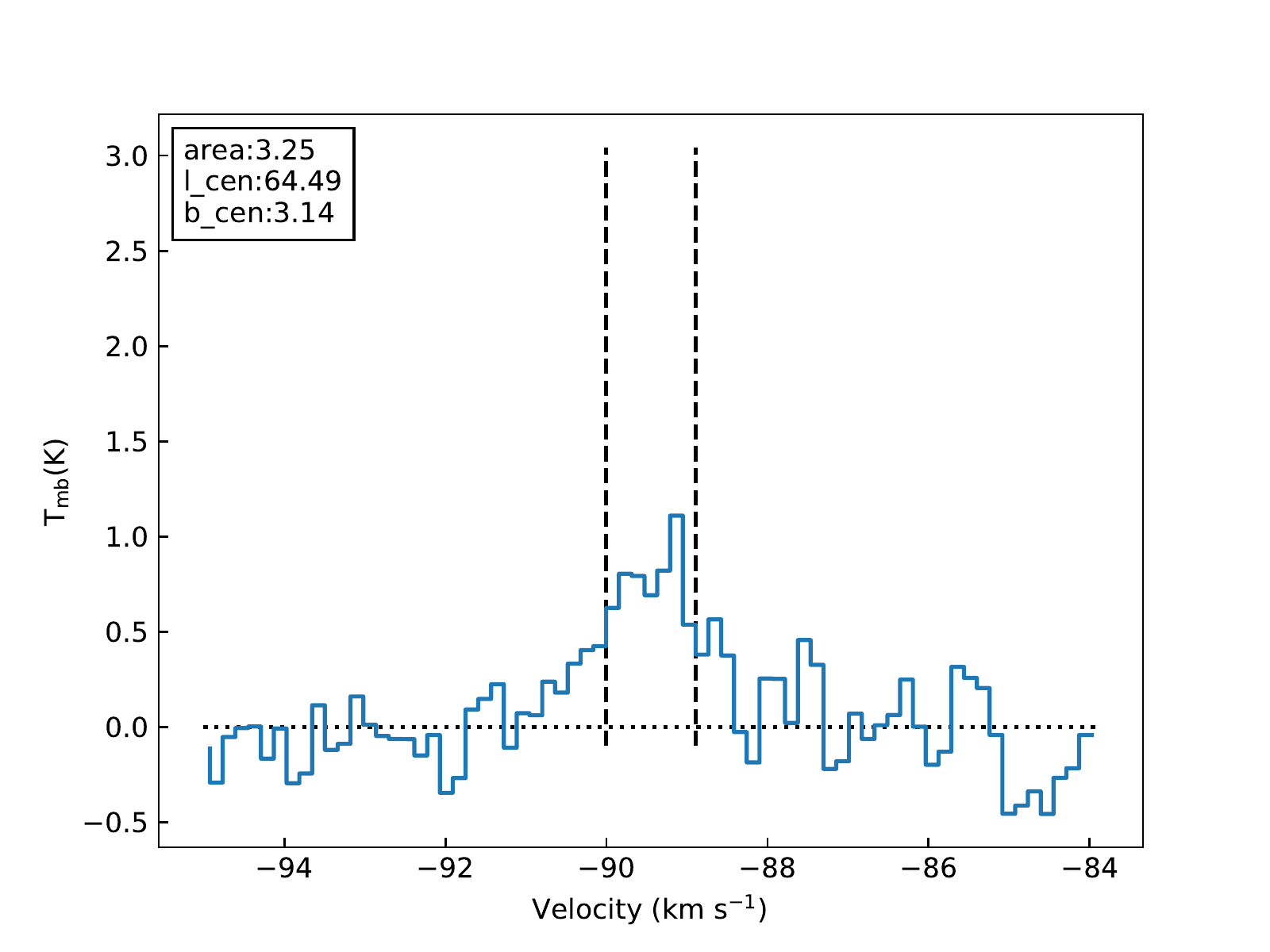}{0.4\textwidth}{(c2)}
          }
\caption{Integrated intensity maps (a1, b1, and c1) and average spectra (a2, b2, and c2) of three distant clouds. The red outlines represent the boundaries of molecular clouds. Two vertical dashed lines in each average spectrum indicate the velocity range of the cloud. The angular sizes in units of arcmin$^2$ and the coordinates of the clouds in degrees are indicated in the top-left corner of the average spectra.}
\label{fig:fig7}
\end{figure*}

Benefitting from the high sensitivity of the MWISP survey, we could try to identify distant molecular clouds in the Galactic plane. From the $^{12}$CO line emission data, we identified 15 extremely $^{12}$CO clouds within kinematic distances of $\sim$[14.72, 17.77] kpc using the algorithm of Density-Based Spatial Clustering of Applications with Noise (DBSCAN; \citealp{Ester1996}). Before the identification, the $^{12}$CO data are masked above 2$\sigma_{RMS}$, which is the standard criterion used in a series of works of the MWISP project (e.g., \citealt{Yan_2020,2022ApJS..262...16M,2022arXiv221202066Y}). The DBSCAN algorithm has two input parameters, i.e., $\epsilon$ and $MinPts$. We used the set of parameters $\epsilon =1$ and $MinPts = 4$ in this work. This settings of the parameters are adopted from \cite{Yan_2020}, which are found most suitable for the MWISP data. The definitions of the two parameters have also been explained in detail in \cite{Yan_2020} and \cite{2022ApJS..262...16M}. Figure \ref{fig:fig7} displays the integrated-intensity maps and average spectra of three of these molecular clouds as an example. The identified molecular clouds are outlined with red lines in Figures \ref{fig:fig7} (a1), (b1), and (c1). The spectra in Figures \ref{fig:fig7} (a2), (b2), and (c2) are averaged within the outlines of the clouds. The two vertical dash lines in Figures \ref{fig:fig7} (a2), (b2), and (c2) represent the velocity spans of the corresponding clouds. Similar maps and spectra for the other 12 extremely distant clouds are given in Figure \ref{fig:figA1_3} in the Appendix. 

For the distances of these clouds, we adopt their kinematic distances assuming a flat Galactic rotation curve, a solar Galactocentric distance R$_{0}$ = 8.15 kpc, and a circular velocity of the Sun of $\Theta_{0}$ = 236 km s$^{-1}$ \citep{2019ApJ...885..131R}. The kinematic distances of these clouds range from 14.72 kpc to 17.77 kpc. We use the X$\rm _{CO}$ factor to convert the integrated intensity of $^{12}$CO emission into the H$_{2}$ column density of clouds, and then calculate their mass according to Eq. \ref{eq1}. The mass of these molecular clouds ranges from 363 to 520 M$_{\odot}$. The centroid velocity, distance, and mass are tabulated in Table \ref{tab:tab2}.

\begin{figure*}[ht!]
\plotone{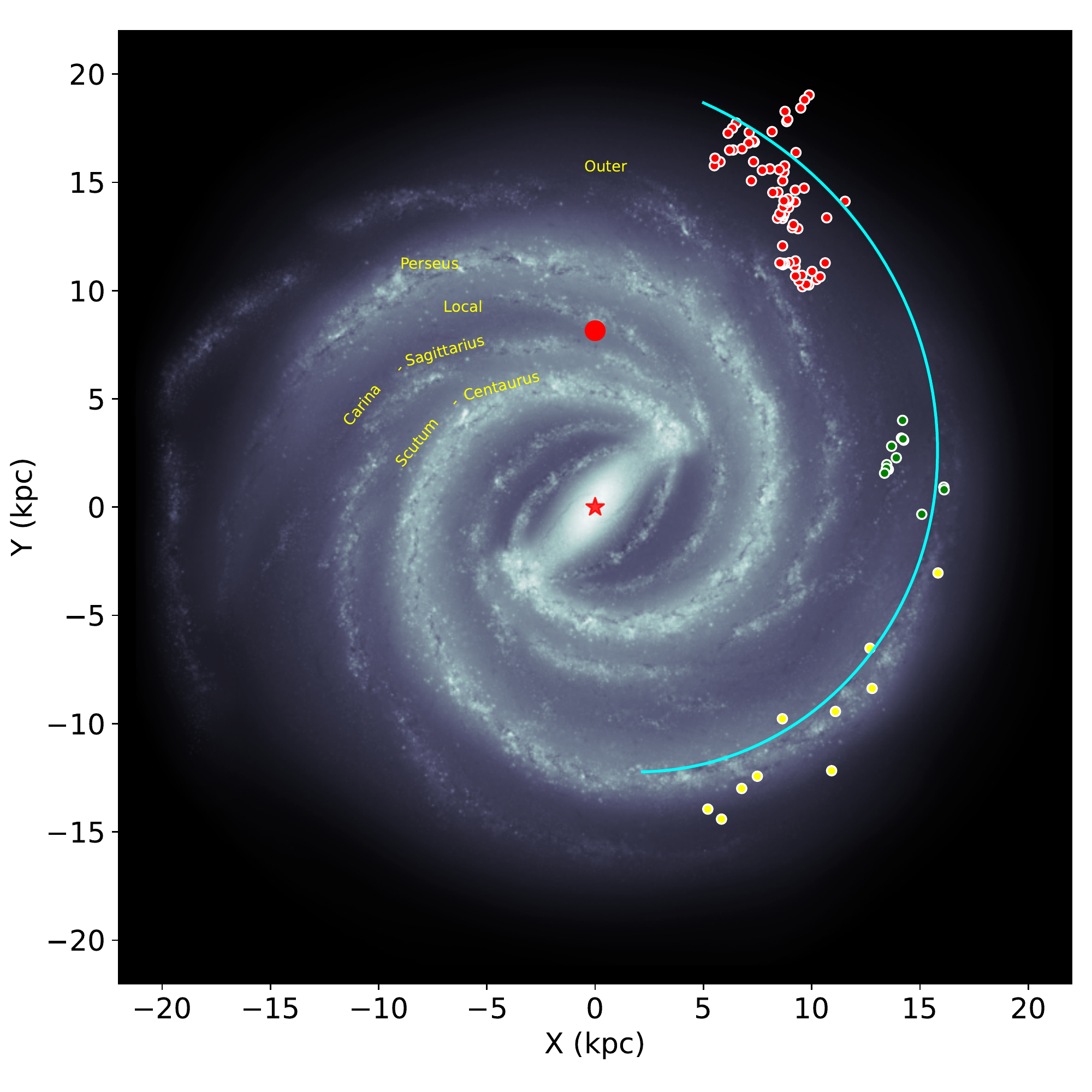}
\caption{Locations of molecular clouds in the OSC arm superposed on the artist's conception of the Milky Way (R.Hurt: NASA/JPL-Caltech/SSC). The green circles, yellow circles, and red circles indicate clouds covering Galactic longitudes from l = 59.75$^\circ$ to 74.75$^\circ$ detected in this work, l = 13$^\circ$ to 55$^\circ$ detected by \cite{2011ApJ...734L..24D}, and l = 100$^\circ$ to 150$^\circ$ detected by \cite{2015ApJ...798L..27S}. The cyan dashed line is a logarithmic spiral line fit to the Scutum-Centaurus arm. \citep{2019ApJ...885..131R}. The Galactic center is marked with a star symbol and the location of the Sun is marked with a large red dot.
\label{fig:fig8}}
\end{figure*}

Combining \ion{H}{1} data and the MWISP data, \cite{2015ApJ...798L..27S} identified 72 clouds with masses of 10$^{2}$-10$^{4}$ M$_{\odot}$ within Galactocentric radii of 15 to 19 kpc, which they found are likely to be the extension of the arm identified by \cite{2011ApJ...734L..24D}. This portion of the arm could be the extention of Scutum-Centaurus arm into the outer Galaxy. We overlaid the molecular clouds from \cite{2011ApJ...734L..24D}, the molecular clouds identified in this work and in \cite{2015ApJ...798L..27S} that are associated with the OSC arm in an artist's conception of the Milky Way\footnote{R.Hurt: NASA/JPL-Caltech/SSC} in Figure \ref{fig:fig8}. The locations of the molecular clouds identified by \cite{2011ApJ...734L..24D} and the molecular clouds identified by \cite{2015ApJ...798L..27S} are indicated with yellow circles and red circles, respectively. The green circles represent the distant molecular clouds identified in this work. We also overlaid the best fitting of the new arm identified by \cite{2011ApJ...734L..24D} and \cite{2015ApJ...798L..27S} in Figure \ref{fig:fig8}. The spiral arm parameters, such as the pitch angle and the reference Galactocentric azimuth, are adopted from \cite{2015ApJ...798L..27S}. The distribution of the distant molecular clouds that we identified in this work can reasonably be thought to follow the best-fitting curve of the OSC arm. Therefore, these 15 distant molecular clouds probably trace the extension of the OSC arm within the Galactic longitude from 60$^\circ$ to 75$^\circ$. However, we note that only three of these 15 molecular clouds are located near the best-fitting curve of the OSC arm, while the other 12 molecular clouds are located in the middle between the Outer and the OSC arms. These 12 molecular clouds constitute a nearly linear filament, which may be an independent structure or a spur, rather than part of an arm, such as the newly discovered Cattail giant \ion{H}{1} filament, which is located between the Galactic longitude from 80 to 95$^\circ$ and well beyond the OSC arm \citep{2021ApJ...918L...2L}.

\section{Discussion: Vertical distribution} \label{sec:Vertical distribution}

Early \ion{H}{1} surveys have found that the Galactic plane of \ion{H}{1} gas warps upward above the $b$ = 0$\arcdeg$ plane for the Galactocentric azimuth range 0\arcdeg $\leq$ $l$ $\leq$ 180$\arcdeg$ and downward below the $b$ = 0$\arcdeg$  plane for the range of 180$\arcdeg$ $\leq$ $l$ $\leq$ 360$\arcdeg$ \citep{1957BAN....13..201W,1982ApJ...263..116H,1988gera.book..295B}. Surveys of molecular gas have shown that the flaring and warping of the molecular gas layer in the outer Galactic varies strongly with Galactocentric azimuth and exhibits a similar structure with that of the \ion{H}{1} gas layer  \citep{1988ApJ...327..139C,1990A&A...230...21W}. Using the CO (J = 1$-$0) data from the CfA-Chile survey, \cite{2006PASJ...58..847N} found that the thickness of the molecular gas disk increases from 48 to 90 pc within the solar circle and then abruptly increases at a  Galactocentric radius of 11 kpc. They also found that the molecular gas disk also shows warping above the Galactic mid-plane in the outer Galaxy within the azimuth angles from 60\arcdeg to 140\arcdeg. With the high-sensitivity CO data from MWISP, we have the chance to analyze the vertical distribution of molecular gas in the section of the Galactic plane studied in this work. 

\begin{figure*}[h]          
\gridline{\fig{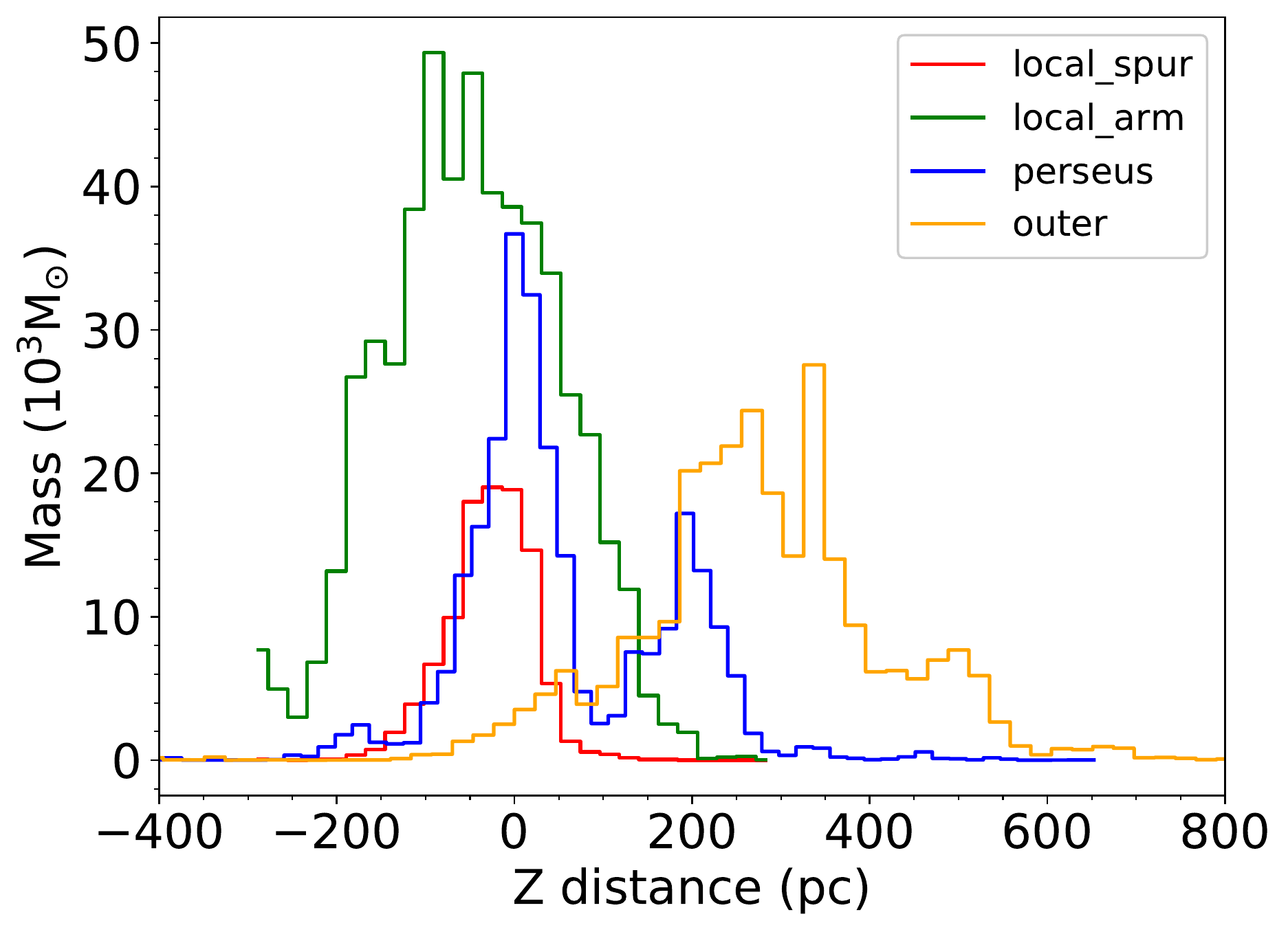}{0.5\textwidth}{(a)}
          \fig{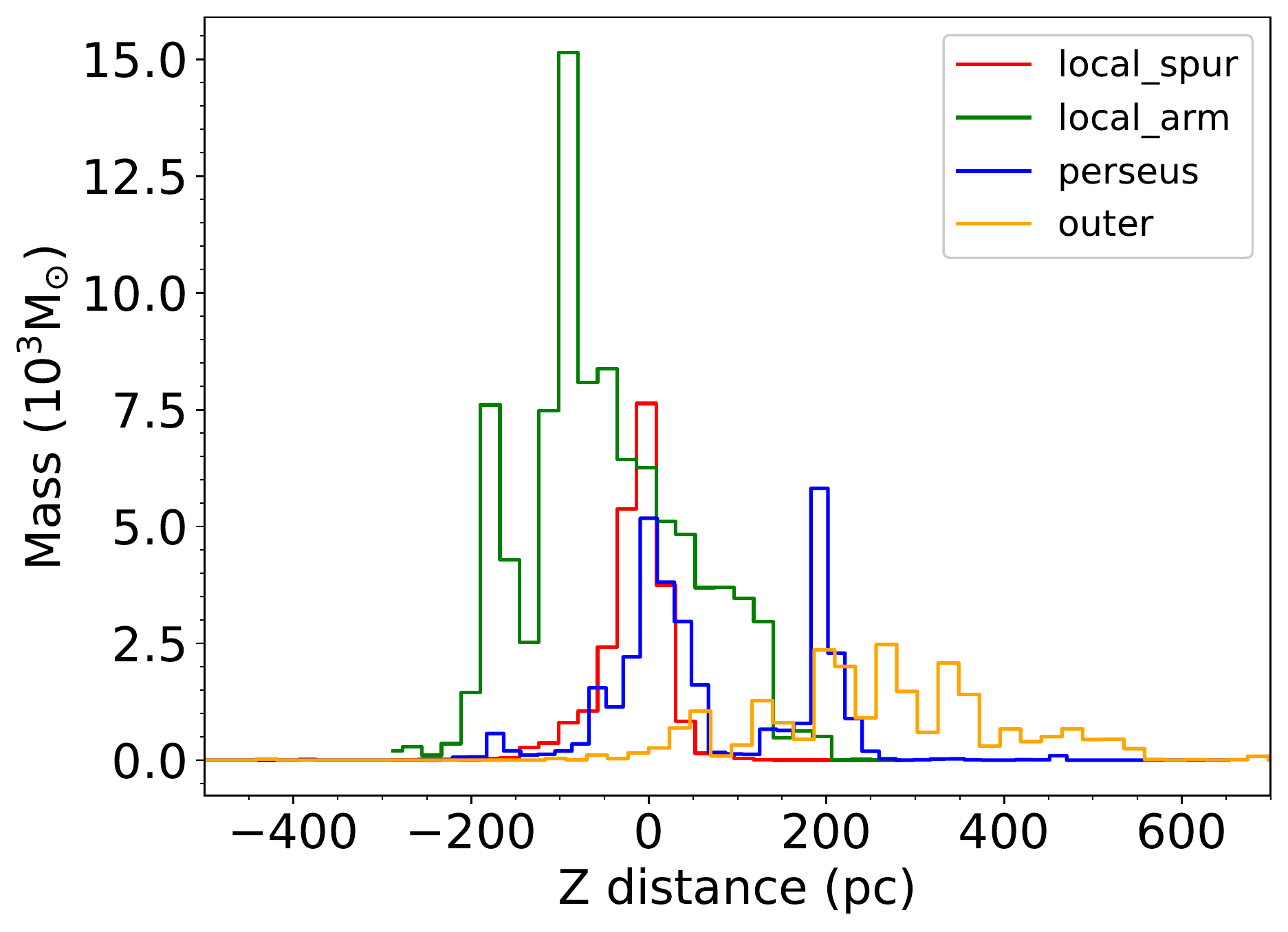}{0.5\textwidth}{(b)}
          } 
          
\gridline{\fig{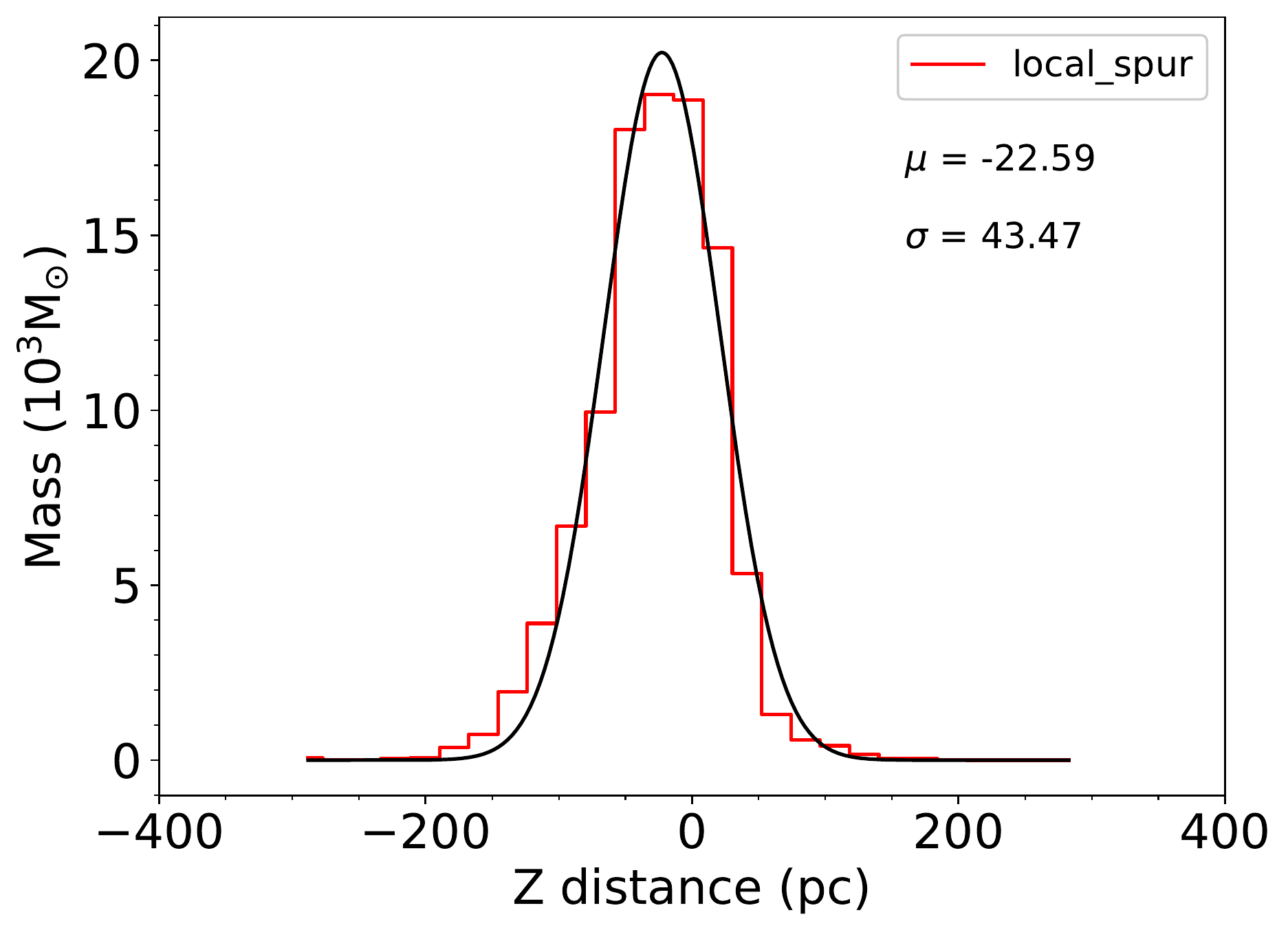}{0.25\textwidth}{(c)}
          \fig{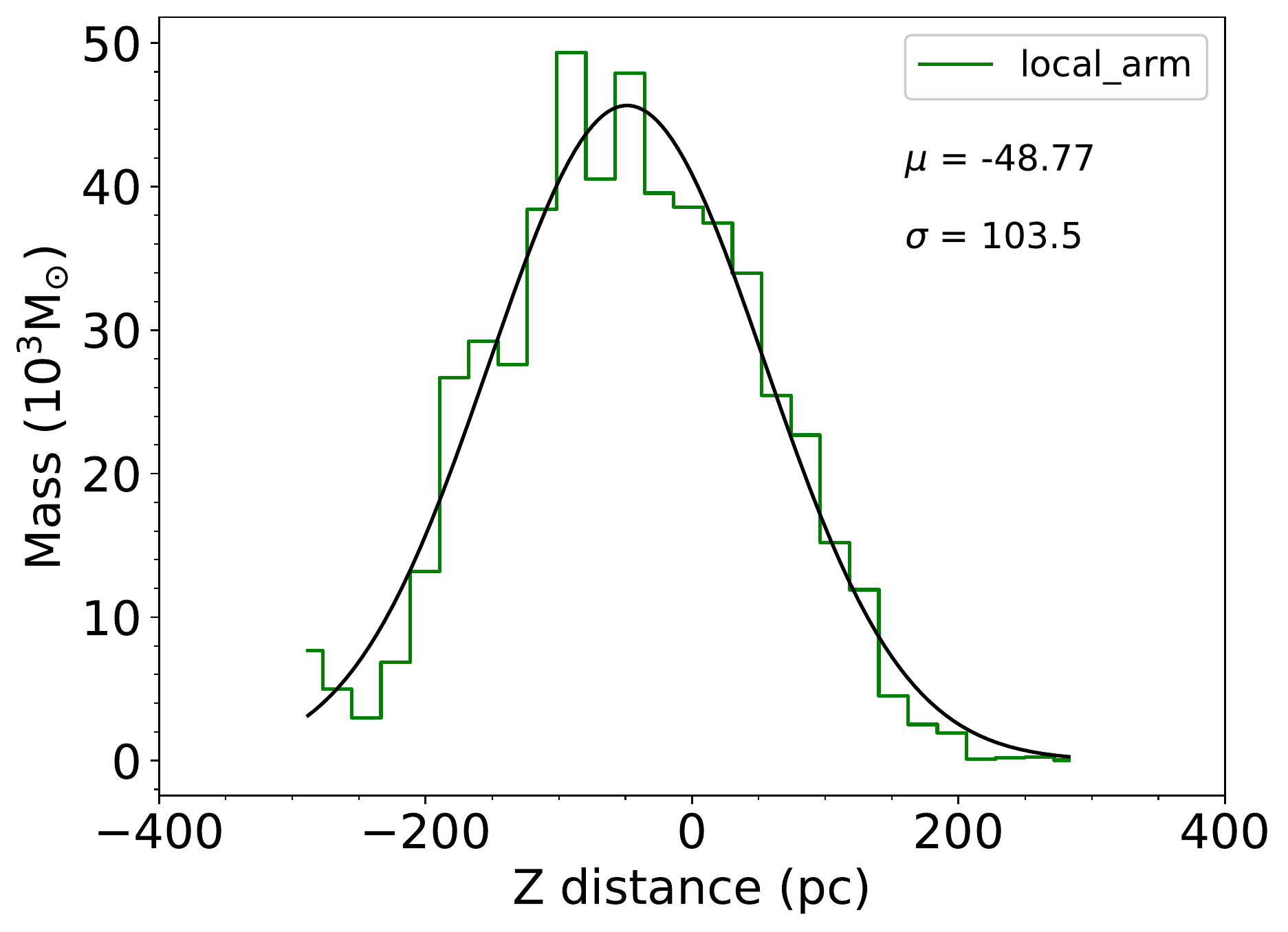}{0.25\textwidth}{(d)}
          \fig{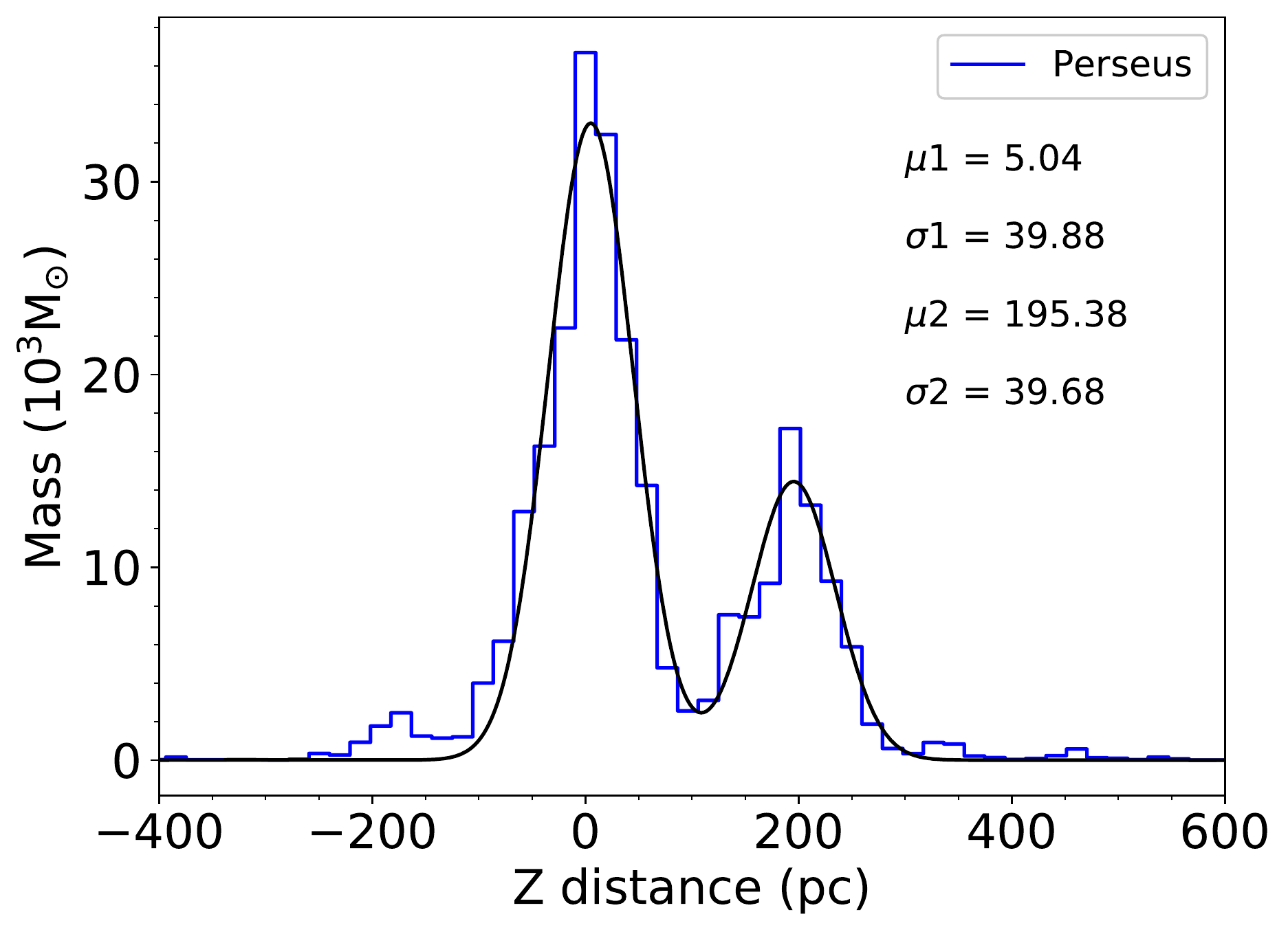}{0.25\textwidth}{(e)}
          \fig{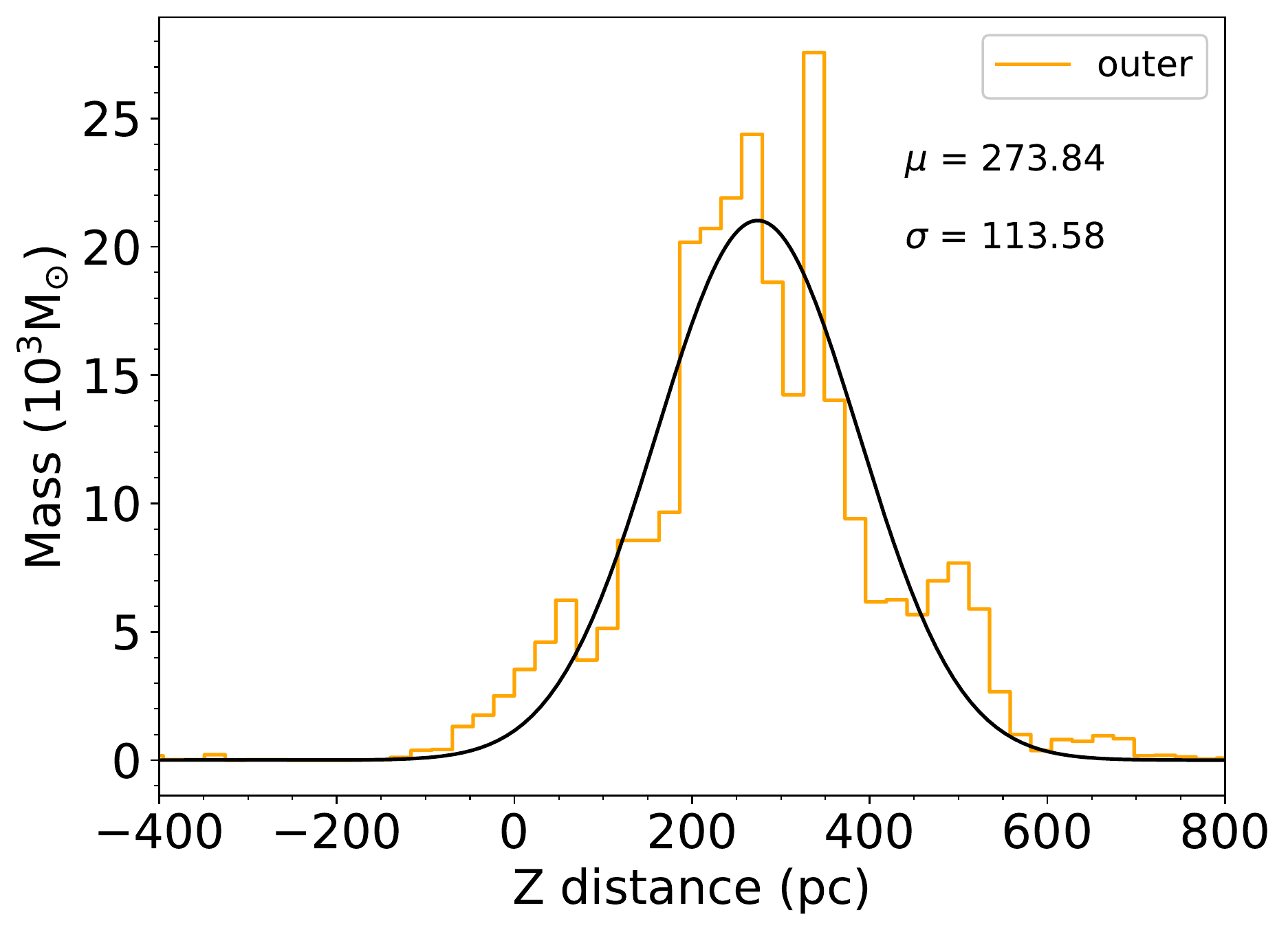}{0.25\textwidth}{(f)}          
          }
          
\gridline{\fig{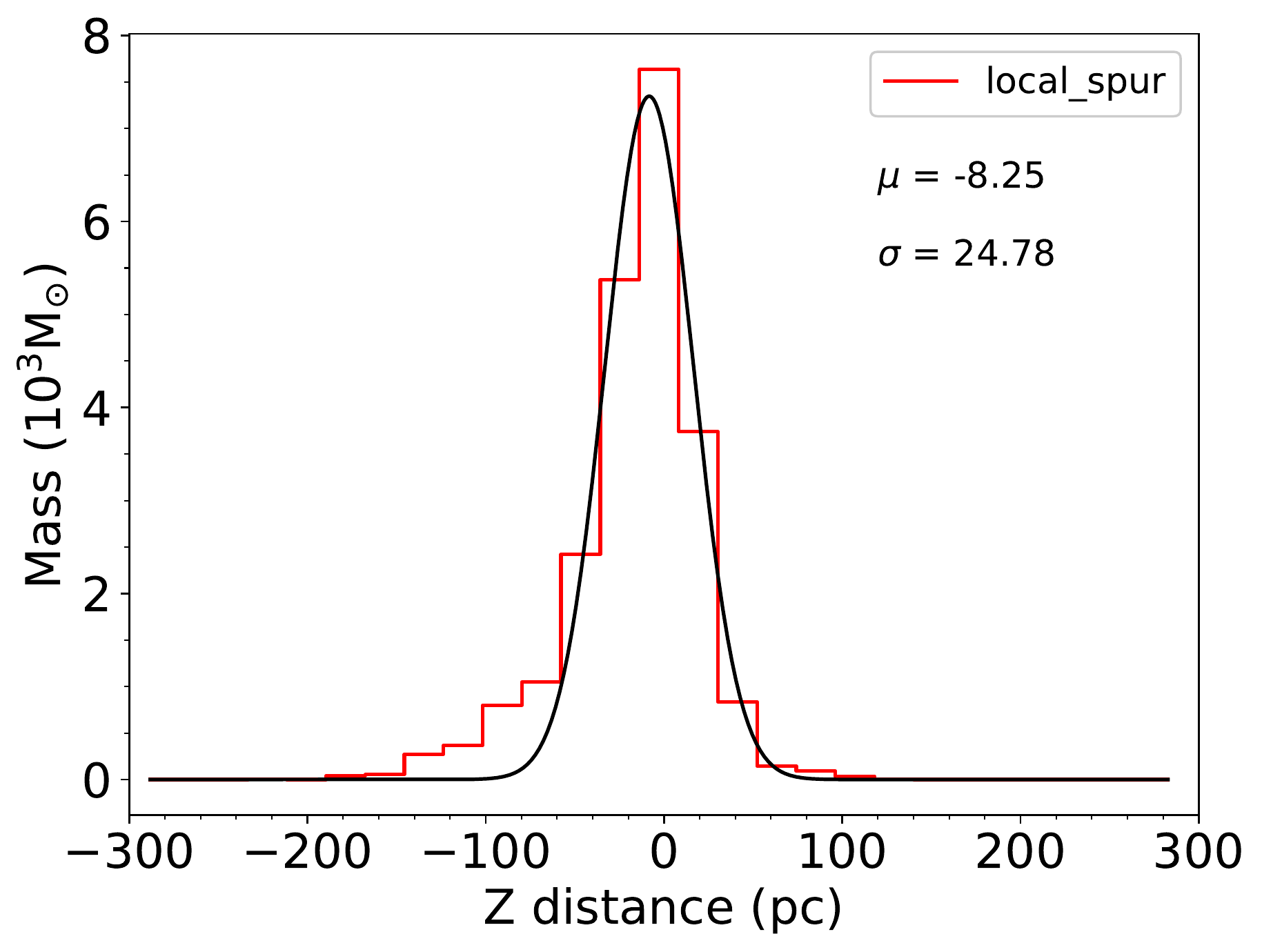}{0.25\textwidth}{(g)}
          \fig{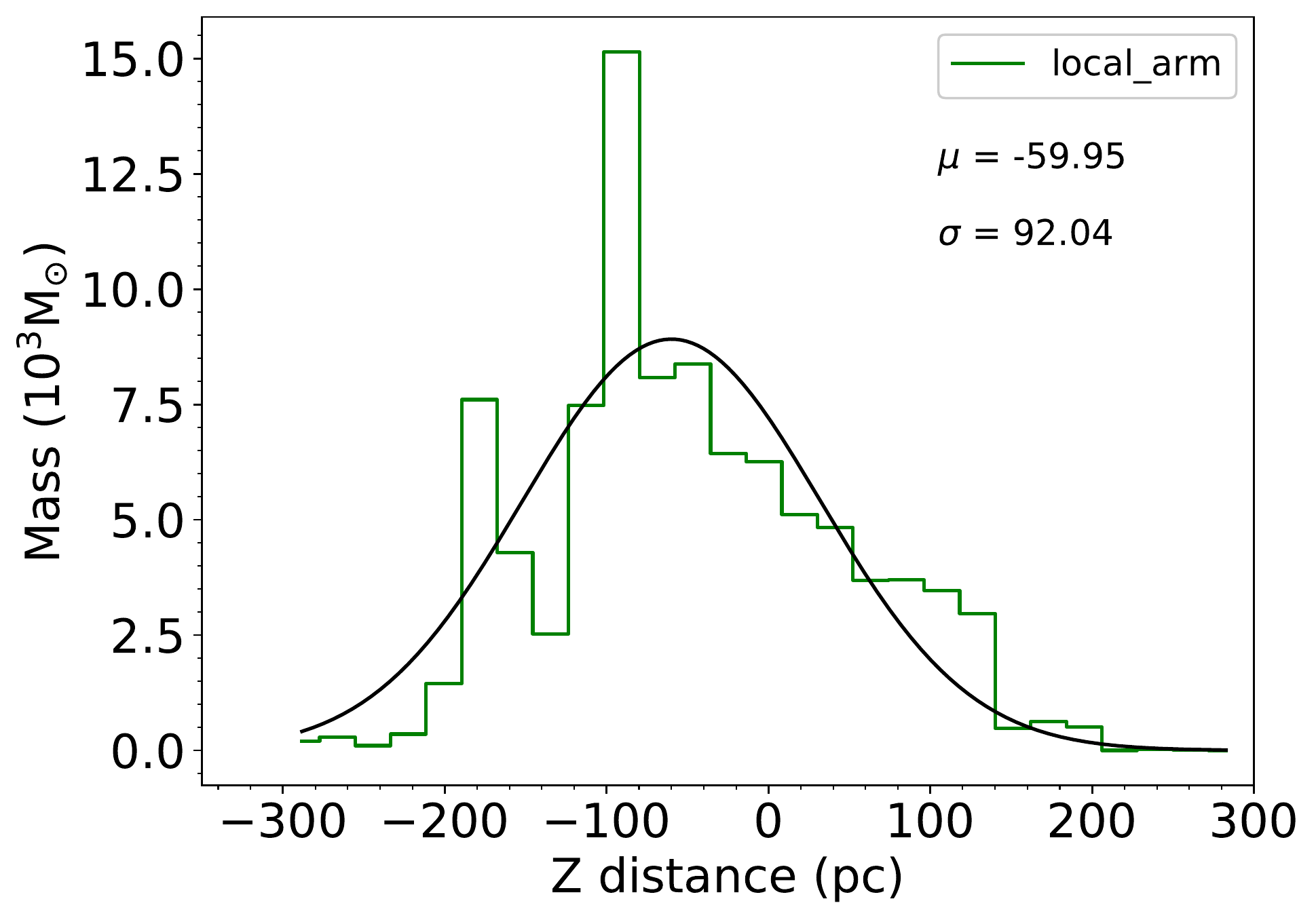}{0.25\textwidth}{(h)}
          \fig{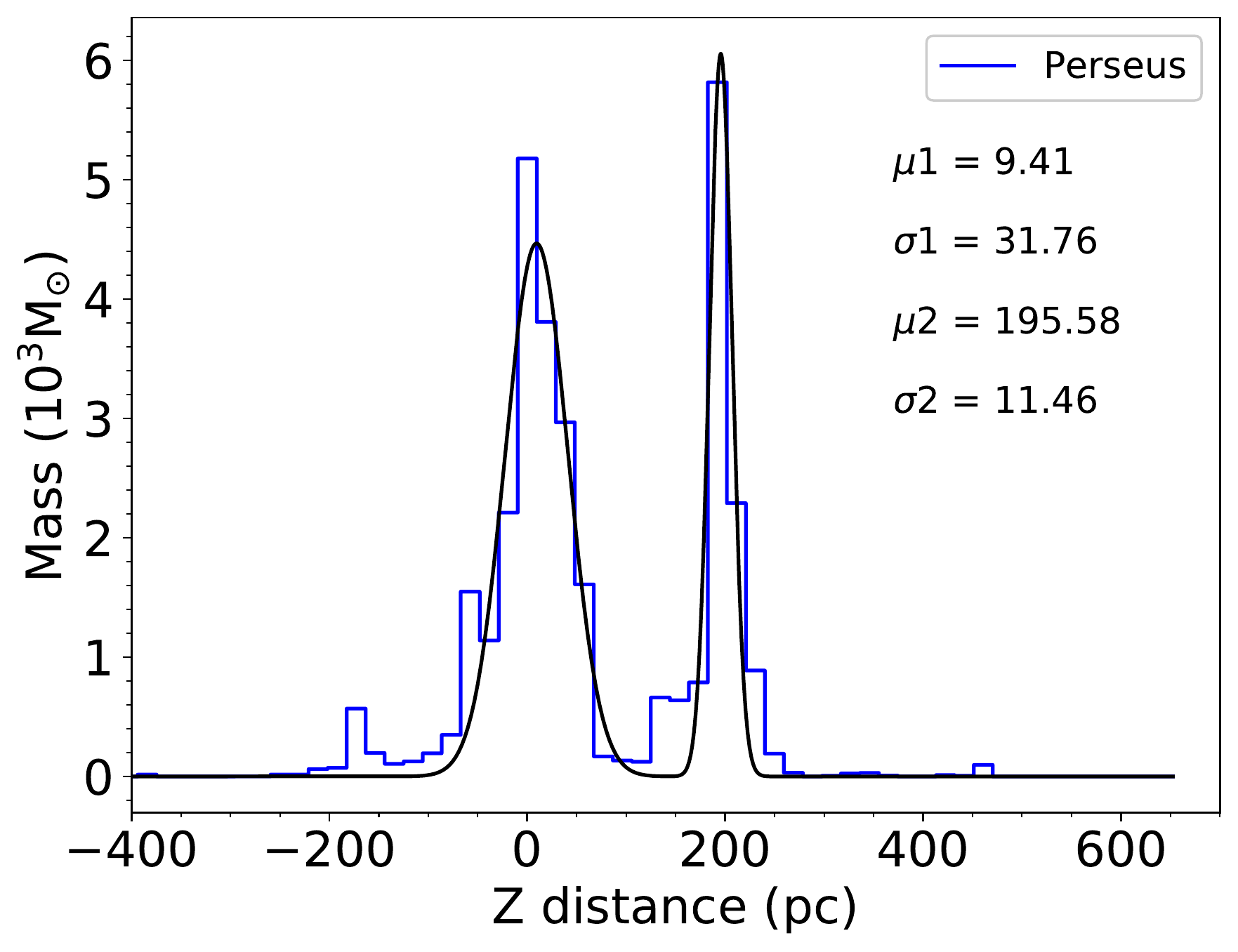}{0.25\textwidth}{(i)}
          \fig{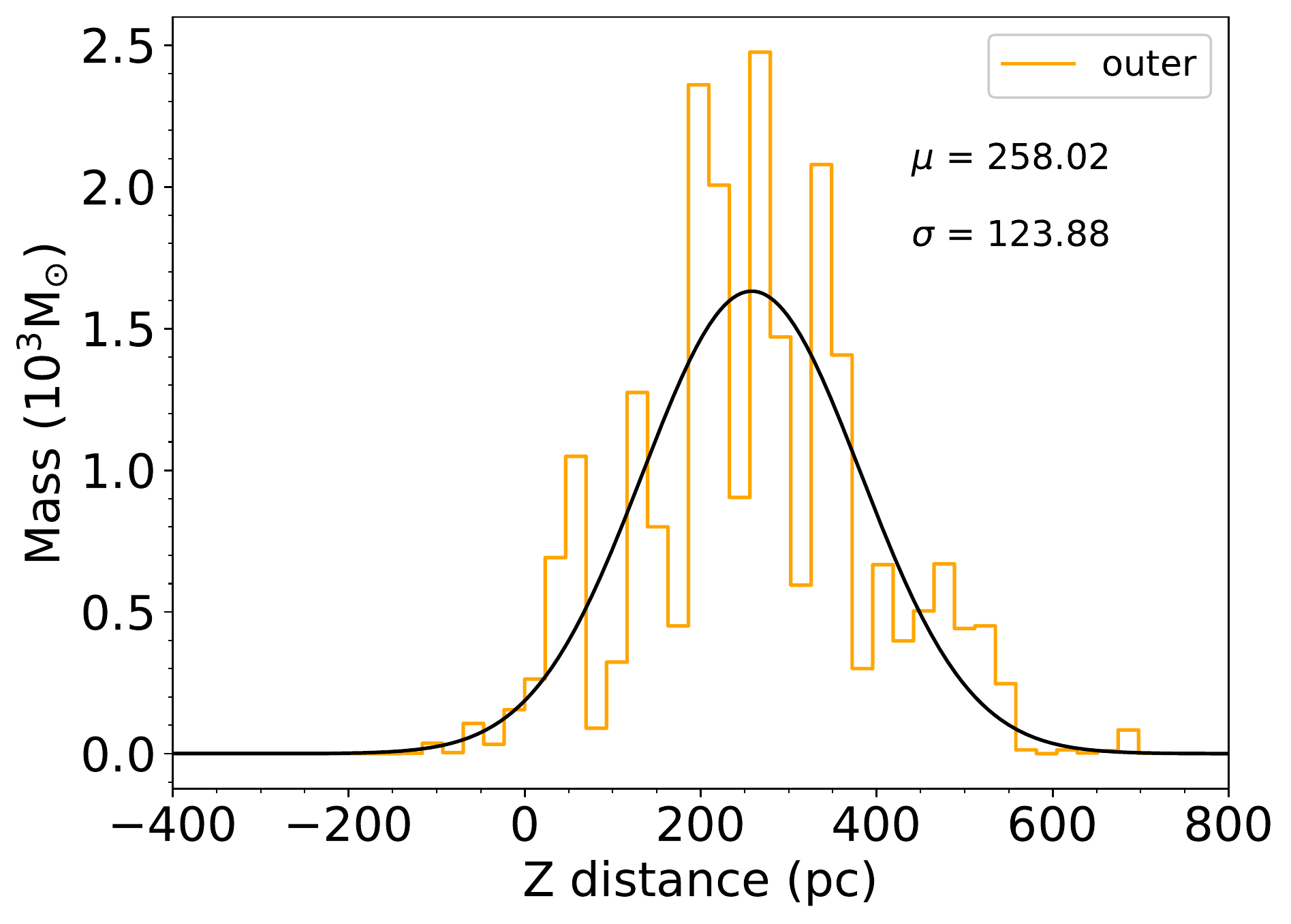}{0.25\textwidth}{(j)}}               
\caption{Mass distributions along the z direction for different spiral arms. The distributions for molecular gas  in the mask 1 and mask 2 regions are presented in (a) and (b), respectively. The Gaussian fittings to the distributions of $^{12}$CO gas (mask 1) are shown in (c)-(f) and those of $^{13}$CO gas (mask 2) in (g)-(j). The fitted parameters, $\mu$ and $\sigma$, are indicated in the top-right corners of the panels.}\label{fig:fig9}
\end{figure*}

We calculated the total mass within each 20 pc bin perpendicular to the Galactic plane. The vertical height of each bin is derived through $z = d sin(b)$, where d is the distance to the Sun of the molecular gas and b is the central Galactic latitude of the bin. Here, the distances estimated for each arm in Section \ref{subsec:Distance} are adopted. Figure \ref{fig:fig9} (a) and (b) show the vertical distribution of gas mass in Mask 1 estimated from $^{12}$CO emission using the X$\rm _{CO}$ factor method and that in Mask 2 from $^{13}$CO emission using the LTE method for different arms, respectively. The red, green, blue, and magenta histograms represent the gas in the Local Spur, Local arm, Perseus arm, and Outer arm, respectively. From Figures \ref{fig:fig9} (a) and \ref{fig:fig9} (b), we can see that the molecular gas in the Local spur and Local arm are generally concentrated around the Galactic mid-plane, with the molecular gas in the Local arm slightly displaced below z = 0 pc. The molecular gas in the Perseus arm is distributed mainly around z = 0 and z = 200 pc. A large positive displacement exists for the molecular gas in the Outer arm, which is quite consistent with the warping feature identified by \cite{2006PASJ...58..847N}. In the G60 region, the warp starts at a distance between the Perseus arm and the Outer arm.

For quantitative description, we also fit the vertical mass distributions in each arm with Gaussian functions, and the results are given in Figures \ref{fig:fig9} (c)$-$(f) for $^{12}$CO emission (Mask 1), and Figures \ref{fig:fig9} (g)$-$(j) for $^{13}$CO emission (Mask 2), respectively. For suitable fittings the bin widths in Figures \ref{fig:fig9} (c)$-$\ref{fig:fig9} (j) have been adjusted to the appropriate values. The thickness of the molecular gas layer does not increase monotonically with the distances of the molecular gas. At similar distances, the molecular gas in the Local arm has an FWHM of 243 pc measured from the $^{12}$CO emission, which is about 2.5 times that in the Local Spur. The FWHM thickness of the $^{13}$CO gas in the Local arm is 216 pc and its ratio to the thickness of the Local Spur is slightly higher than that obtained from $^{12}$CO emission. The vertical distribution of molecular gas in the Perseus arm can be well-fitted with two components of Gaussian functions. Combined with the spatial distribution of molecular gas in Figures \ref{fig:fig4} and \ref{fig:fig5}, the two Gaussian components correspond to two giant filaments parallel to the Galactic plane. The two Gaussian components have a vertical distance of $\sim$ 190 and 186 pc as seen from the $^{12}$CO and $^{13}$CO emission, respectively. The FWHM thickness of the z$\sim$200 pc component obtained from the $^{13}$CO emission is much smaller than that obtained from the $^{12}$CO emission. This feature of two parallel components in the G60 section of the Milky Way is first observed in this work. At the distance of $\sim$13 kpc, the thickness of the Outer arm gas is about 265 pc. Along the directions of l = 60$-$75\arcdeg, the distances 13 kpc corresponds to Galactocentric distance of 11.3$-$13.4 kpc. Therefore, the thickness of the Outer arm that we obtained is 1.6 times the thickness of the molecular disk found by in \cite{2006PASJ...58..847N} at a comparable Galactocentric radius. From Figures \ref{fig:fig9} (f) and (j), we can see that the molecular gas disk at the Galactocentric distances of 11.3$-$13.4 kpc, as traced by the Outer arm, warps upward with respect to the b = 0$\arcdeg$ plane with an amplitude of 273 pc from the $^{12}$CO emission, which is similar to that (258 pc) from $^{13}$CO emission.

\begin{deluxetable*}{ccclrlrc}
\tablenum{2}
\tablecaption{Parameters of Molecular Clouds in the OSC Arm\label{tab:tab2}}
\tablewidth{3pt}
\tablehead{
\colhead{Number} & \colhead{l}  & \colhead{b} &
\colhead{V$\rm _{lsr}$} & \colhead{T$\rm _{peak}$} & \colhead{Area} & \colhead{d}& \colhead{Mass} \\
\colhead{} & \colhead{($\circ$)} & \colhead{($\circ$)} & \colhead{(km s$^{-1}$)} &
\colhead{(K)} & \colhead{(arcmin$^{2}$)} & \colhead{(kpc)} & \colhead{(M$_{\odot }$)}
}
\decimalcolnumbers
\startdata
1 & 60.66 & 4.58 & $-$99.35 & 3.25 & 5.75 & 17.36 &497\\
2 & 63.78 & 3.18 & $-$87.7 & 4.11 & 11.50 & 14.92& 367\\
3 & 64.49 & 3.14 & $-$89.41 & 2.94 & 3.25 &  14.95 &368\\
4 & 64.64 & 3.01 & $-$89.46 & 2.38 & 5.00 & 14.92& 367\\
5 & 64.67 & 2.88 & $-$90.18 & 2.35 & 2.25 & 15.00 &371\\
6 & 64.69 & 2.93 & $-$89.37 & 2.26 & 3.75 & 14.90&366 \\
7 & 65.32 & 2.94 &$-$90.17 & 2.39 & 2.25 & 14.84 &363\\
8 & 65.49 & 4.60 & $-$110.94 & 2.93 & 3.50 & 17.77& 520\\
9 & 65.83 & 4.78 & $-$111.19 & 2.79 & 4.50 & 17.71&517 \\
10 & 67.12 & 3.21 & $-$95.87 & 2.80 & 4.50 & 15.12 & 377 \\
11 & 68.70 & 3.31 & $-$95.72 & 3.23 & 6.75 & 14.72 & 357 \\
12 & 70.46 & 3.04 & $-$102.14 & 3.38 & 14.00 & 15.14 & 378 \\
13 & 70.62 & 2.94 & $-$102.08 & 2.72 & 2.50 & 15.09 & 375 \\
14 & 70.67 & 3.10 & $-$101.57 & 4.33 & 7.25 & 15.01 & 371 \\
15 & 73.73 & 3.07 & $-$105.42 & 2.46 & 4.75 & 14.81 & 361 \\
\enddata
\tablecomments{Column 1: cloud number which is organized by increasing galactic longitude. Columns 2$-$6 give the centroid position, the centroid velocity, the peak intensity, and the angular area of the clouds. The distance and the mass of the clouds are presented in columns 7$-$8. }
\end{deluxetable*}

\section{Summary} \label{sec:Summary}
Using the multi-line MWISP survey data, we performed a comprehensive study of the physical properties of the molecular gas in the range of l = 59.75$^\circ$ $\sim$ 74.75$^\circ$ and b = $-$5.25$^\circ$ $\sim$ 5.25$^\circ$. The large-scale distribution of the molecular gas is presented. We used the DBSCAN algorithm to identify molecular clouds traced by $\rm ^{12}CO$ emission in the velocity range of $-$119 $\sim$ $-$87 km s$^{-1}$. We analyed the vertical distribution of the molecular gas in this region. The main results of this study are presented as follows:
\begin{itemize}
    \item[1.] The molecular gas in the observed region can be attributed to one structure and three arms, i.e., the Local spur, Local arm, Perseus arm, and Outer arm. We present the distribution and velocity structures of the molecular gas in this part of the Galactic plane in detail for the first time. Detailed statistics of the properties of the molecular gas such as excitation temperature, optical depth, and column density are calculated for each arm and the Local spur. 
    \item[2.] We identified 15 extremely distant molecular clouds with kinematic distances of 14.72$-$17.77 kpc and masses of 363$-$520 M$_{\odot}$, which we find could be part of the Scutum-Centaurus arm in the outer Galaxy identified by \cite{2011ApJ...734L..24D} and \cite{2015ApJ...798L..27S}. There is also a possibility that 12 of these extremely molecular clouds may be an independent structure between the Outer and the OSC arms, or a spur.

    \item[3.] We observed a warp for the molecular gas in the Outer spiral arm and a two-component distribution of the molecular gas in the Perseus spiral arm. The two Gaussian components in the vertical distribution of the molecular gas in the Perseus arm in the G60 region are observed for the first time, and they correspond to two giant filaments parallel to the Galactic plane. 
\end{itemize}

\clearpage

\begin{acknowledgments}
We thank the anonymous referee for the constructive suggestions that helped to improve this manuscript. This research made use of the data from the Milky Way Imaging Scroll Painting (MWISP) project, which is a multi-line survey in $\rm {}^{12}{CO}$/$\rm{}^{13}{CO}$/$\rm{C}{}^{18}{O}$ along the northern Galactic plane with the PMO-13.7m telescope. We are grateful to all the members of the MWISP working group, particularly the staff members at PMO-13.7m telescope, for their long-term support. MWISP was sponsored by National Key R${\rm \&}$D Program of China with grant 2017YFA0402701 and  CAS Key Research Program of Frontier Sciences with grant QYZDJ-SSW-SLH047. This publication utilizes the GALFA-\ion{H}{1} survey data set obtained with the Arecibo L-band Feed Array (ALFA) on the Arecibo 305 m telescope. The Arecibo Observatory is operated by SRI International under a cooperative agreement with the National Science Foundation (AST-1100968), and in alliance with Ana G. Méndez-Universidad Metropolitana and the Universities Space Research Association. The GALFA-\ion{H}{1} surveys have been funded by the NSF through grants to Columbia University, the University of Wisconsin, and the University of California. We acknowledge the support of NSFC grants 11973091 and 12173090. C.L. acknowledges the supports by China Postdoctoral Science Foundation No.2021M691532, and Jiangsu Postdoctoral Research Funding Program No. 2021K179B. Y.S. acknowledges support by the Youth Innovation Promotion Association, CAS (Y2022085), and Light of West China Program, CAS (2022-XBJCTD-003). 
\end{acknowledgments}
%


\bibliography{sample631}{}
\bibliographystyle{aasjournal}

\begin{figure*}[htb]
\gridline{\fig{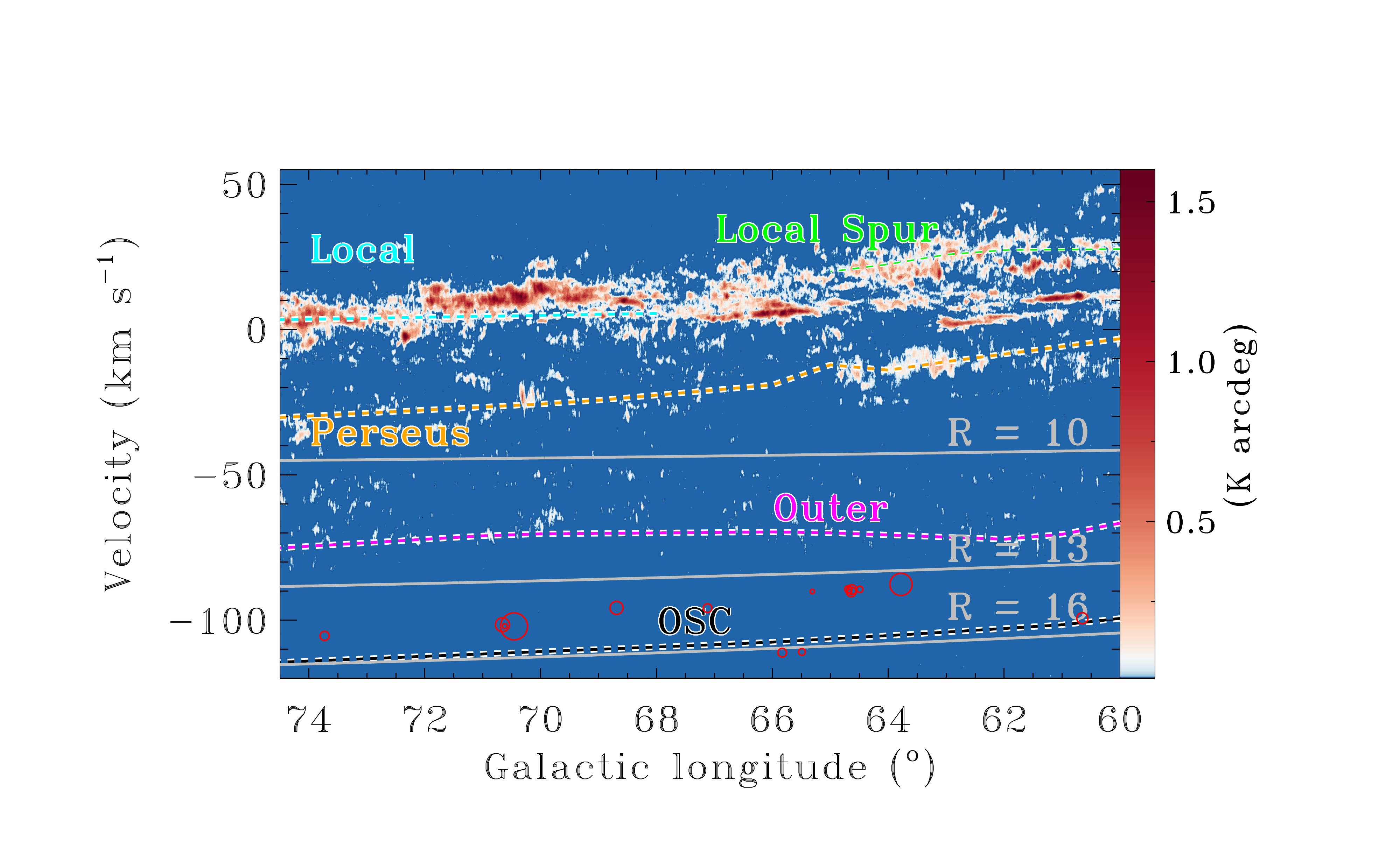}{0.85\textwidth}{(a)}}
{\vskip -5em}
\gridline{\fig{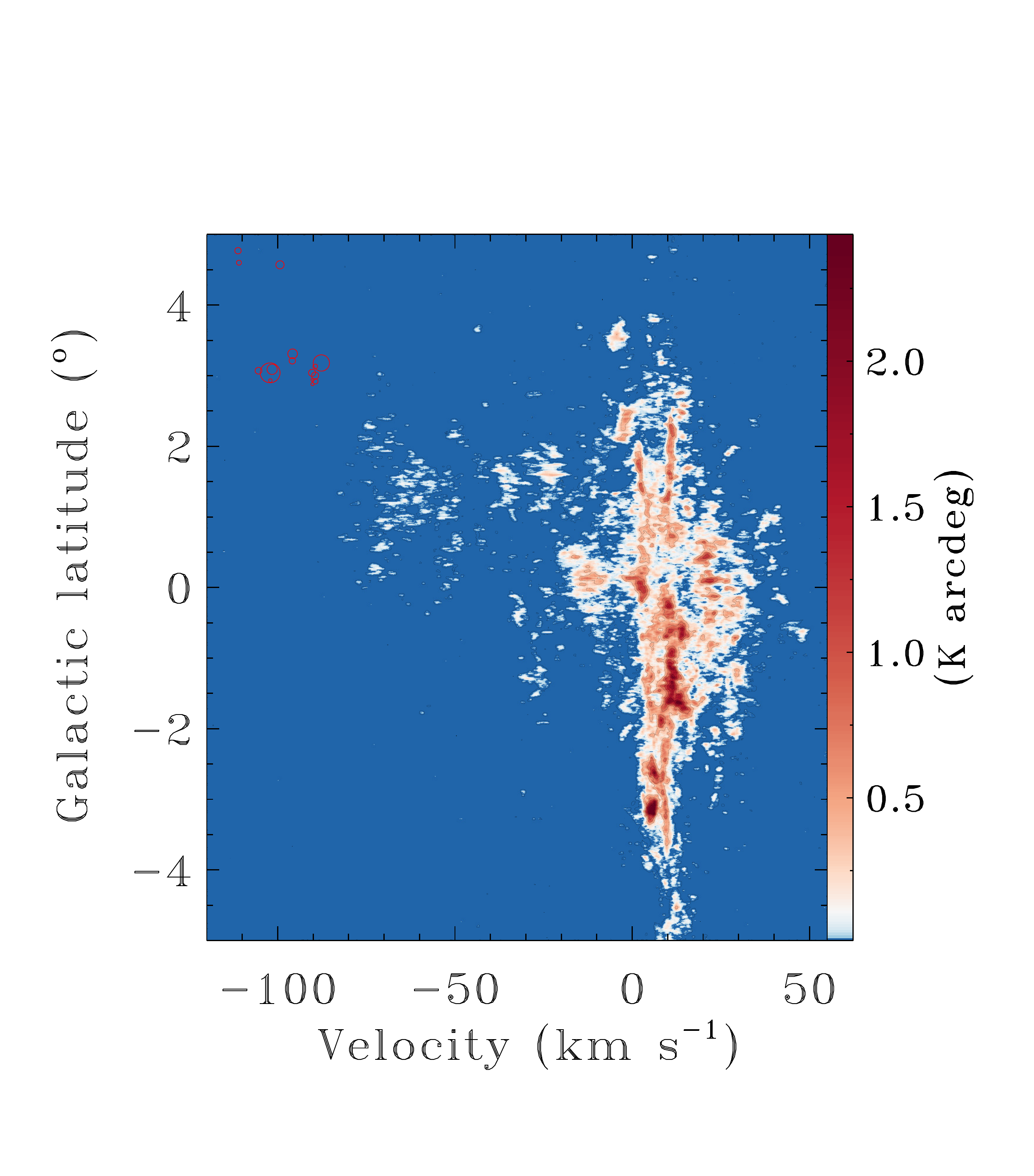}{0.57\textwidth}{(b)}}            
\caption{Same as Figure~\ref{fig:fig2} but for $^{13}$CO line emission.  }
\label{fig:figA2}
\end{figure*}

\begin{figure*}[ht]
\gridline{\fig{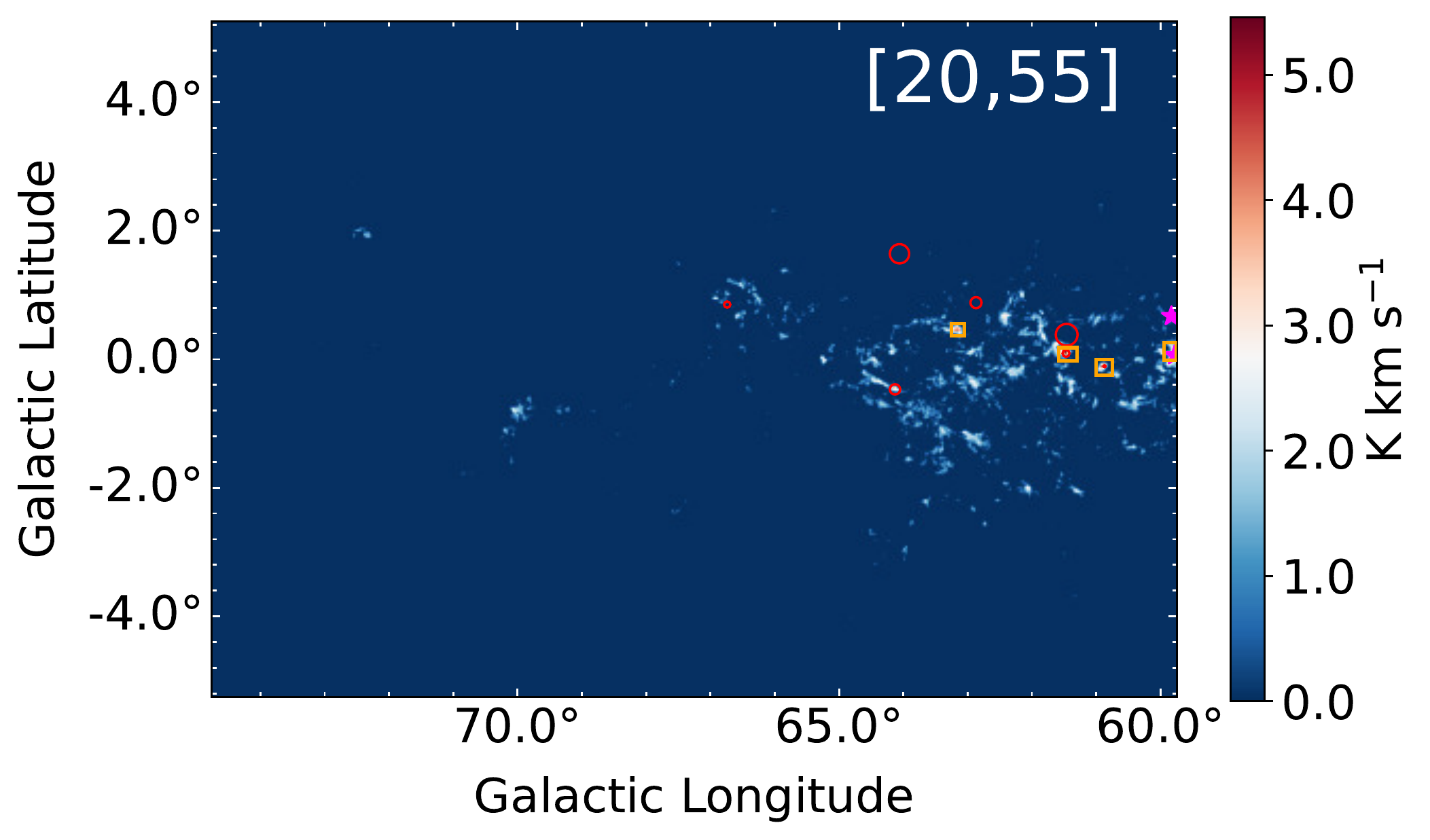}{0.45\textwidth}{(a)}
          \fig{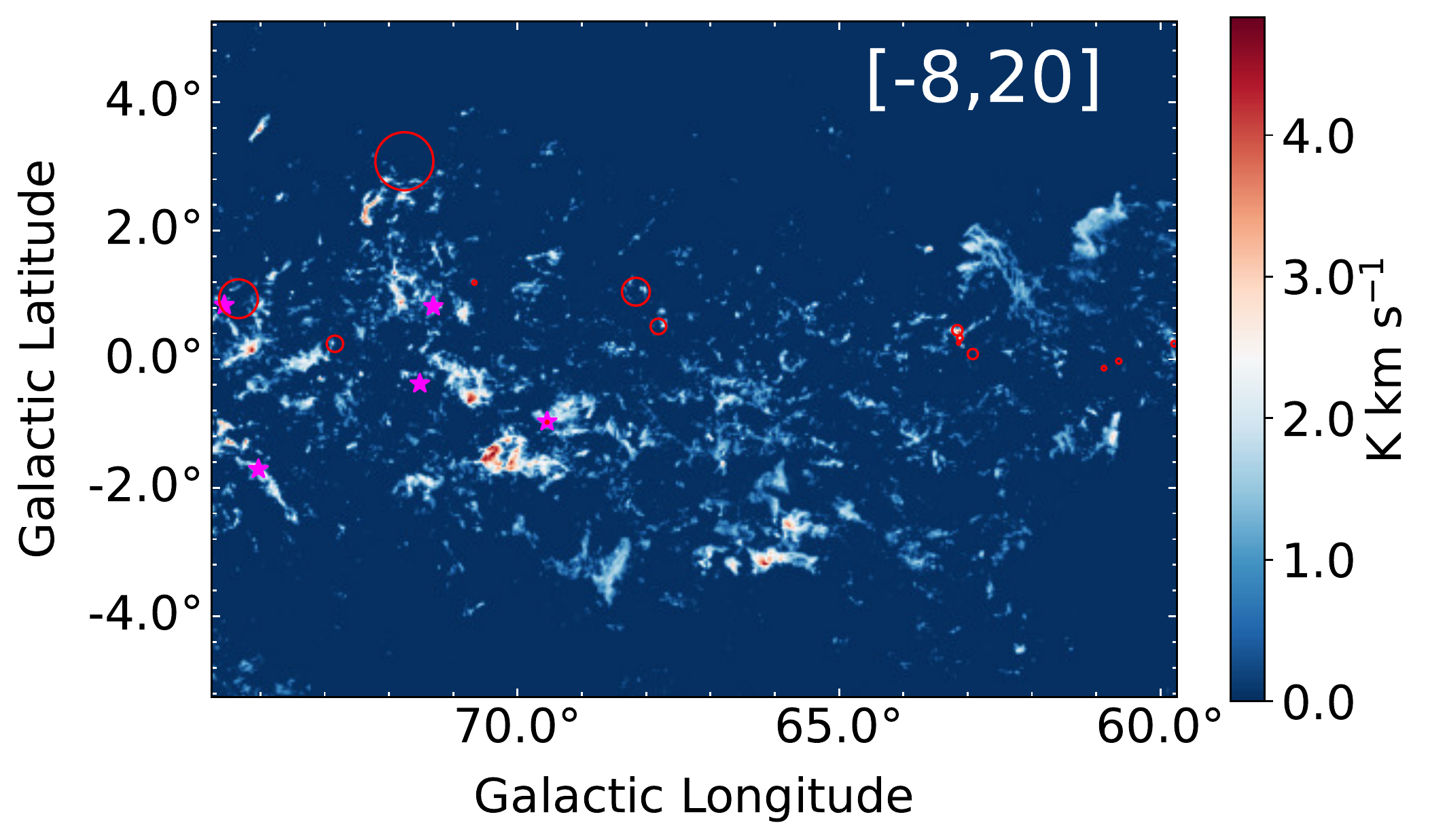}{0.45\textwidth}{(b)}
          }
          
\gridline{\fig{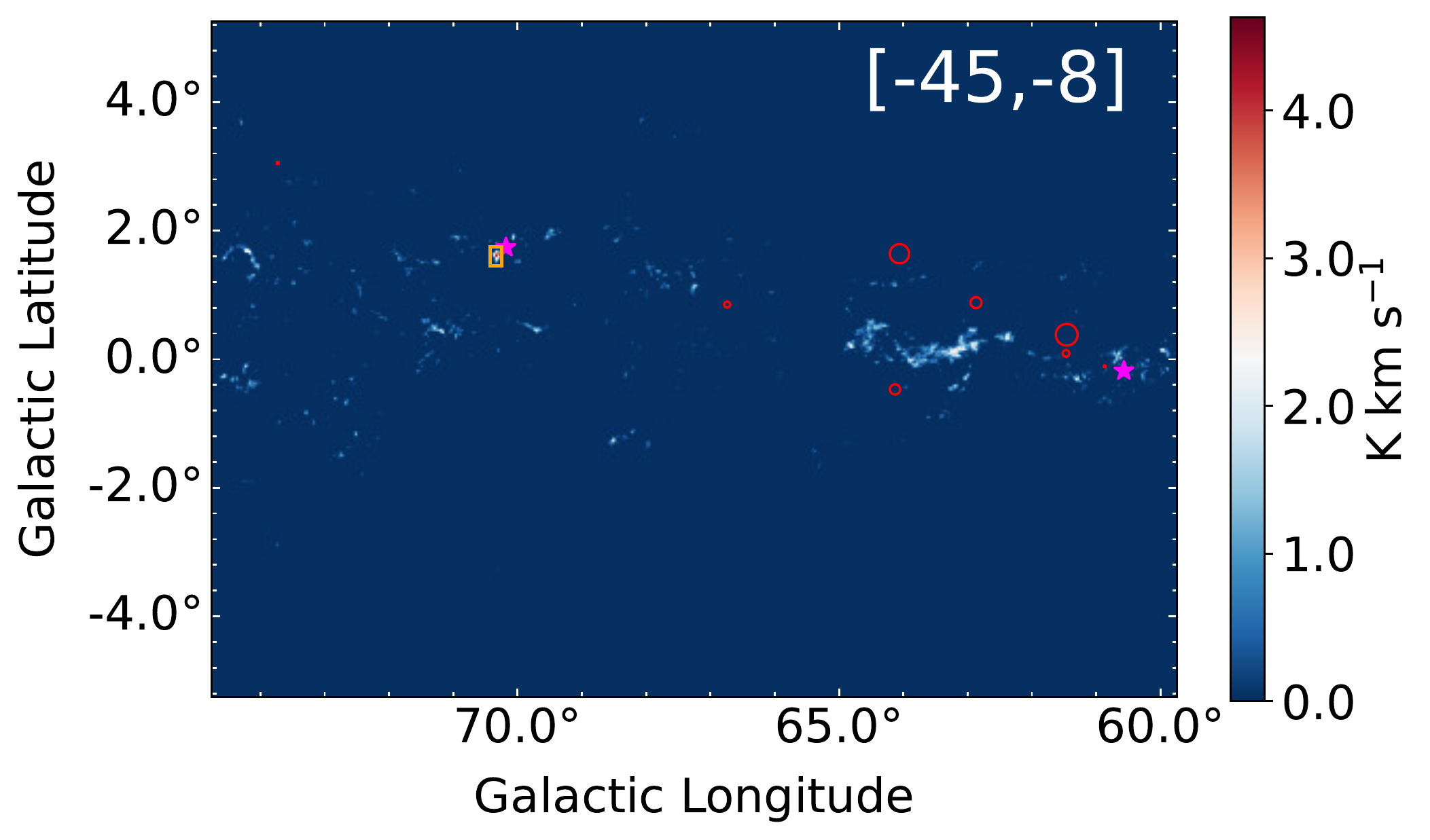}{0.45\textwidth}{(c)}
          \fig{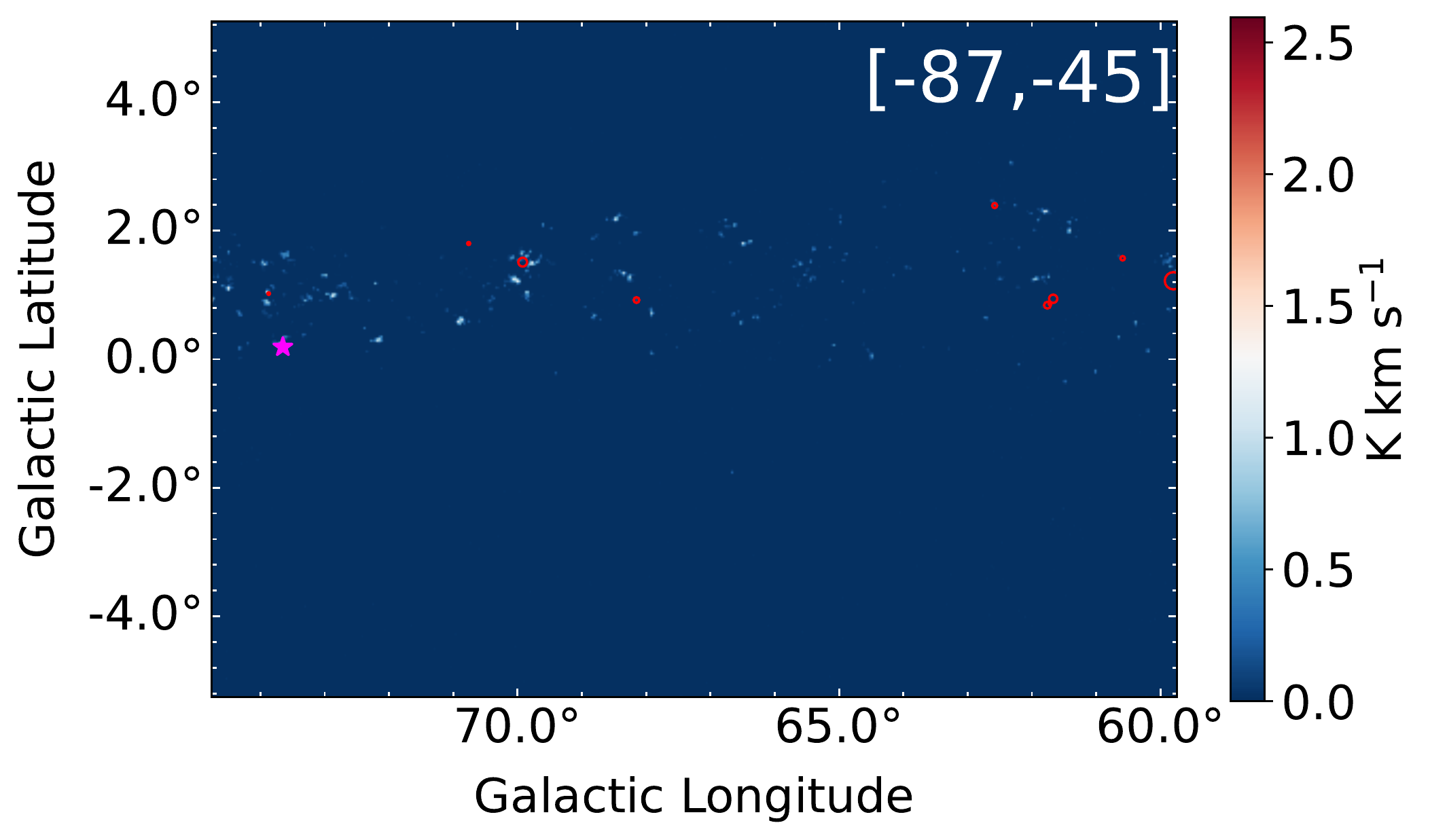}{0.45\textwidth}{(d)}
          }
\caption{Same as Figure \ref{fig:fig4} but for $^{13}$CO line emission. The orange boxes in (a) and (c) mark the regions with high-column density.}
\label{fig:fig4_13CO}
\end{figure*}

\begin{figure*}[ht!]
\includegraphics[width=14.0cm, trim = -2cm -2cm 1cm -0.5cm]{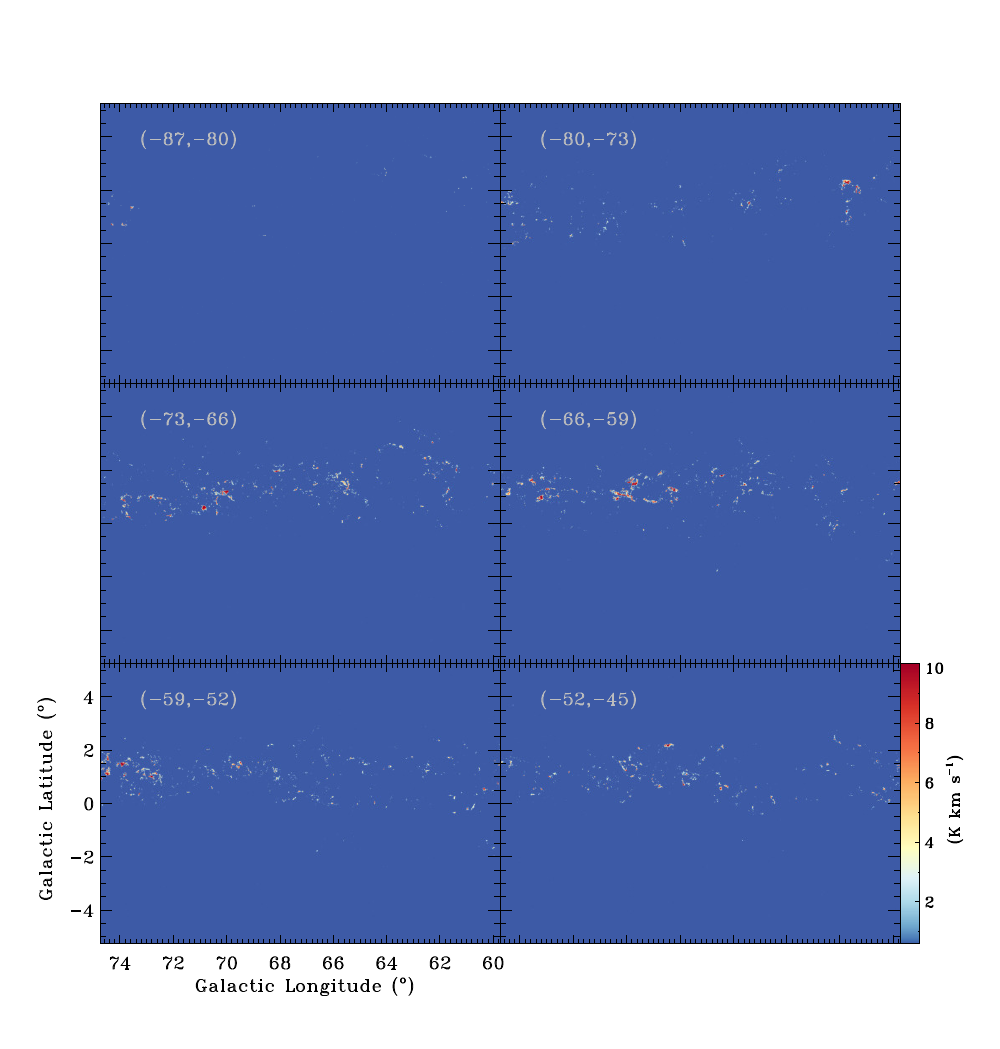}
\caption{Velocity channel maps of $\rm^{12}$CO from ${-}$87 to $-$45 km s$^{-1}$ of the Outer arm. The velocity range of each channel is shown at the top-left corner of each panel.
\label{fig:fig33channel}}
\end{figure*}

\begin{figure*}[ht!]
\includegraphics[width=13.0cm, trim = -2cm 1cm 1.5cm 0.5cm]{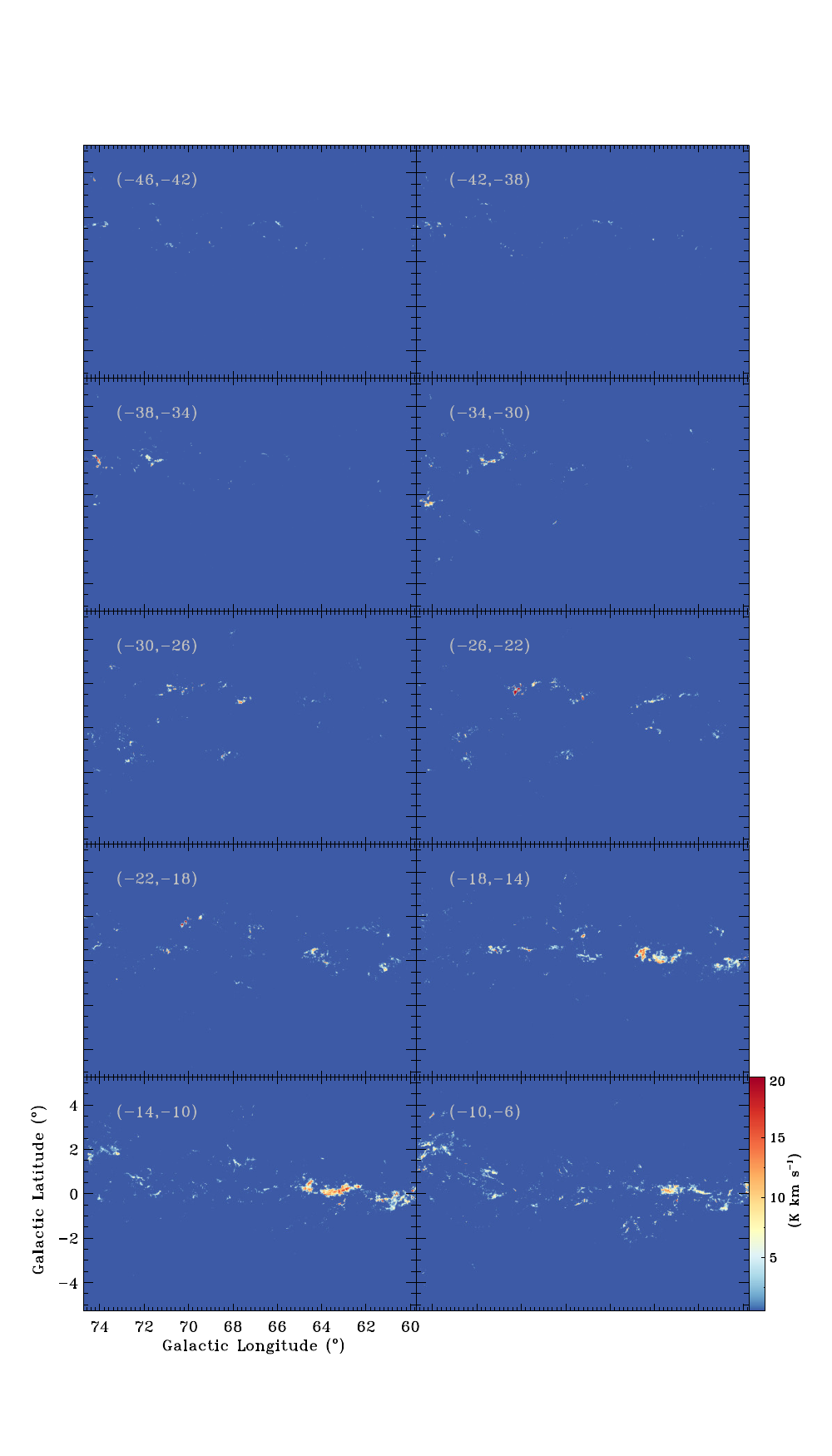}
\caption{Velocity channel maps of $\rm^{12}$CO from ${-}$45 to $-$8 km s$^{-1}$ of the Perseus arm. The velocity range of each channel is shown at the top-left corner of each panel.
\label{fig:fig33channel_1}}
\end{figure*}

\begin{figure*}[ht!]
\includegraphics[width=12.0cm, trim = -2.5cm 2cm 1.3cm 2.5cm]{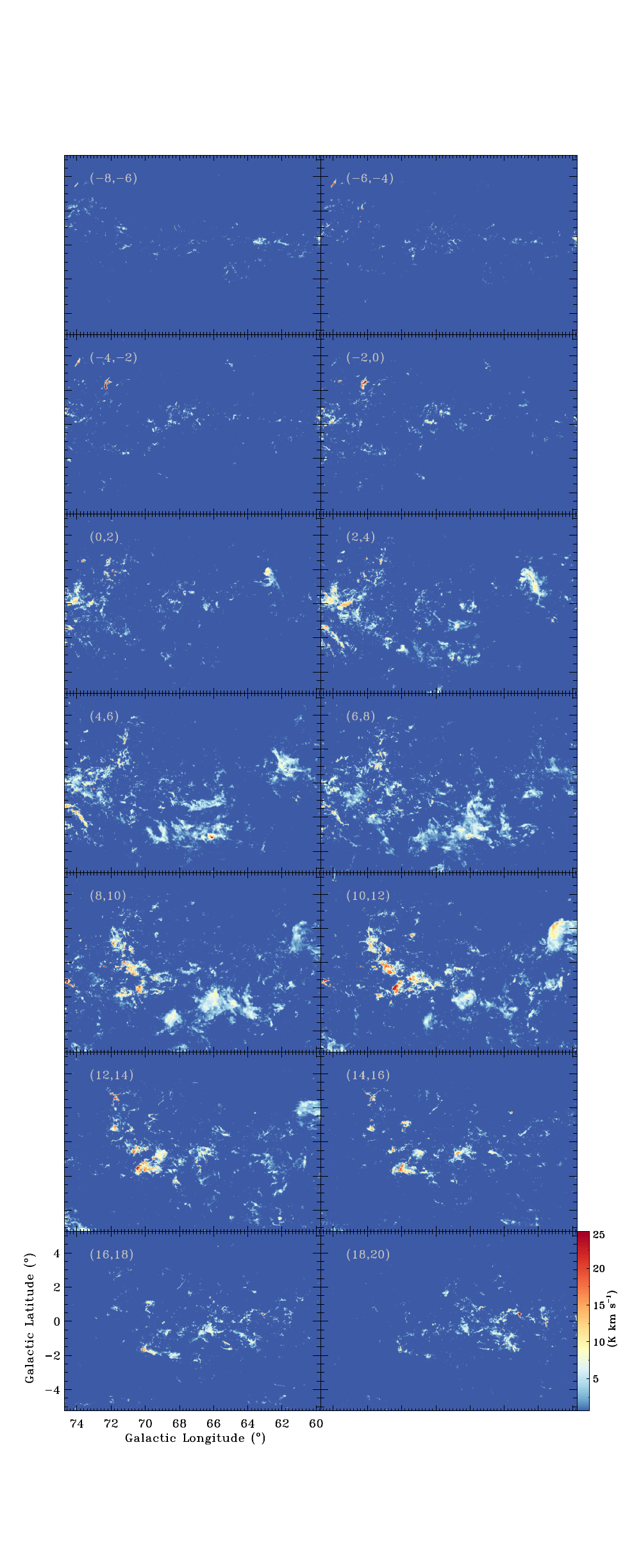}
\caption{Velocity channel maps of $\rm^{12}$CO from ${-}$8 to 20 km s$^{-1}$ of the Local arm. The velocity range of each channel is shown at the top-left corner of each panel.
\label{fig:fig33channel_2}}
\end{figure*}

\begin{figure*}[ht!]
\includegraphics[width=12.0cm, trim = -1.0cm 1.5cm 1.5cm 1.5cm]{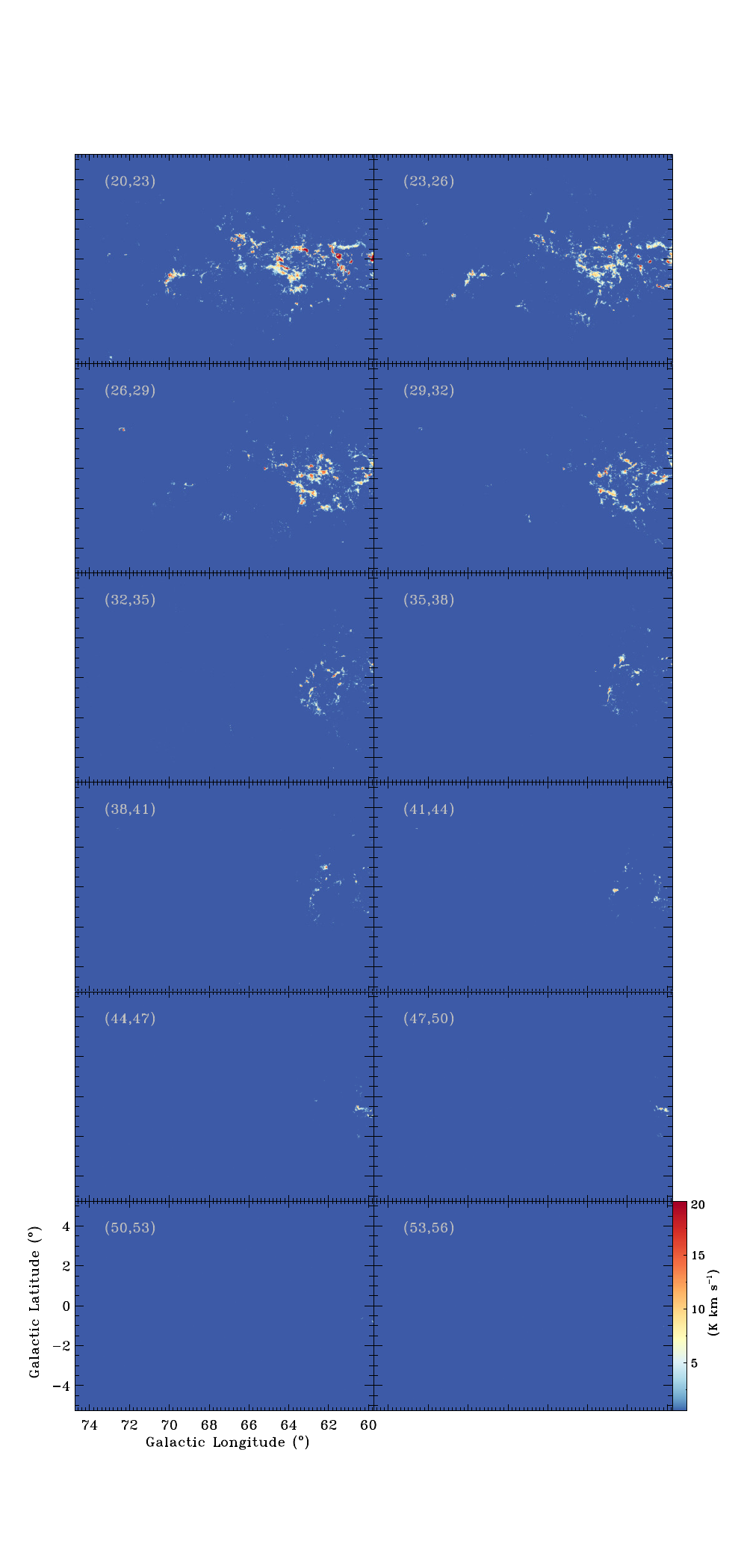}
\caption{Velocity channel maps of $\rm^{12}$CO from 20 to 55 km s$^{-1}$ of the Local Spur. The velocity range of each channel is shown at the top-left corner of each panel.
\label{fig:fig33channel_3}}
\end{figure*}

\begin{figure*}[h]         
          \gridline{\fig{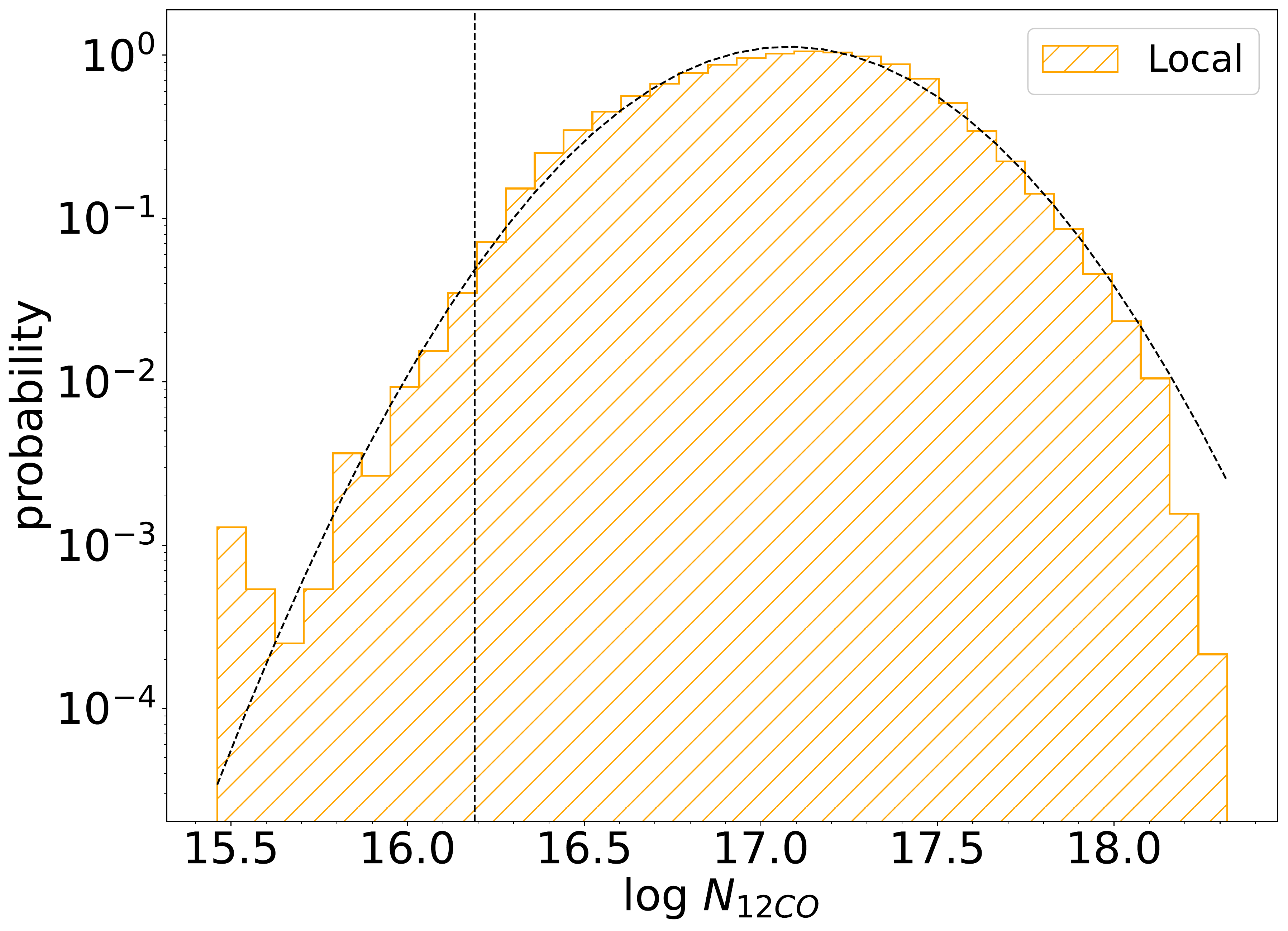}{0.25\textwidth}{(a)}
          \fig{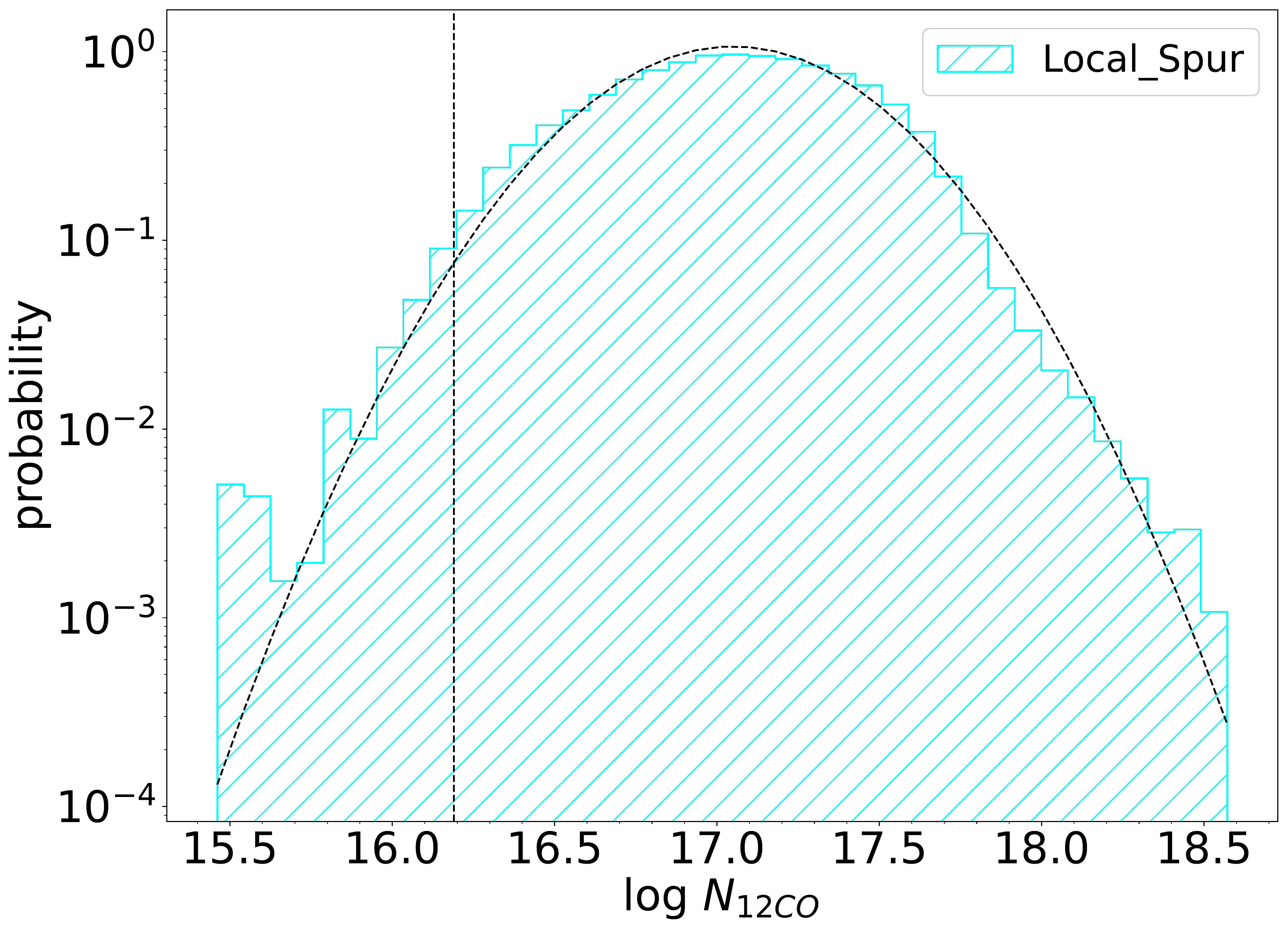}{0.25\textwidth}{(b)}
          \fig{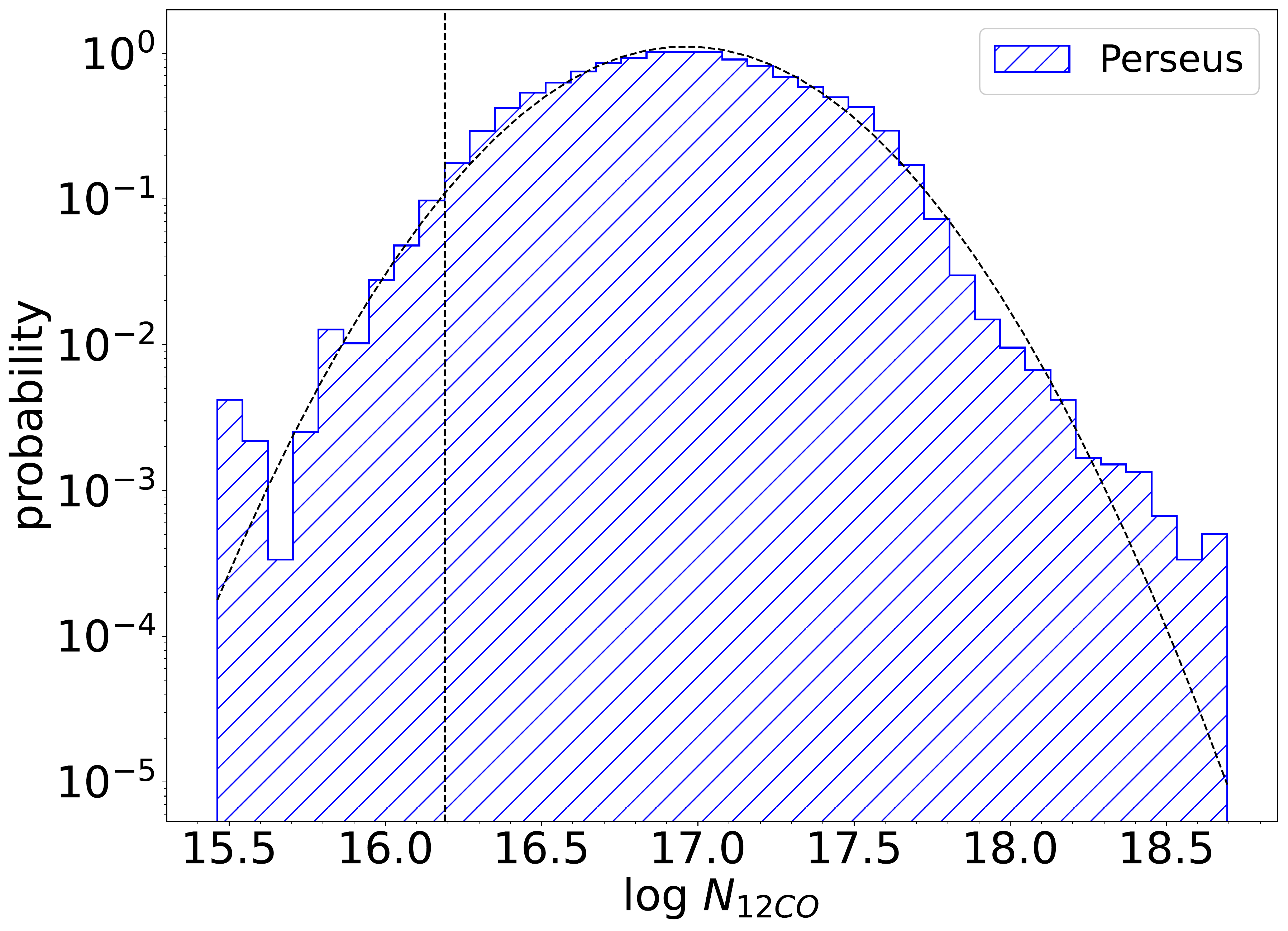}{0.25\textwidth}{(c)}
          \fig{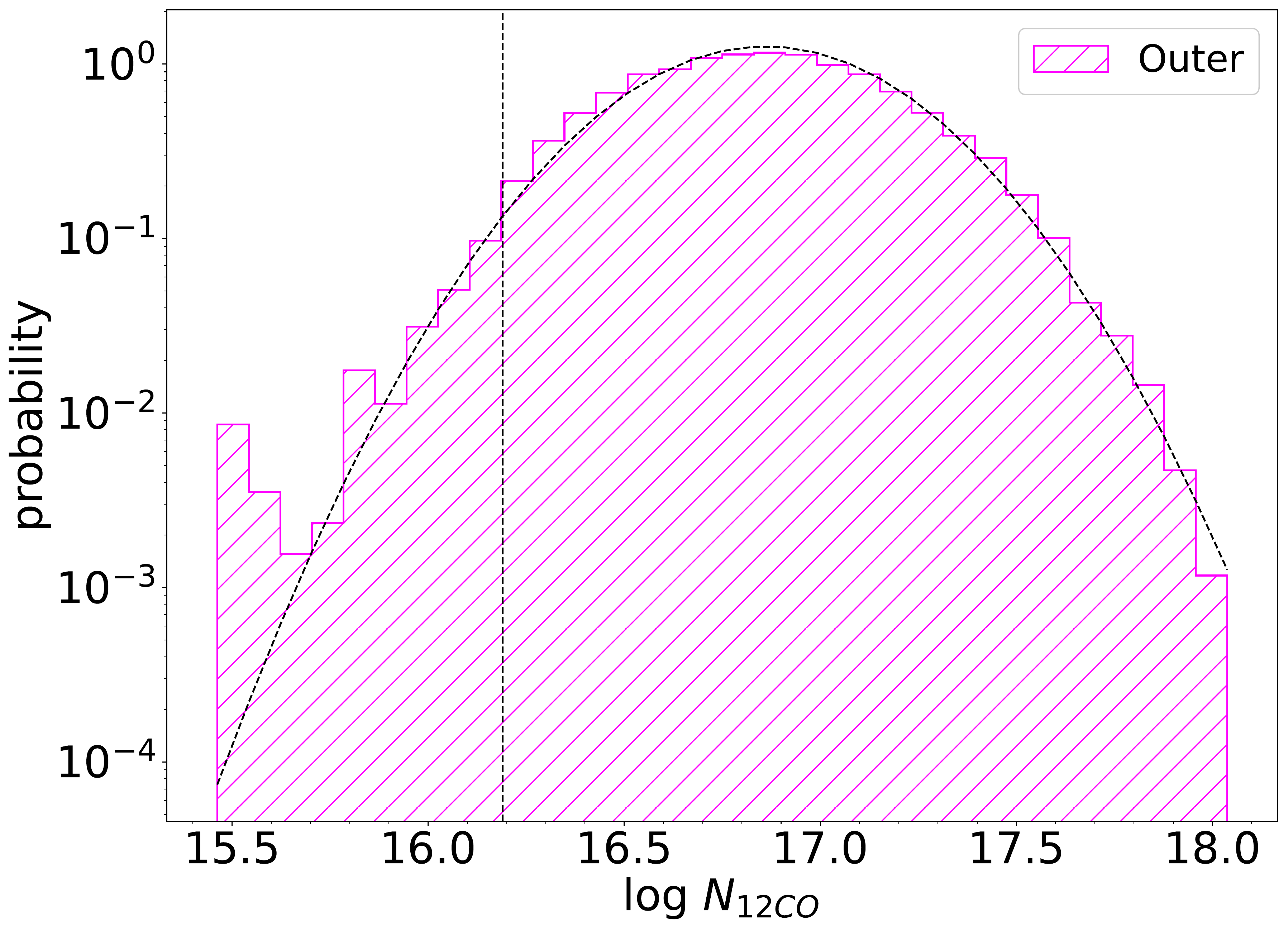}{0.25\textwidth}{(d)}          
          }
          
\gridline{\fig{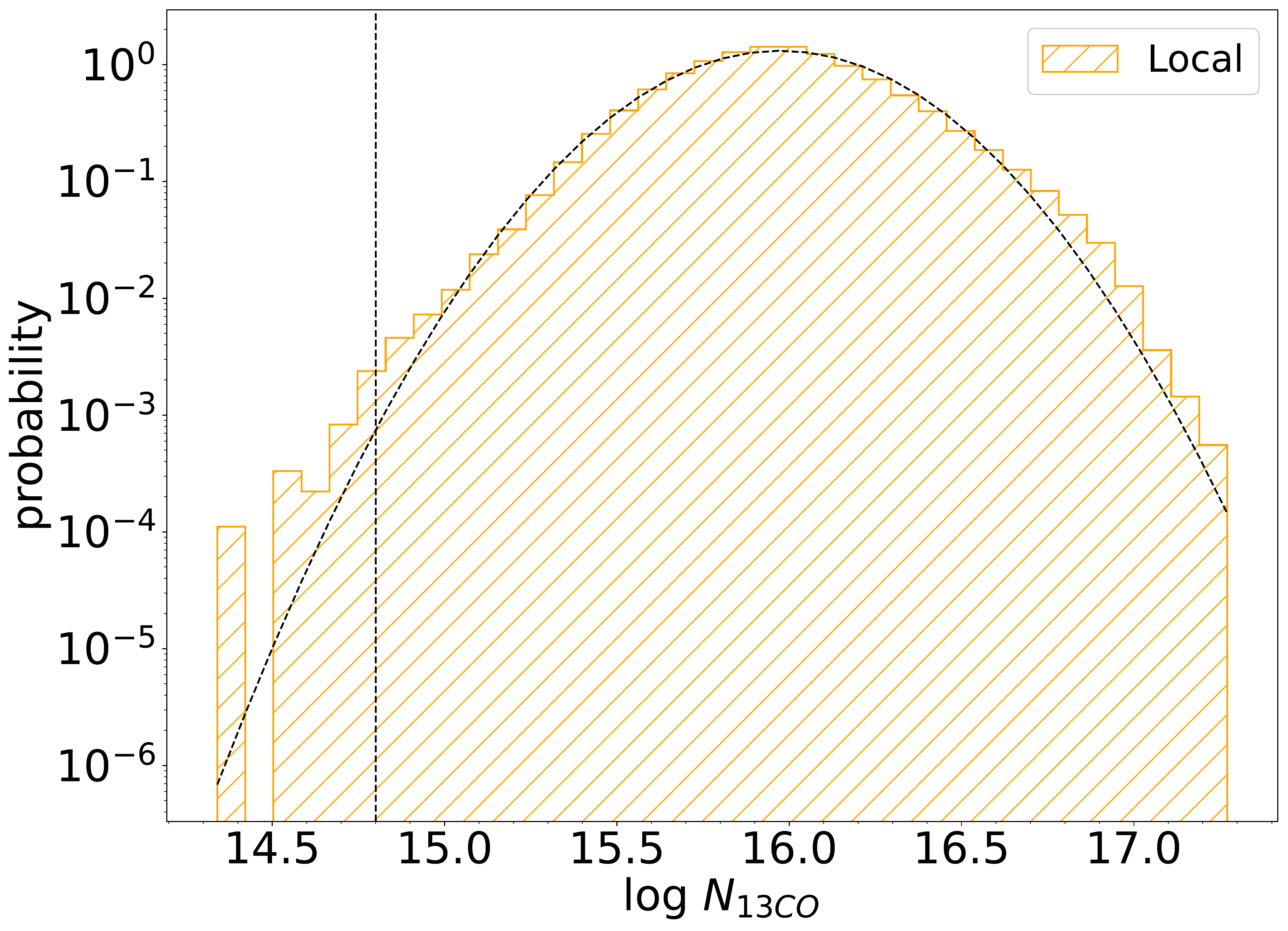}{0.25\textwidth}{(e)}
          \fig{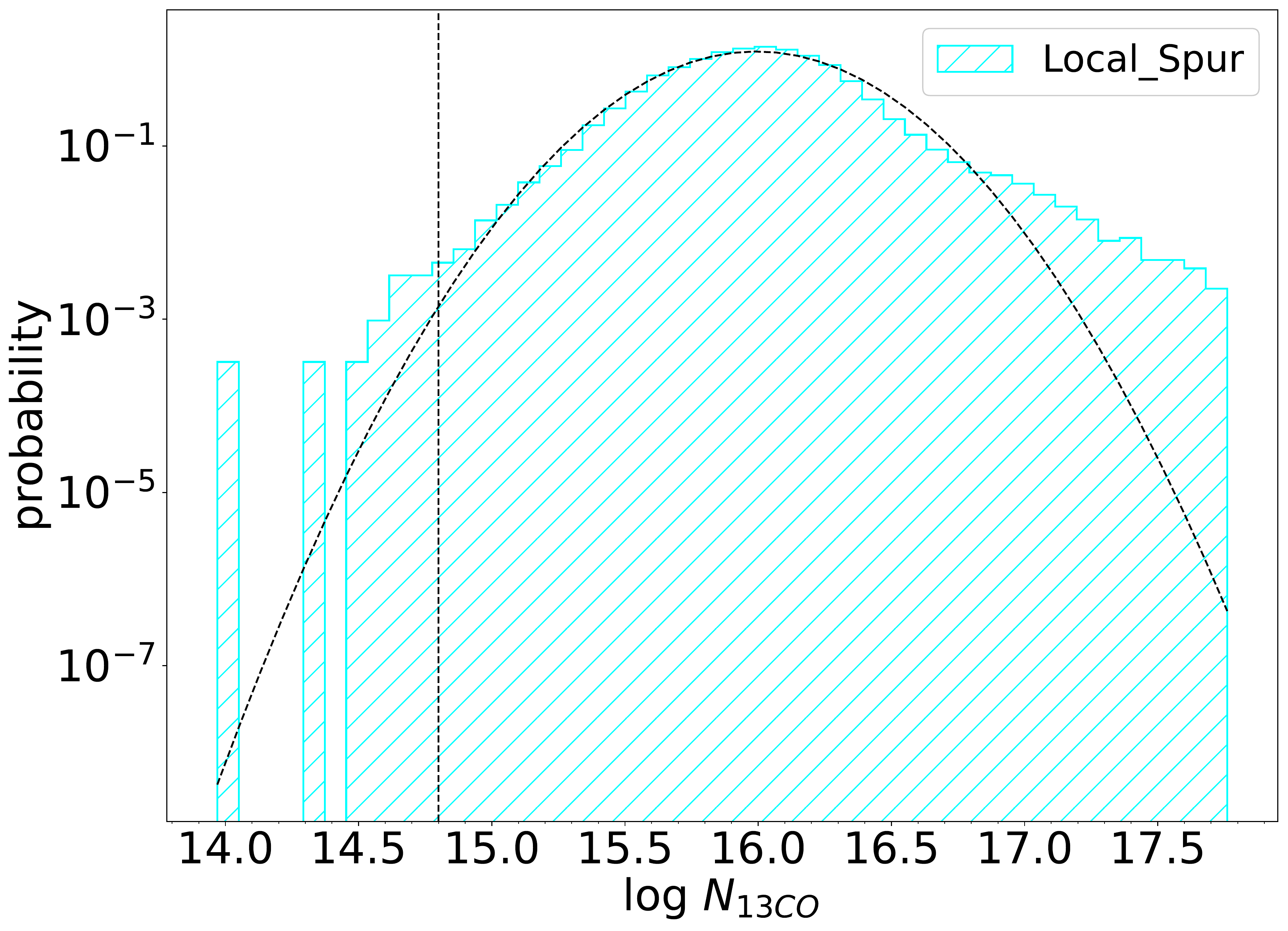}{0.25\textwidth}{(f)}
          \fig{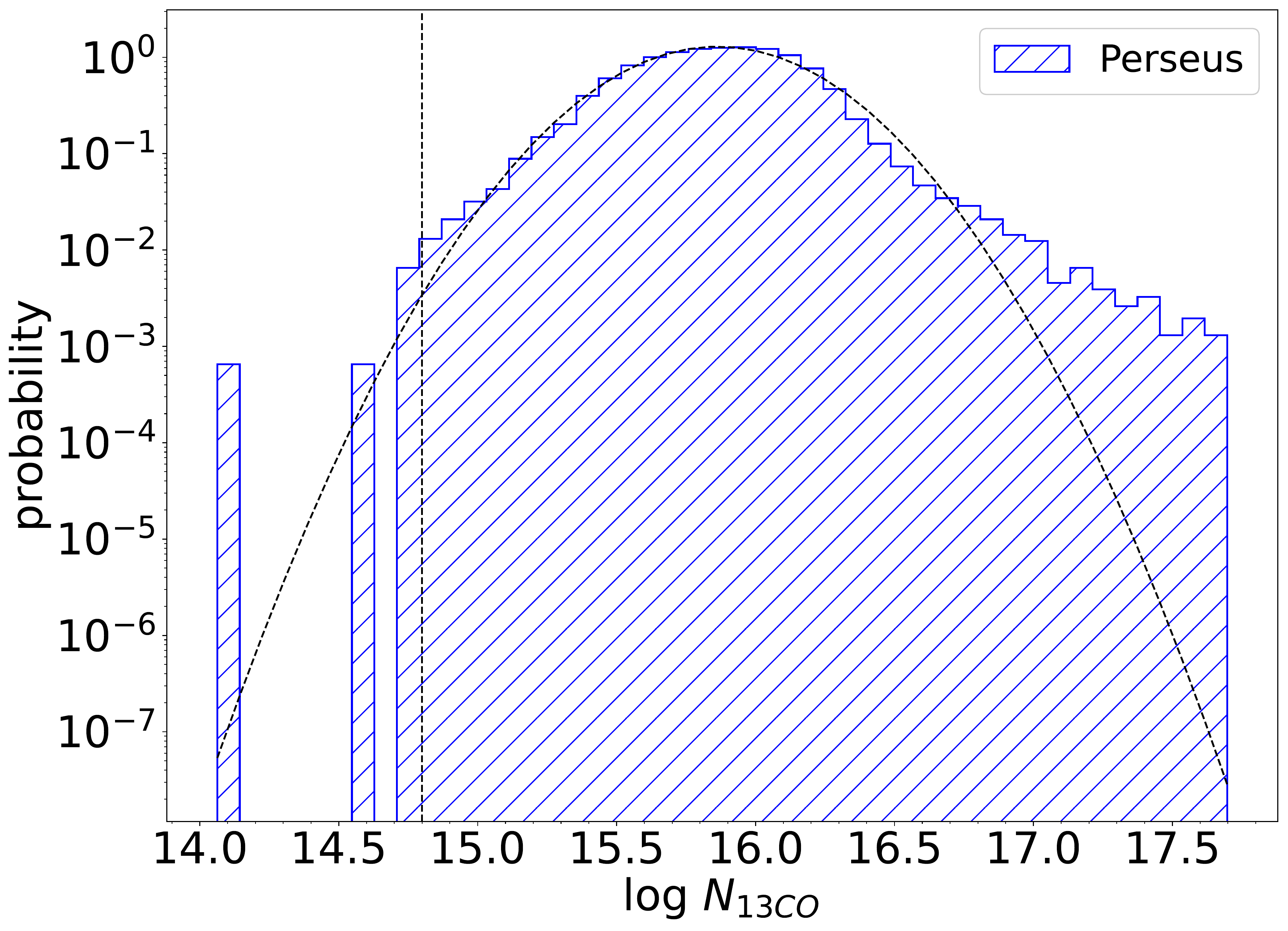}{0.25\textwidth}{(g)}
          \fig{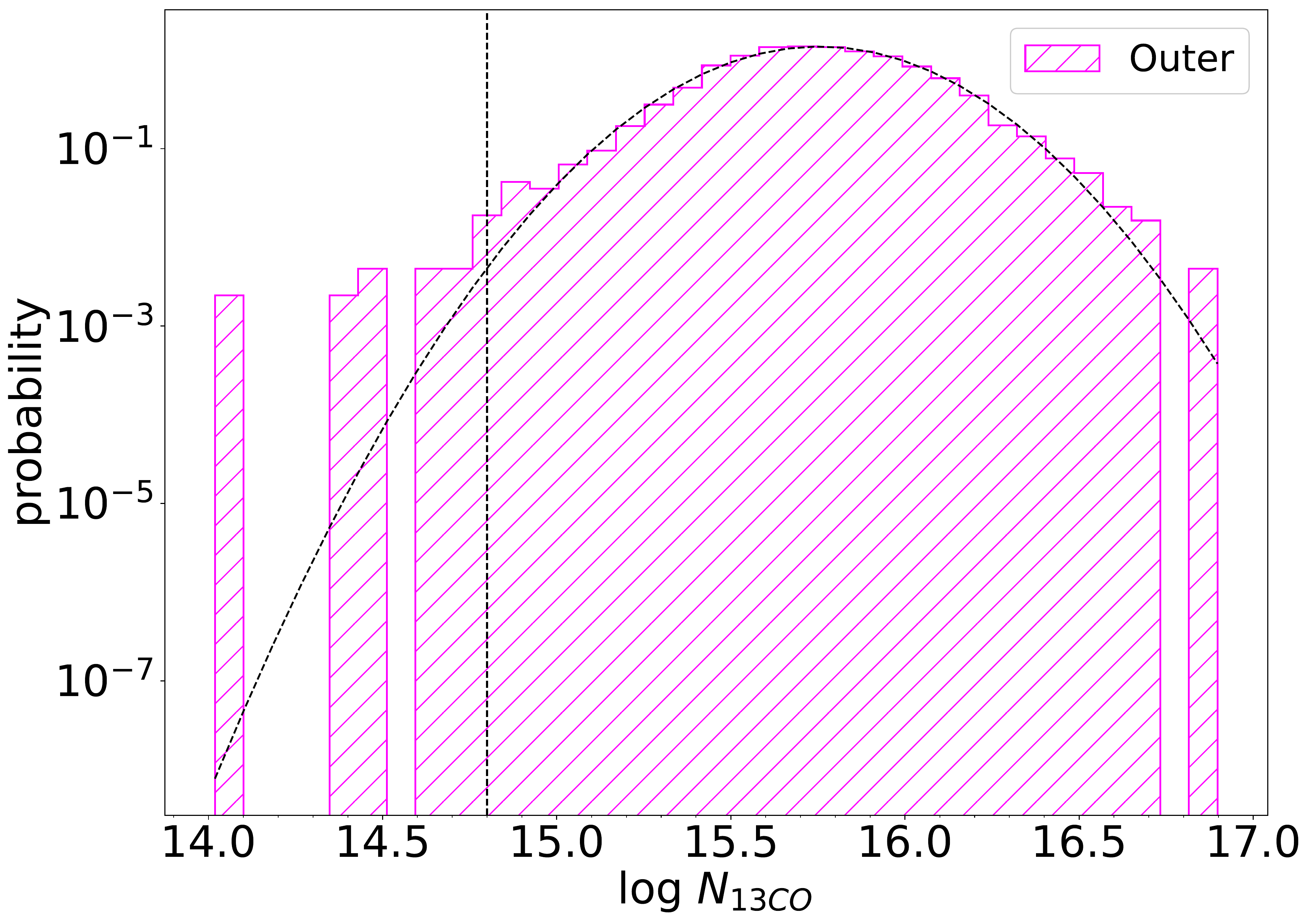}{0.25\textwidth}{(h)}}          
             
\caption{Distributions of $^{12}$CO (top row) and $^{13}$CO (bottom row) column densities for each spiral arm individually in the log-log space, overlaid with log-normal functions fit to the data. The vertical dashed line represents the reference detection limit of the molecular gas column density. }\label{fig:figA3}
\end{figure*}

\appendix
\label{appendix}
The position-velocity diagrams and the integrated intensity maps of the $^{13}$CO line emission are presented in Figure \ref{fig:figA2} and \ref{fig:fig4_13CO}. The velocity channel maps of the $^{12}$CO line emission in each spiral arm and Local Spur are presented in Figures \ref{fig:fig33channel}, \ref{fig:fig33channel_1}, \ref{fig:fig33channel_2}, and \ref{fig:fig33channel_3}. Figure \ref{fig:figA3} presents the distributions of $^{12}$CO and $^{13}$CO column densities for each arm and the fittings with log-normal function are overlaid. Figure \ref{fig:figA1_3} is similar to Figure \ref{fig:fig7}, but for the other 12 extremely distant molecular clouds.

\begin{figure*}          
    \gridline{\fig{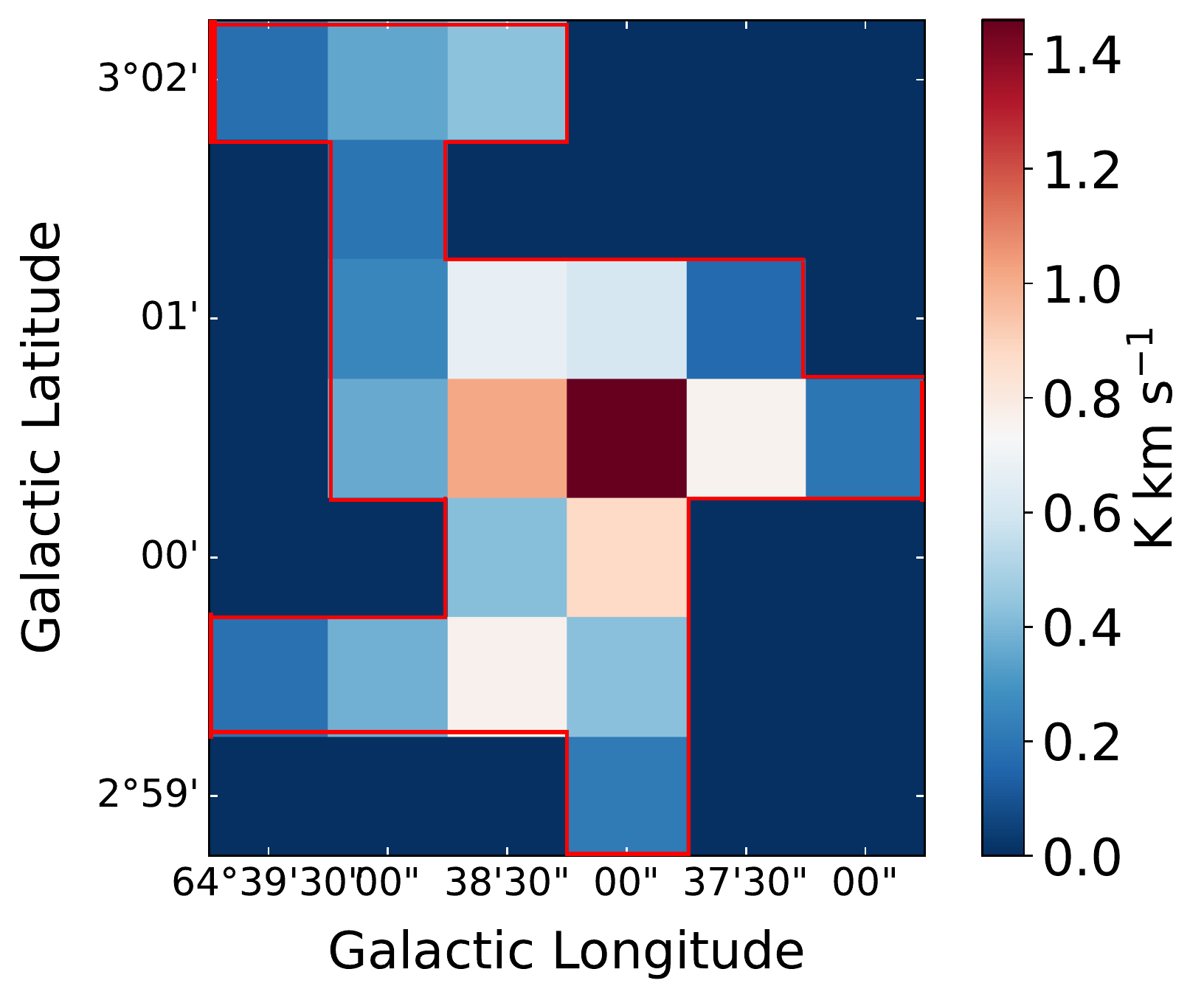}{0.45\textwidth}{(d1)}
              \fig{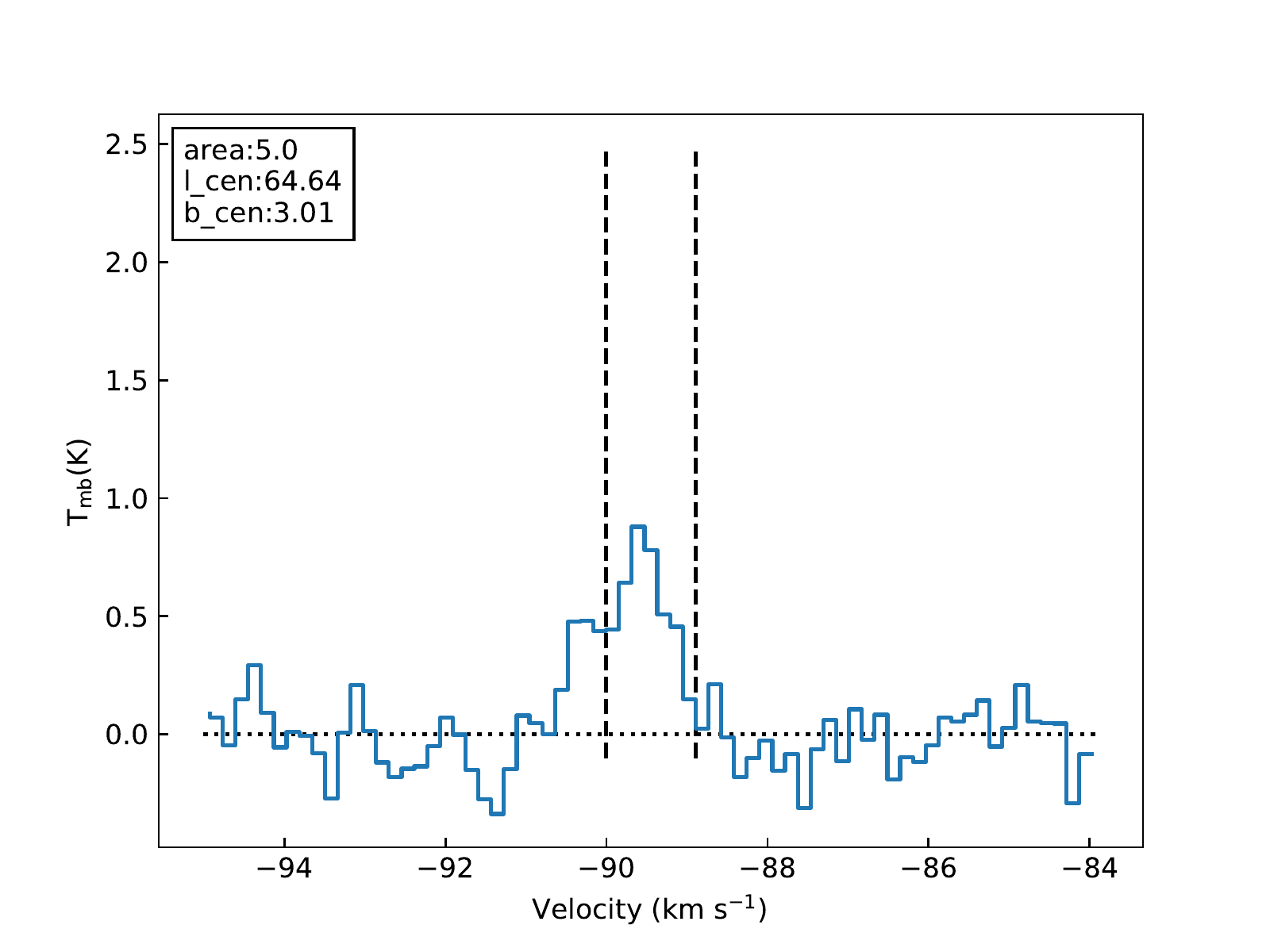}{0.5\textwidth}{(d2)}
              }
              
    \gridline{\fig{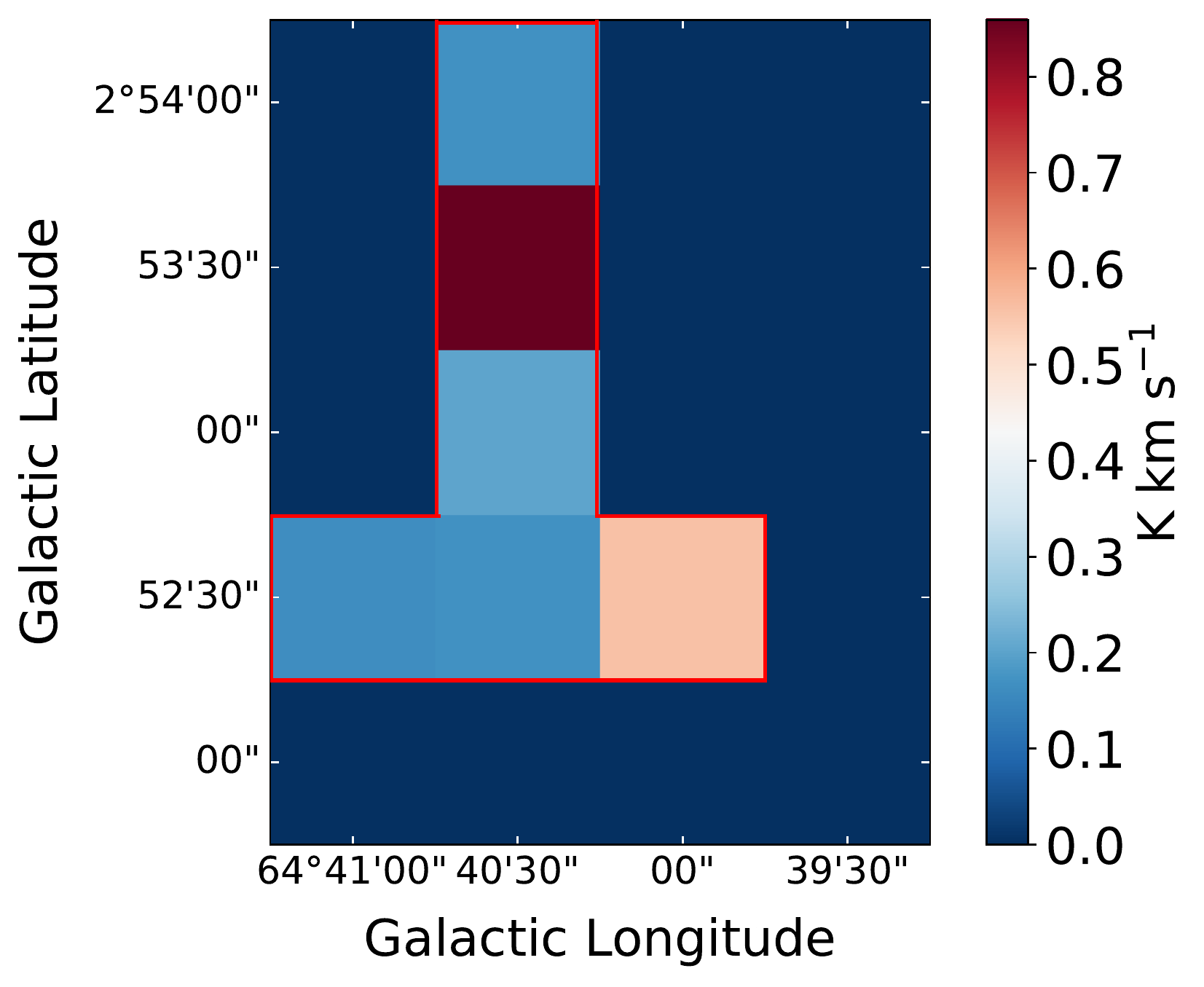}{0.45\textwidth}{(e1)}
              \fig{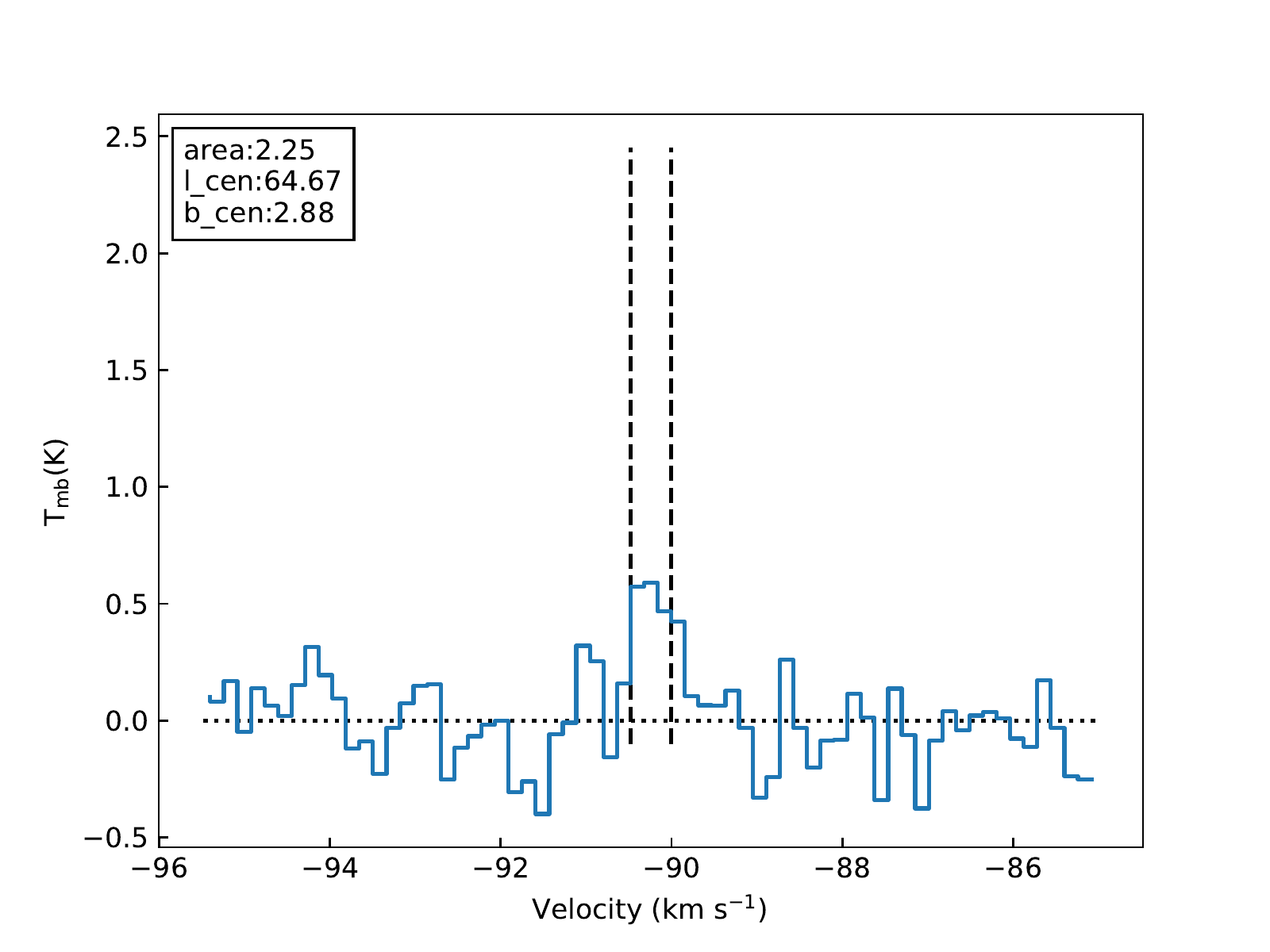}{0.5\textwidth}{(e2)}
              }
              
    \gridline{\fig{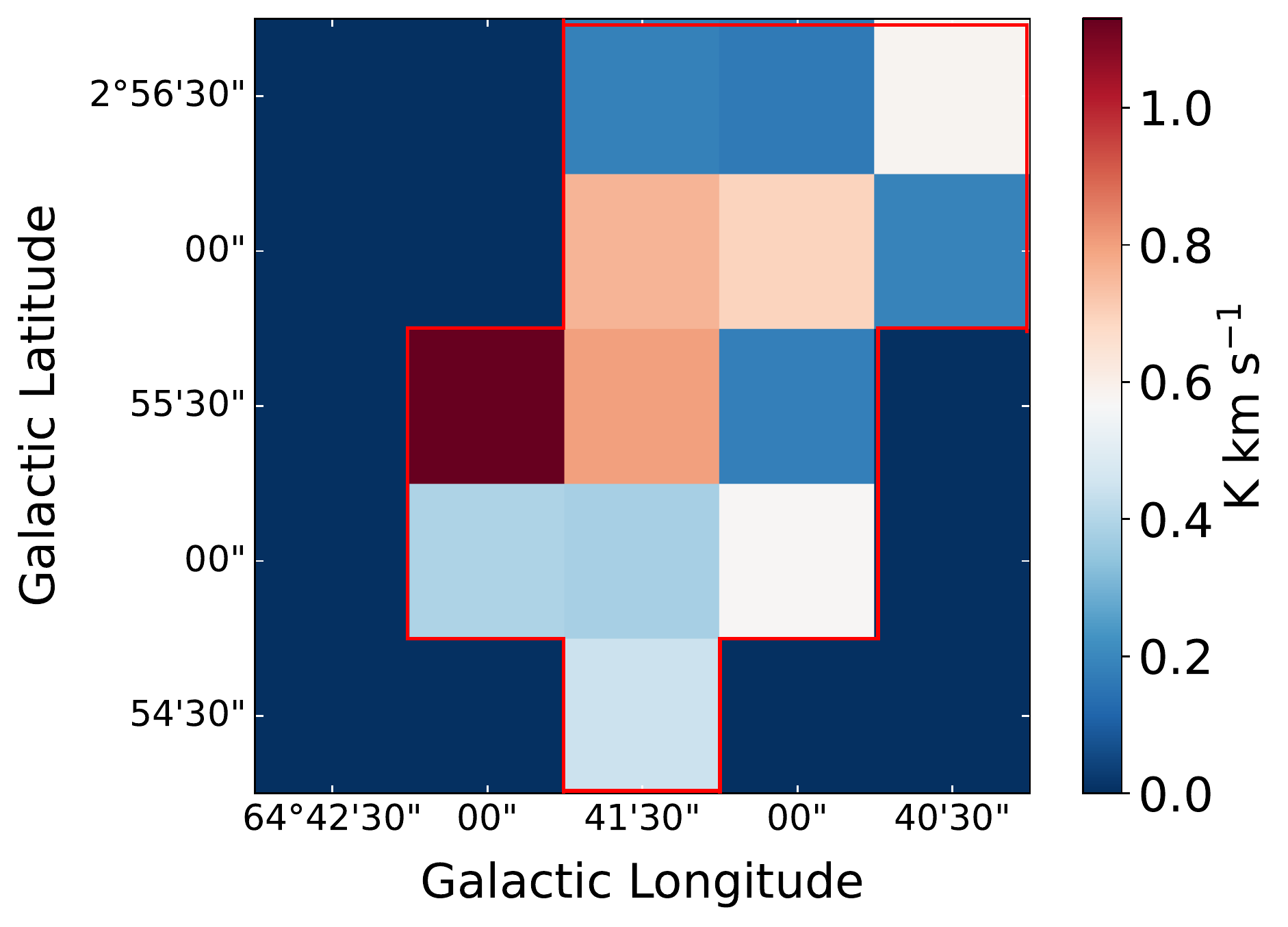}{0.45\textwidth}{(f1)}
              \fig{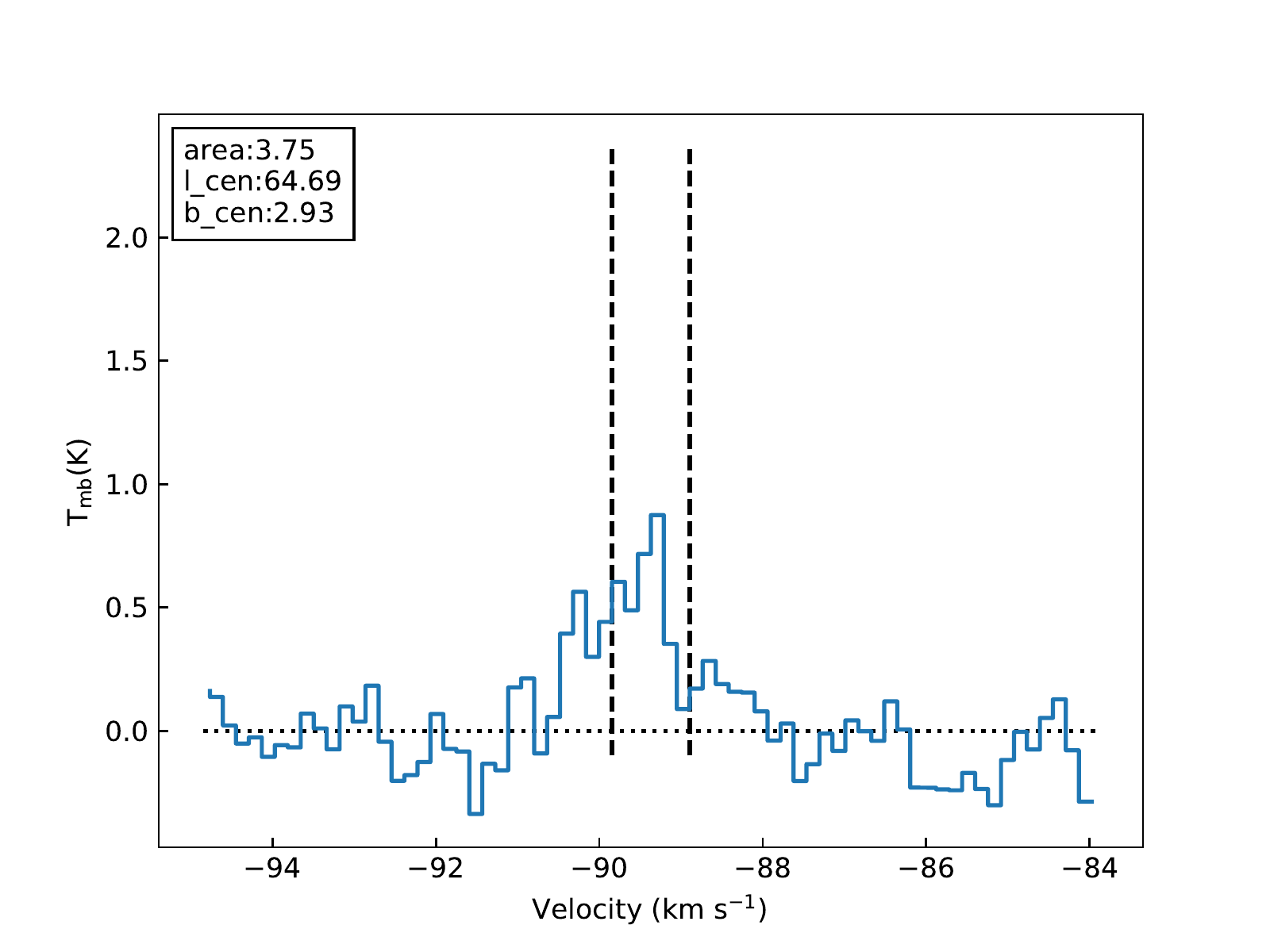}{0.5\textwidth}{(f2)}
              }
    \label{fig:figA1}
    \caption{Same as Figure~\ref{fig:fig7}, but for the other 12 extremely distant molecular clouds in the OSC arm.}
    \end{figure*}

    \addtocounter{figure}{-1}

    \begin{figure*}          
    \gridline{\fig{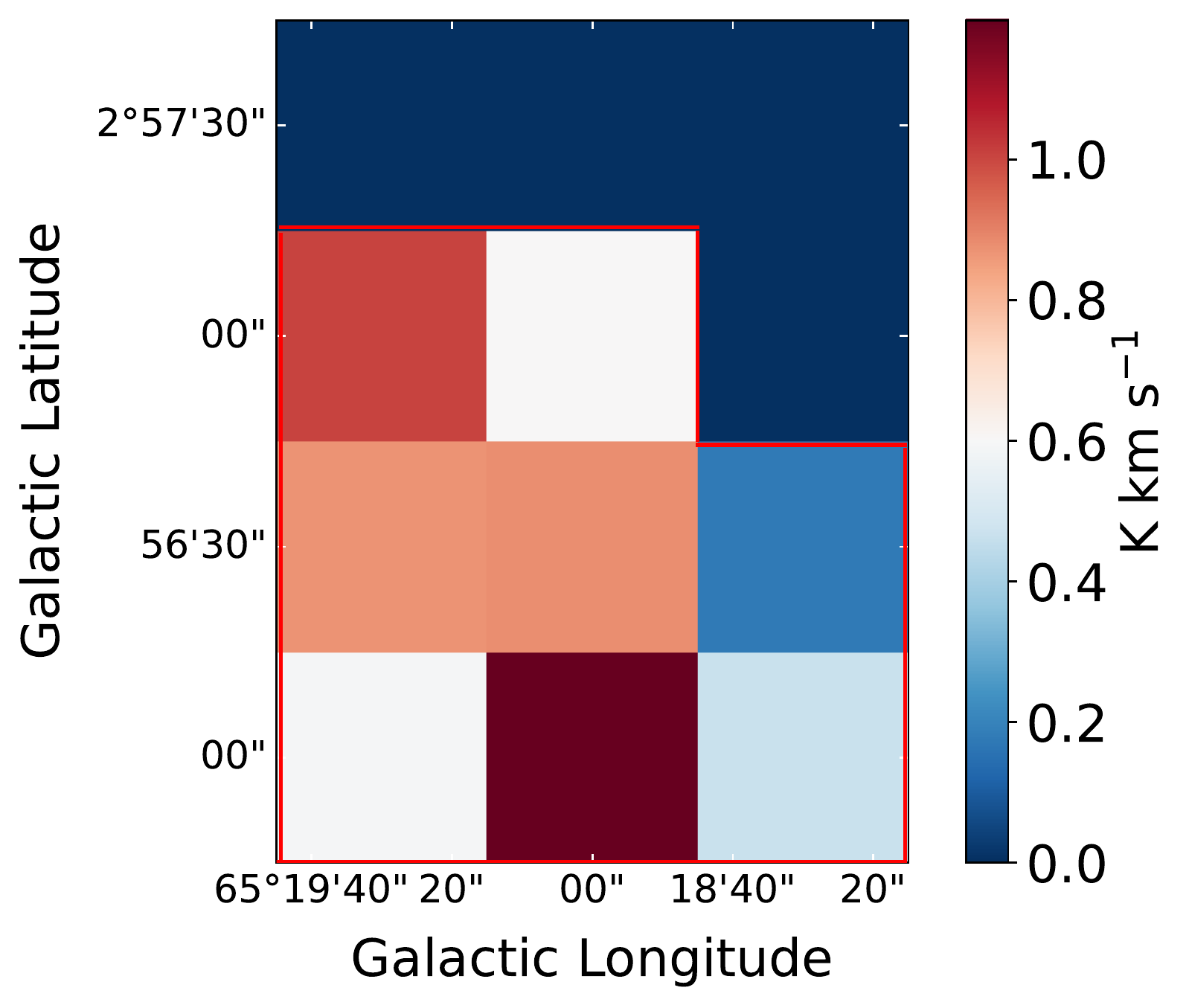}{0.45\textwidth}{(g1)}
              \fig{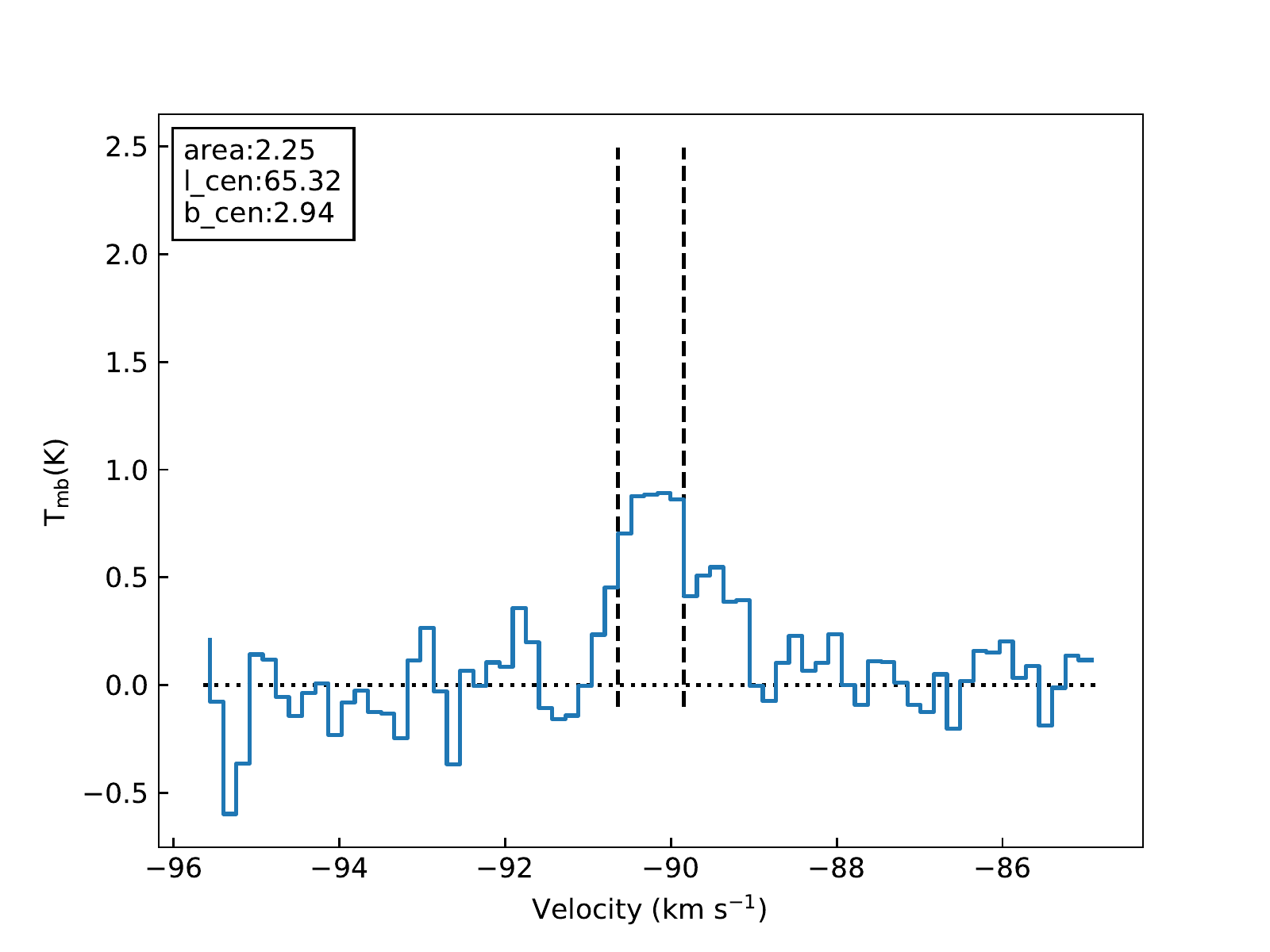}{0.5\textwidth}{(g2)}
              }
              
    \gridline{\fig{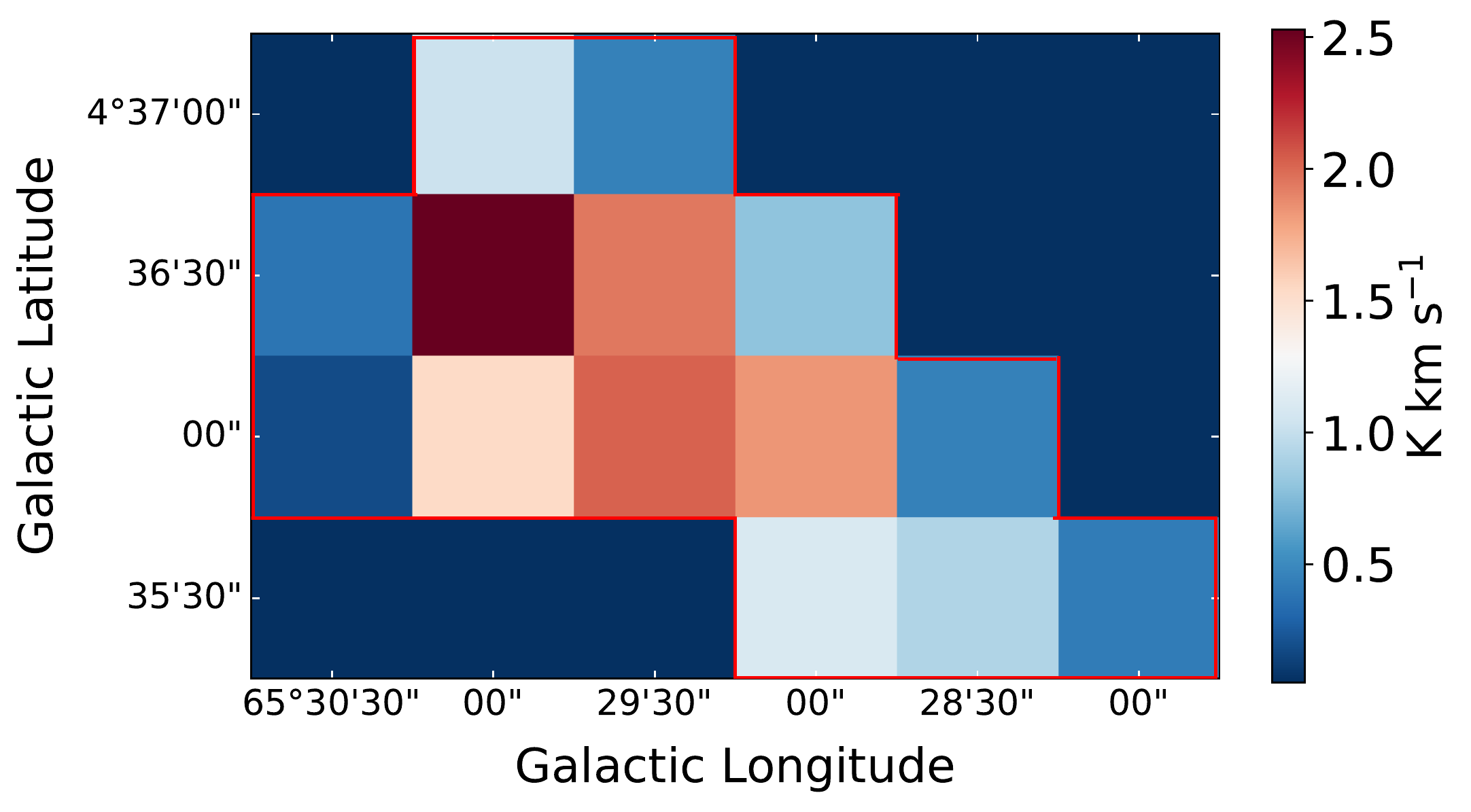}{0.45\textwidth}{(h1)}
              \fig{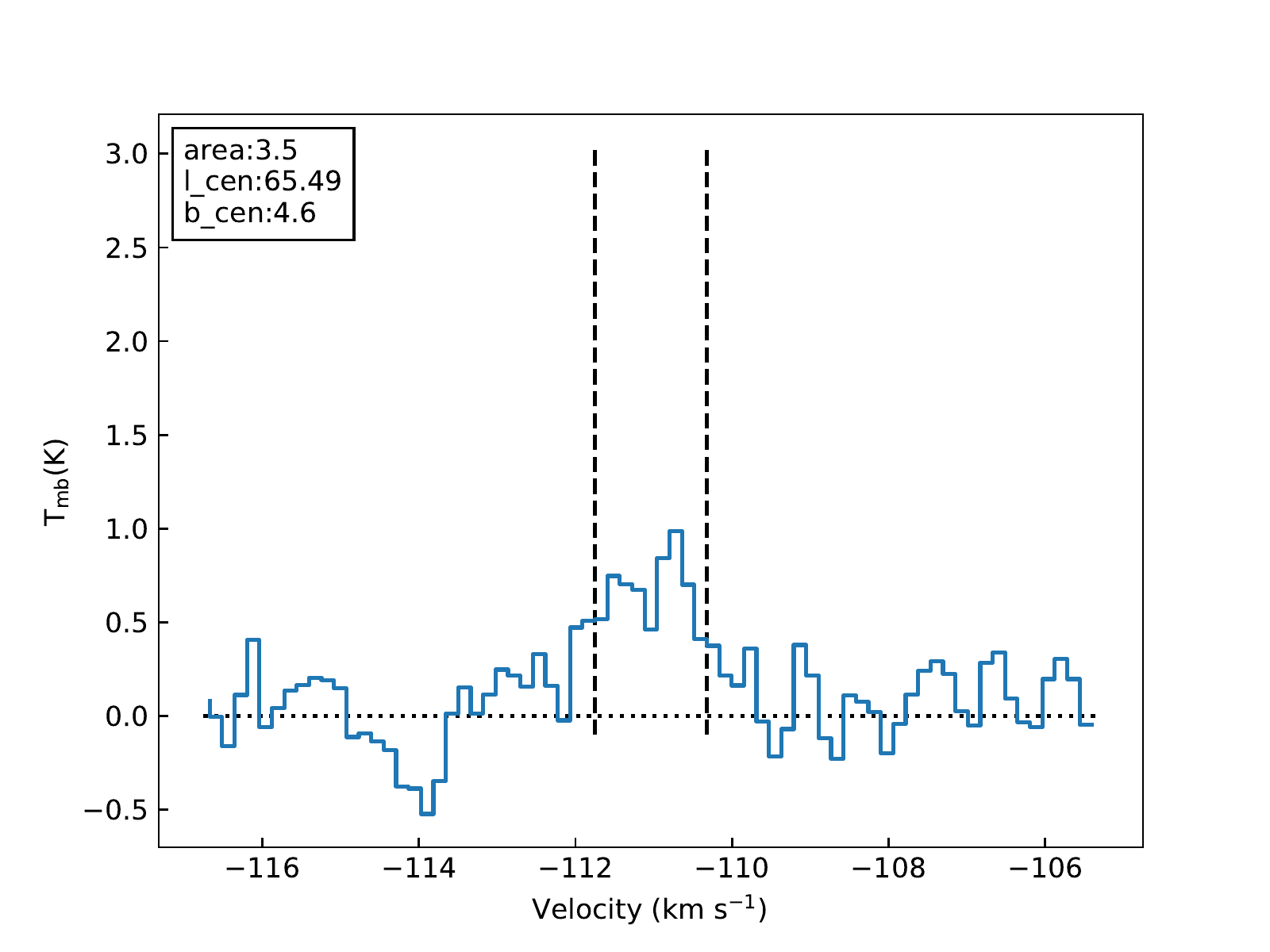}{0.5\textwidth}{(h2)}
              }
              
    \gridline{\fig{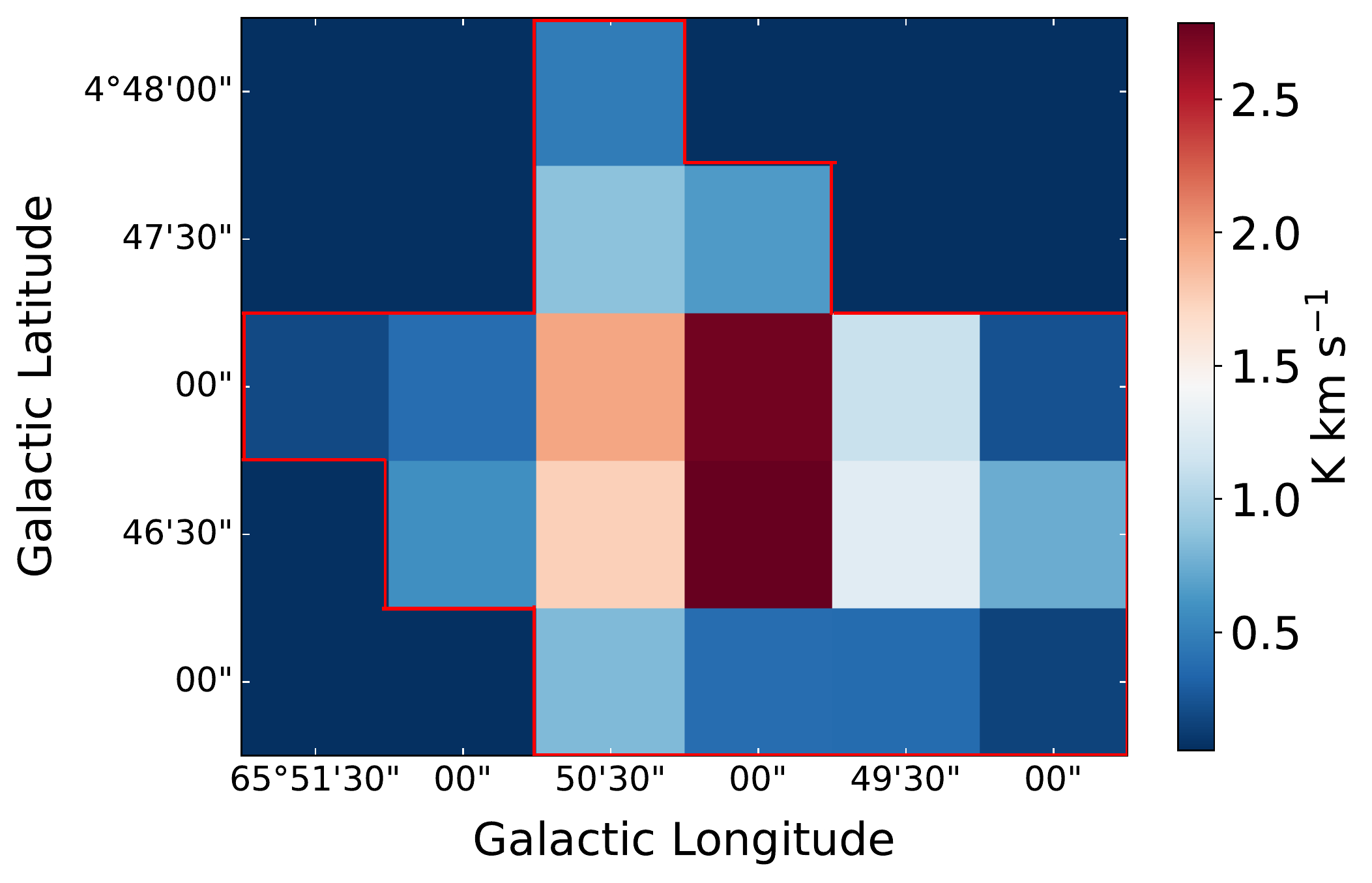}{0.45\textwidth}{(i1)}
              \fig{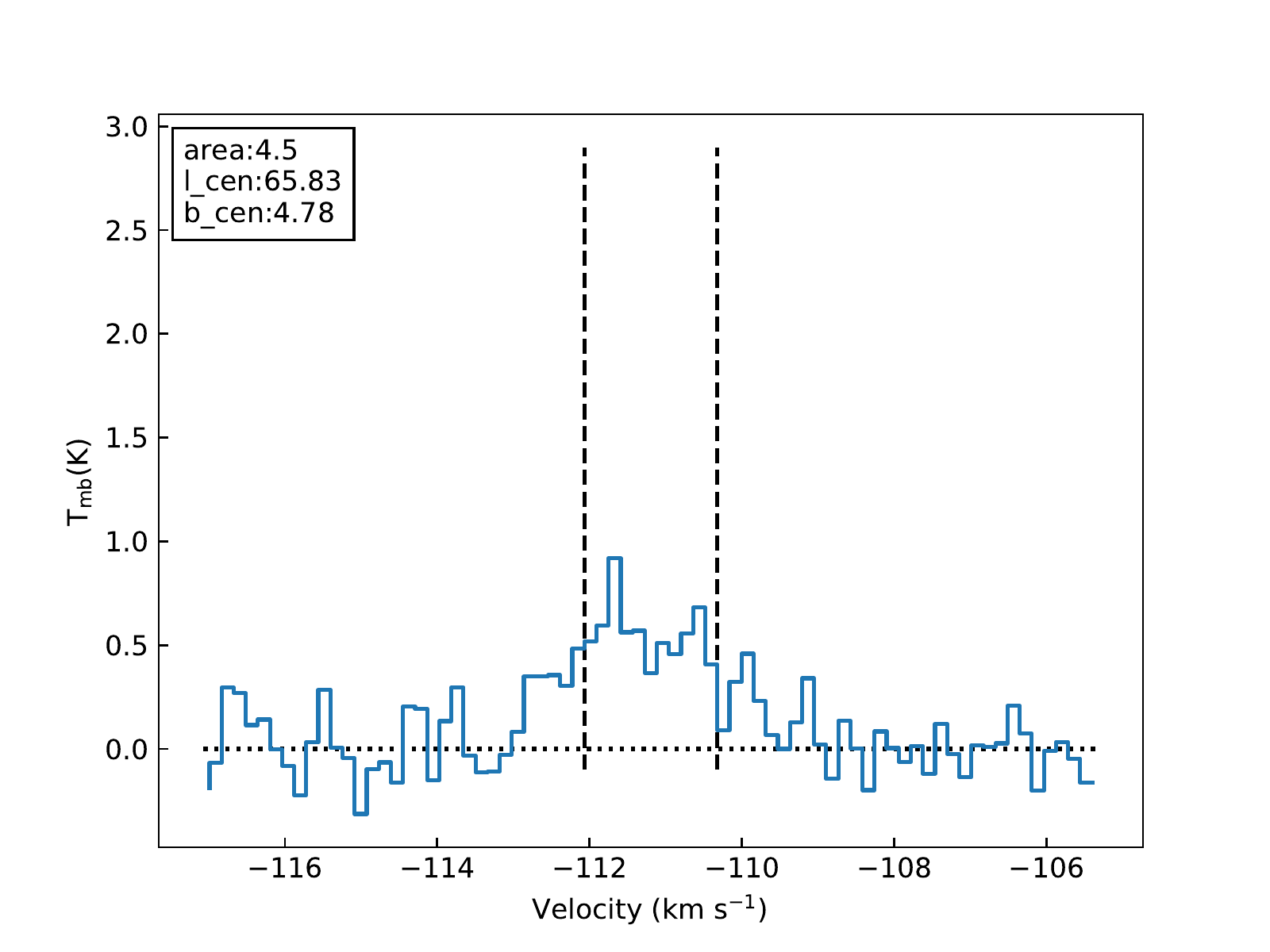}{0.5\textwidth}{(i2)}
              } 
    \label{fig:figA1_1}
    \caption{Continued}
    \end{figure*}

    \addtocounter{figure}{-1}

    \begin{figure*}          
    \gridline{\fig{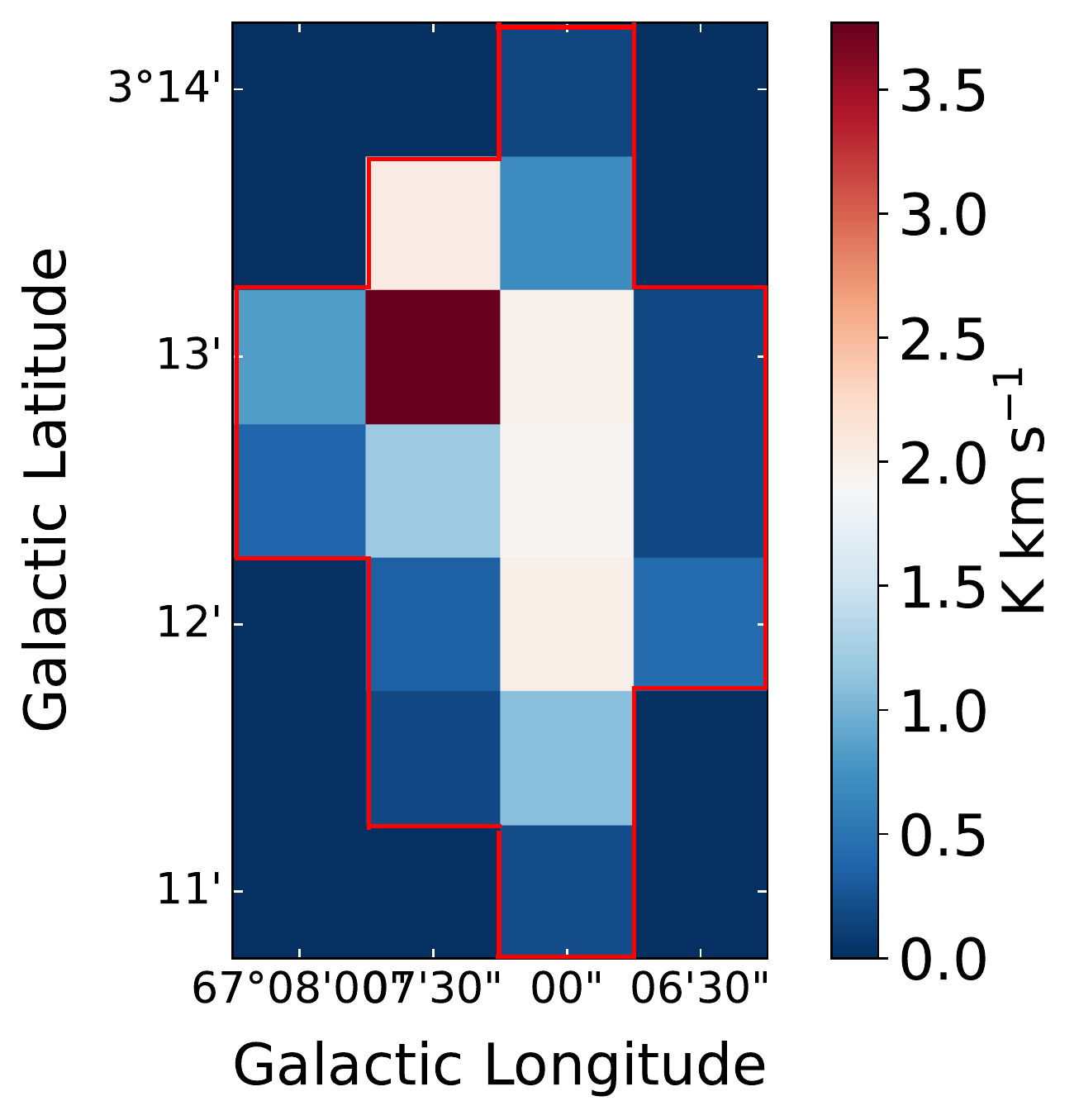}{0.35\textwidth}{(j1)}
              \fig{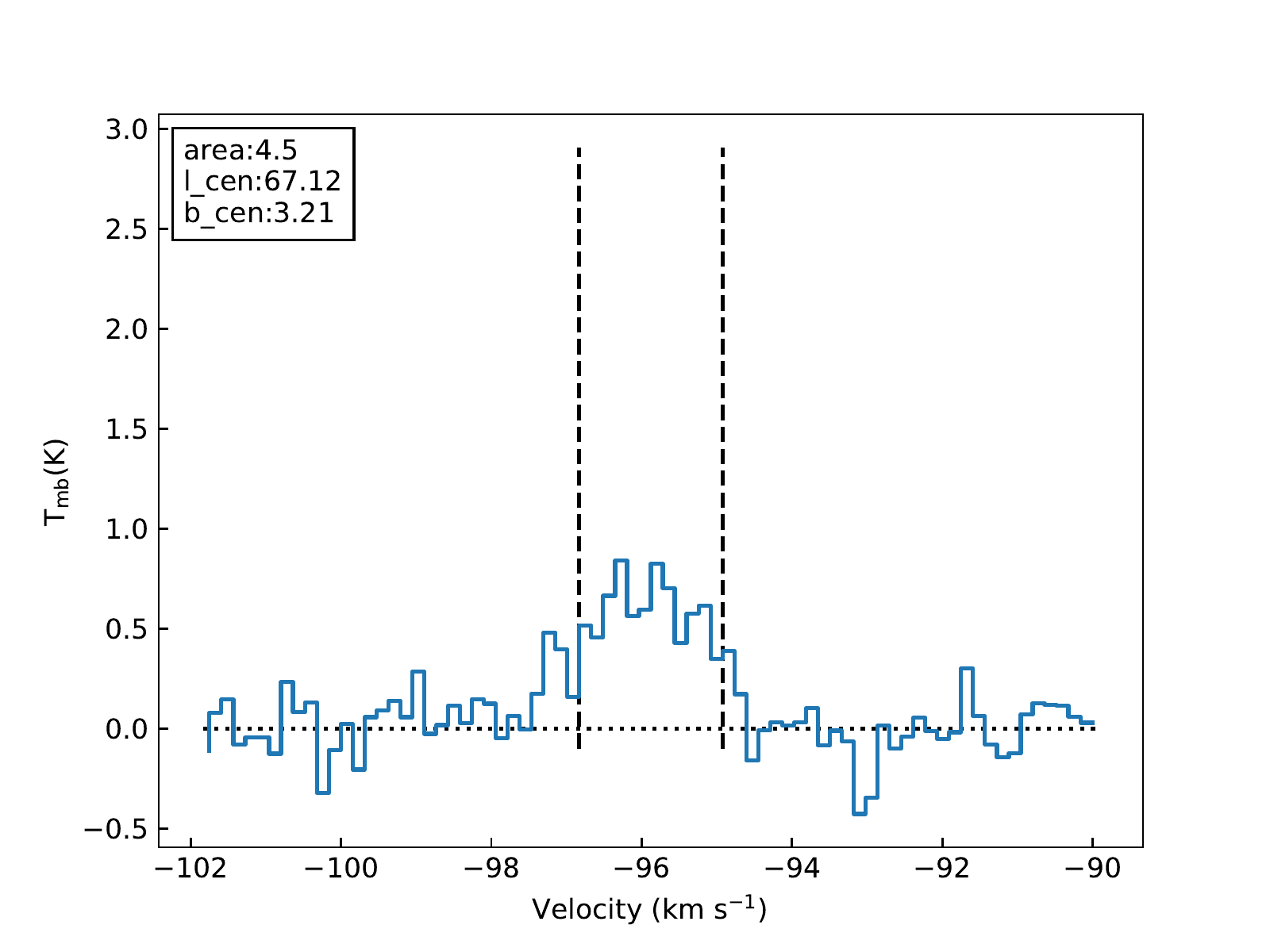}{0.5\textwidth}{(j2)}
              }
              
    \gridline{\fig{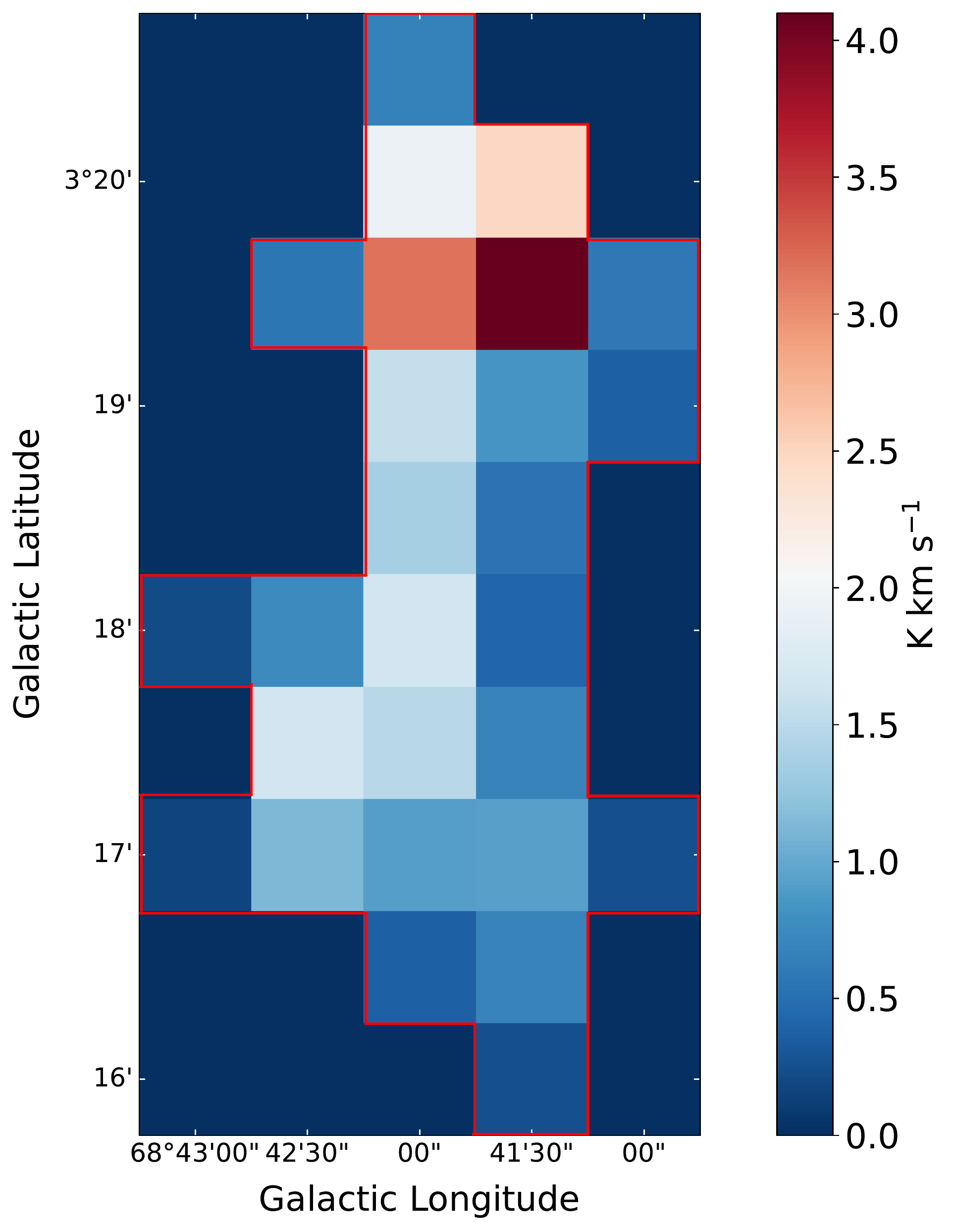}{0.3\textwidth}{(k1)}
              \fig{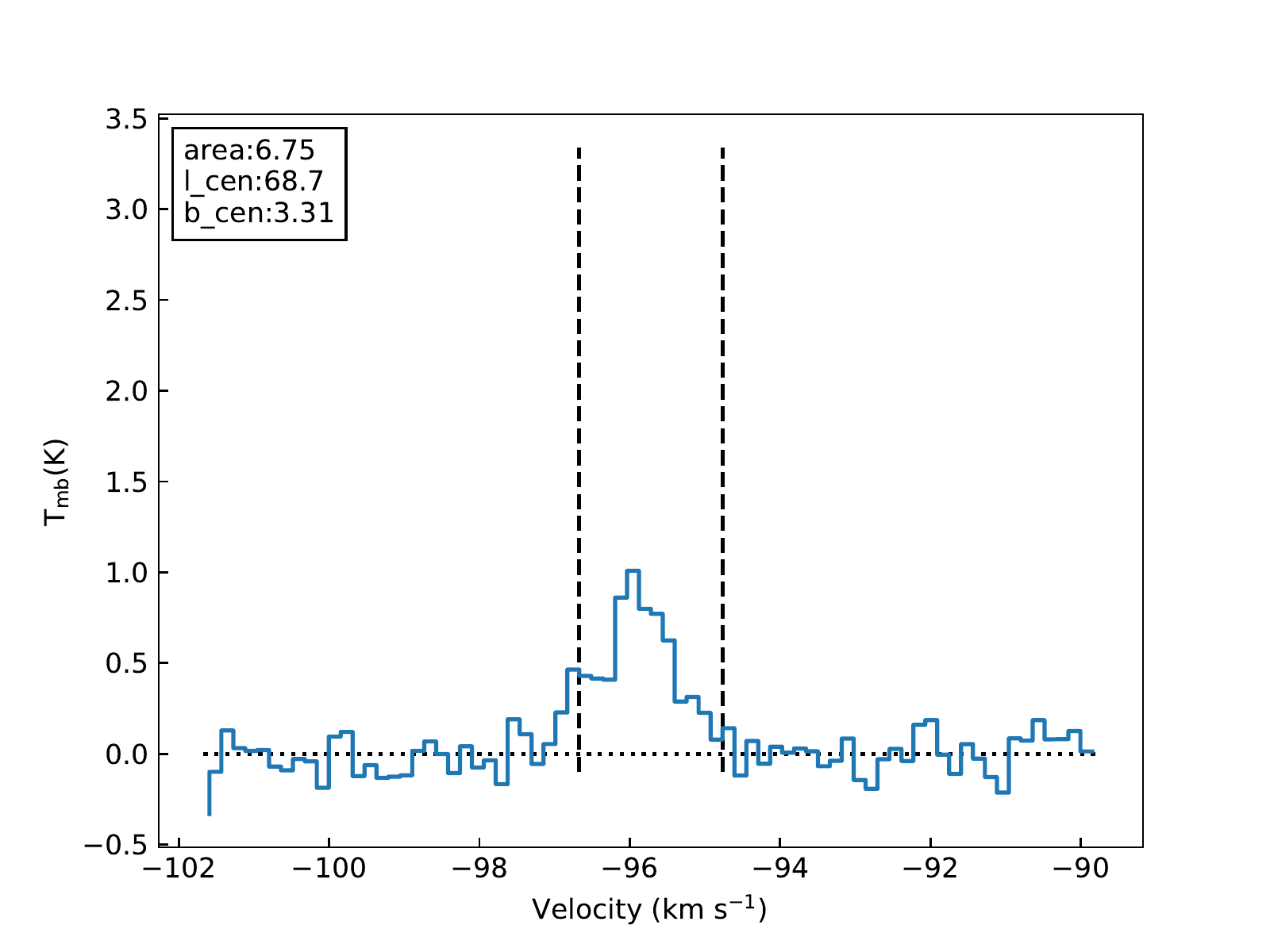}{0.5\textwidth}{(k2)}
              }
        
    \gridline{\fig{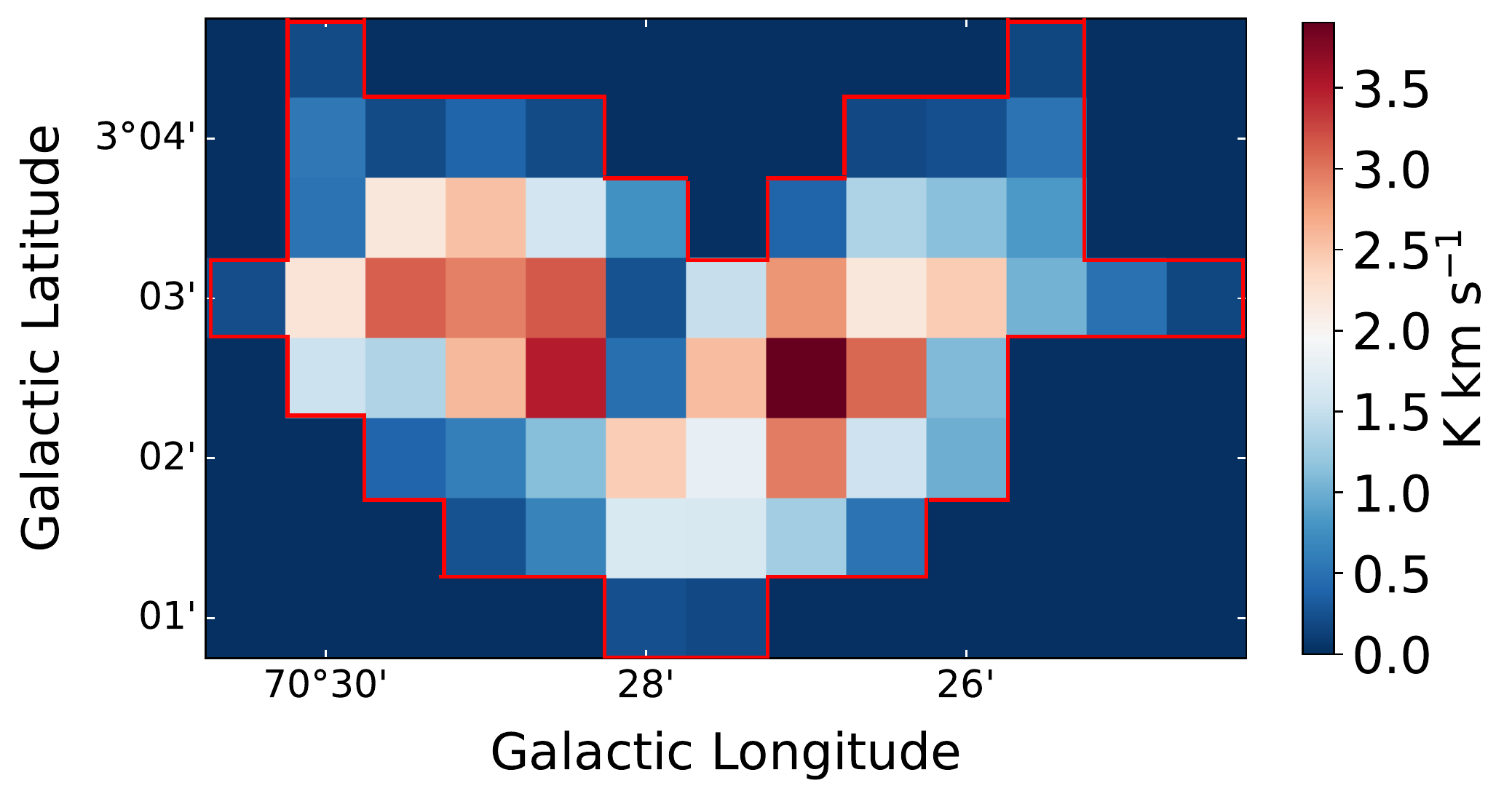}{0.5\textwidth}{(l1)}
              \fig{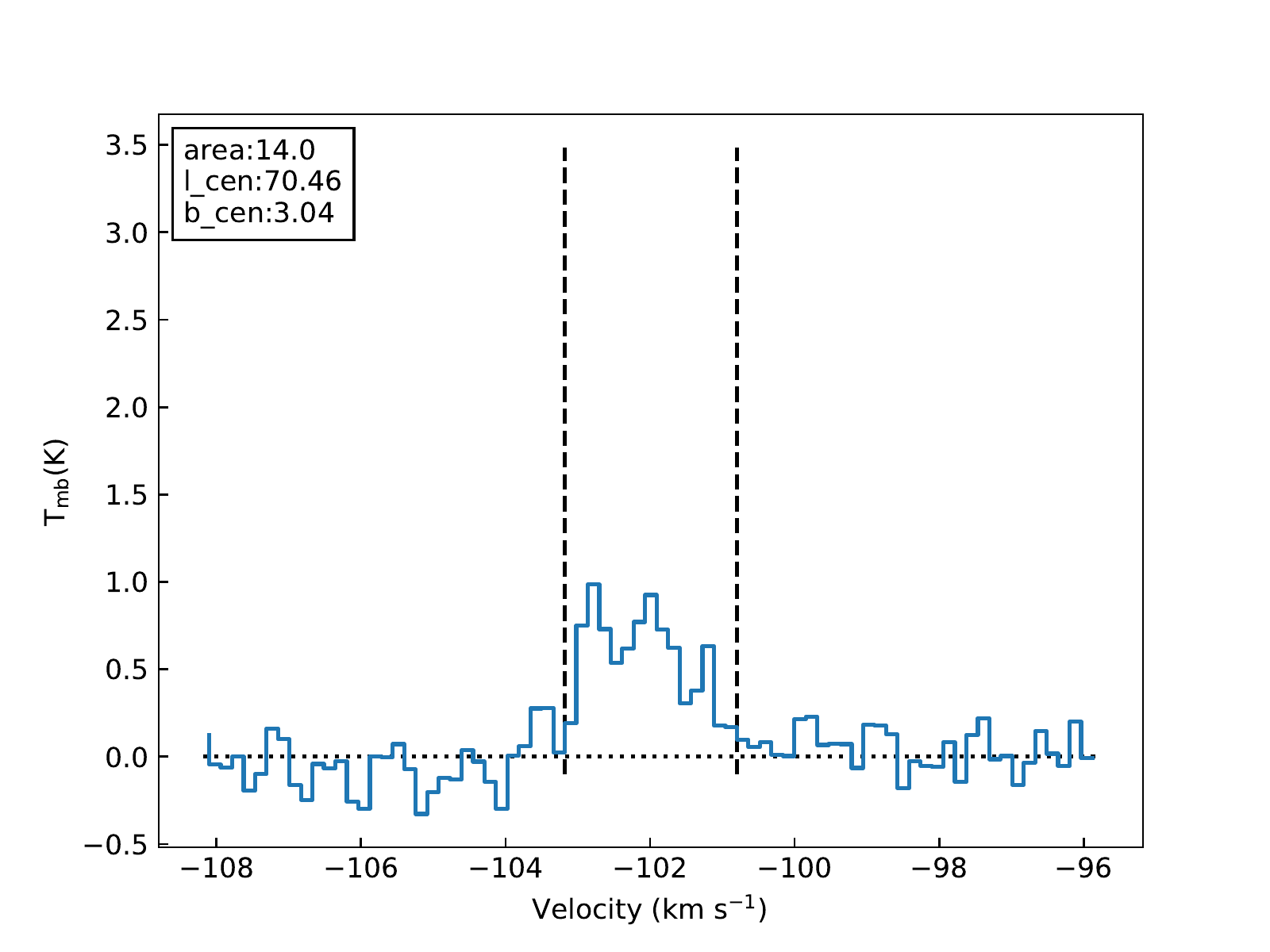}{0.5\textwidth}{(l2)}
              }
    \label{fig:figA1_2}
    \caption{Continued}
    \end{figure*}

    \addtocounter{figure}{-1}

    \begin{figure*}          
    \gridline{\fig{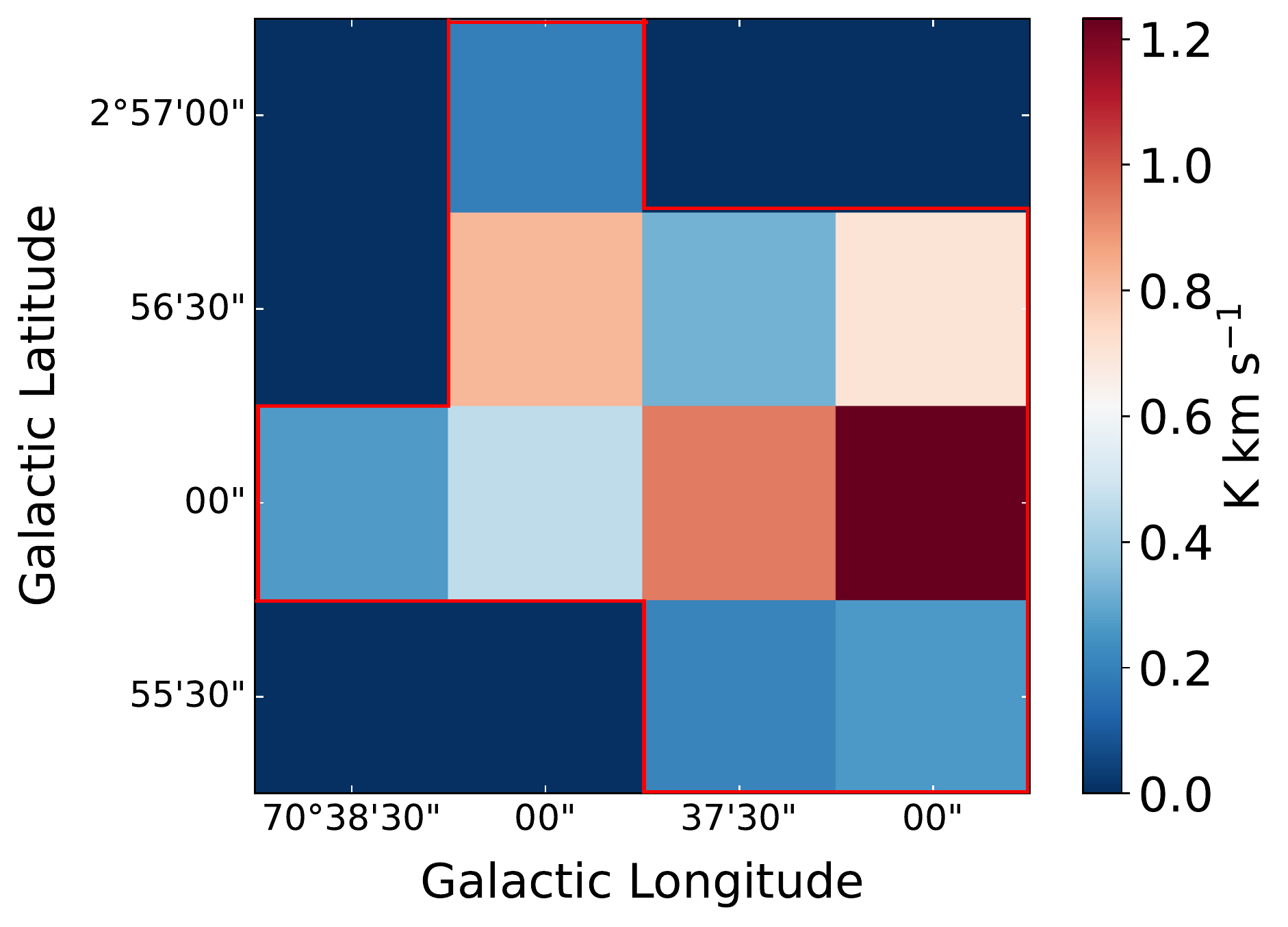}{0.45\textwidth}{(m1)}
              \fig{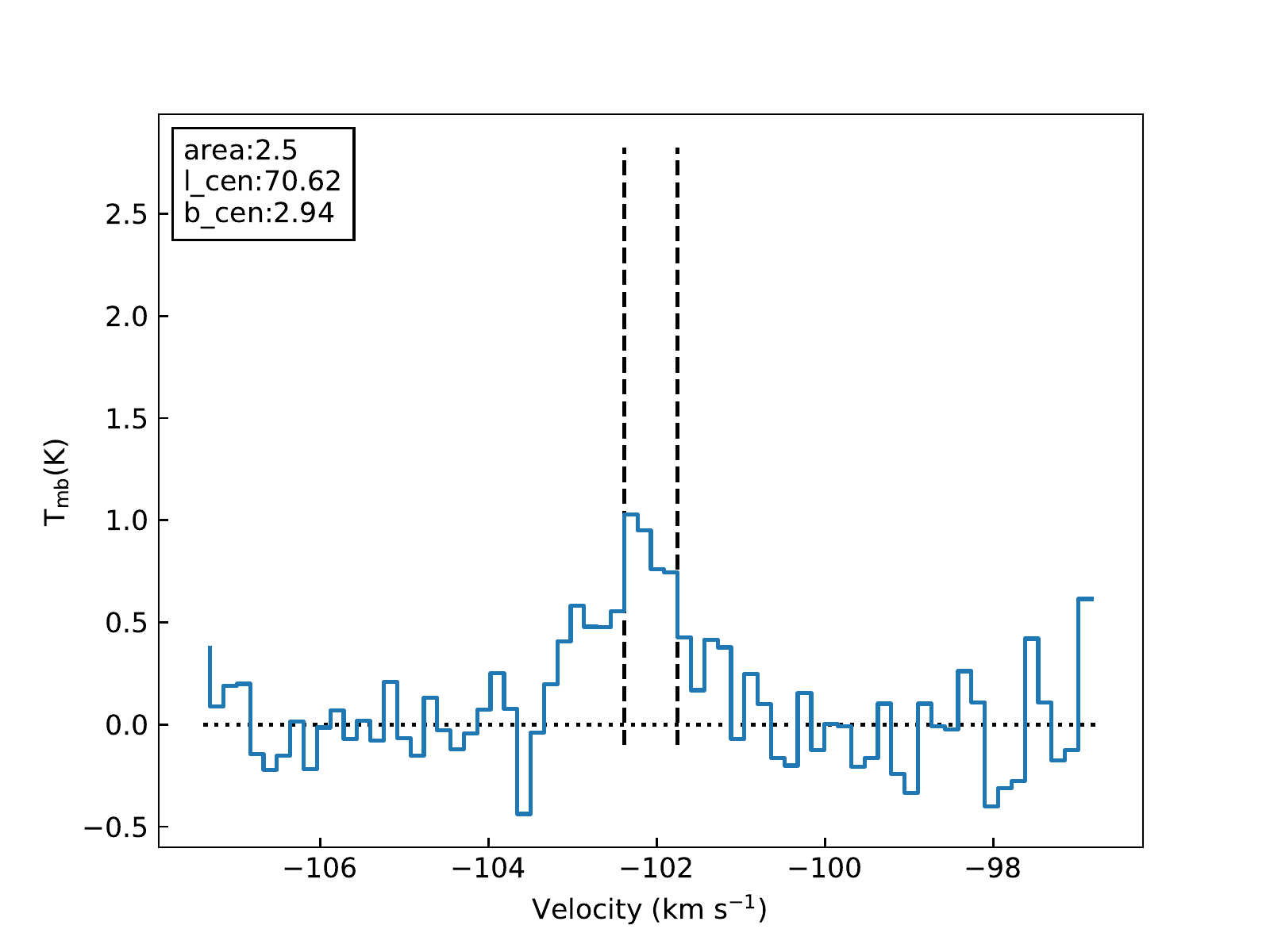}{0.5\textwidth}{(m2)}
              } 
              
    \gridline{\fig{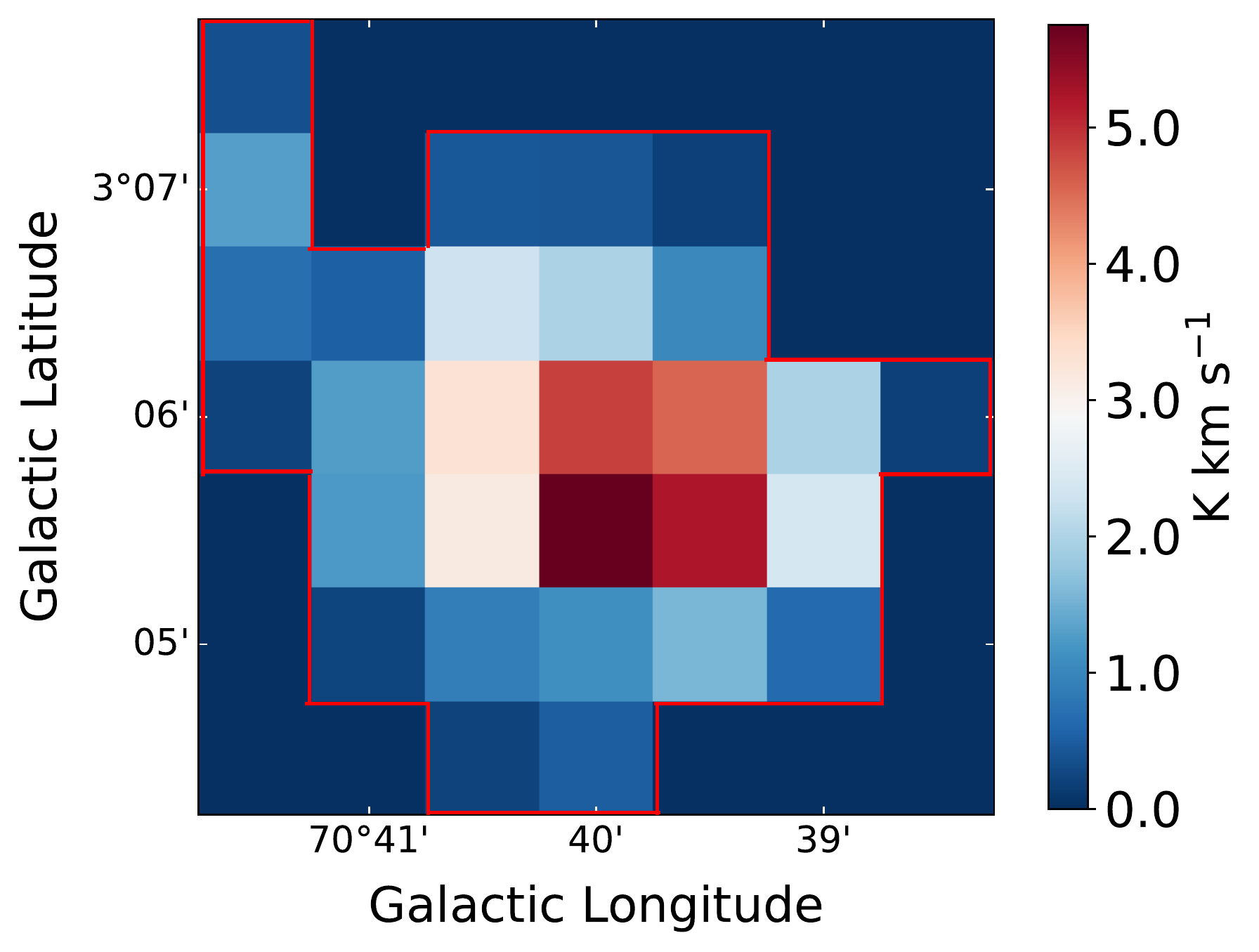}{0.45\textwidth}{(n1)}
              \fig{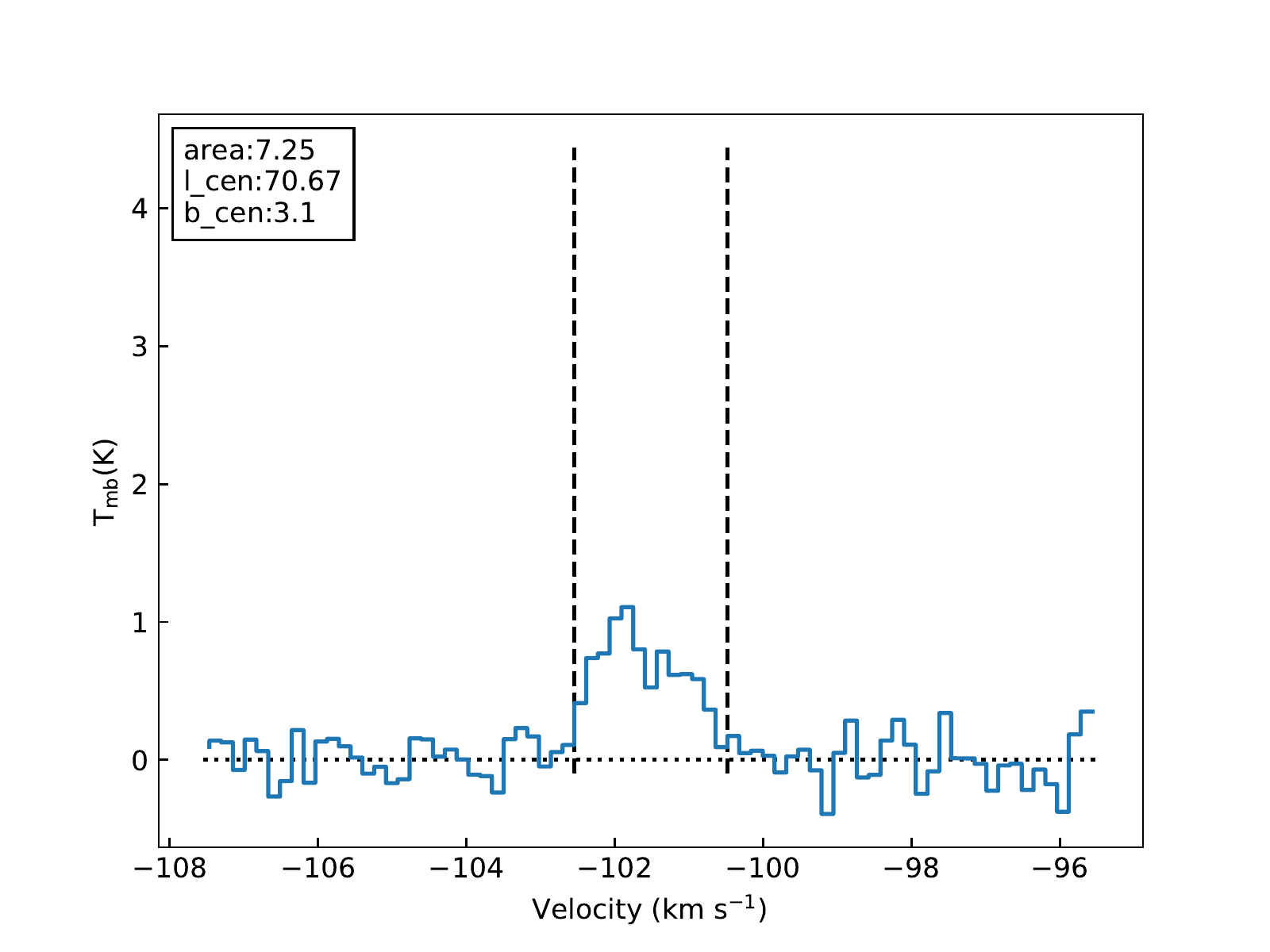}{0.5\textwidth}{(n2)}
              }
              
    \gridline{\fig{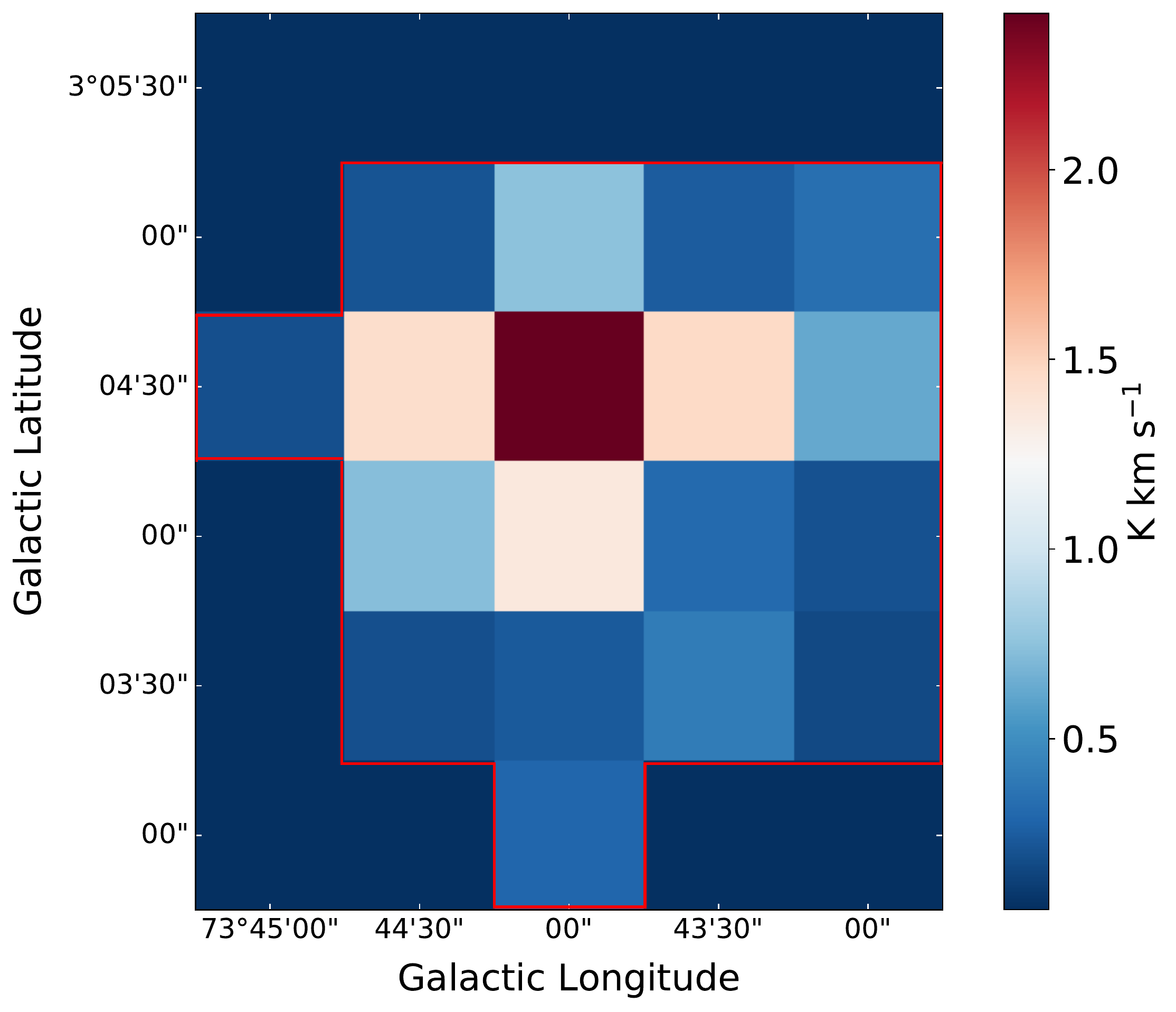}{0.45\textwidth}{(o1)}
              \fig{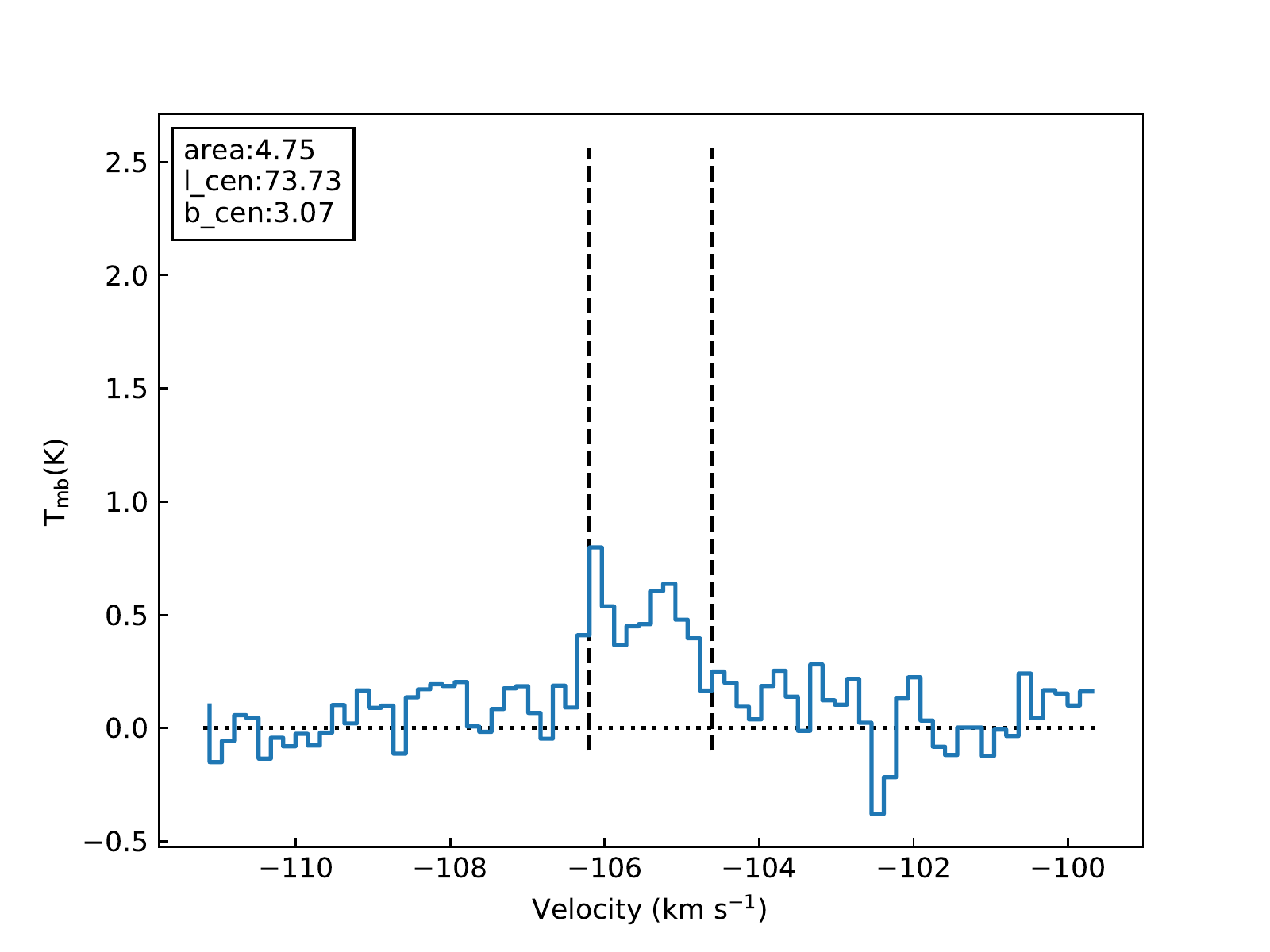}{0.5\textwidth}{(o2)}
              }          
                                                                   
    \caption{Continued}
    \label{fig:figA1_3}
    \end{figure*}

\clearpage





\end{document}